\documentclass[final,3p,times]{elsarticle}
\makeatletter
\def\ps@pprintTitle{%
  \let\@oddhead\@empty
  \let\@evenhead\@empty
  \let\@oddfoot\@empty
  \let\@evenfoot\@oddfoot
}
\makeatother

\usepackage{amssymb}
\usepackage{latexsym}
\usepackage{amsmath}
\usepackage{gensymb}
\usepackage{mathrsfs}
\usepackage{algorithm}
\usepackage{algorithmic}
\usepackage{setspace} % for vertical spacing in algorithm
\usepackage{textcomp} % for arrow
\usepackage{graphicx}
\usepackage{dsfont} % for indicator rv icon
\usepackage{centernot} % to negate any symbol
\usepackage{sectsty}
\usepackage{multirow}
\usepackage{amsthm}
\usepackage[clockwise]{rotating}

\usepackage[title]{appendix}
\usepackage{subcaption} % for subfigures
\usepackage{enumerate}
\usepackage{array} % for Table custom column width
\usepackage[usestackEOL]{stackengine} % for top aligning the table headings
\usepackage[normalem]{ulem} % for striking out texts
\usepackage[dvipsnames]{xcolor}

\newcommand{\bS}{\textbf{S}}

\biboptions{sort&compress}

\usepackage{hyperref}
\hypersetup{
    colorlinks=true,
    linkcolor=blue,
    filecolor=magenta,      
    urlcolor=cyan,
    pdftitle={Bayesian thermal digital twin for a space habitat subjected to an impact event},
    pdfpagemode=FullScreen,
    citecolor=blue,
    pdfborder={0 0 0}
    }

\begin{document}

\begin{frontmatter}



\title{Bayesian thermal digital twin for a space habitat subjected to an impact event}

\author[label1]{Sreehari Manikkan}
\ead{smanikka@purdue.edu}
\author[label1]{Seungho Rhee}
\ead{rhee18@purdue.edu}
\author[label3]{Herta Montoya}
\ead{herta.montoya@utsa.edu}
\author[label1]{Davide Ziviani}
\ead{dziviani@purdue.edu}
\author[label1,label2]{Shirley J. Dyke}
\ead{sdyke@purdue.edu}
\author[label1]{Ilias Bilionis\corref{cor1}}
\ead{ibilion@purdue.edu}
\affiliation[label1]{organization={School of Mechanical Engineering},
            addressline={Purdue University},
            city={West Lafayette},
            postcode={47907},
            state={IN},
            country={USA}}
\affiliation[label2]{organization={Lyles School of Civil Engineering},
            addressline={Purdue University},
            city={West Lafayette},
            postcode={47907},
            state={IN},
            country={USA}}
\affiliation[label3]{organization={Department of Mechanical, Aerospace and Industrial Engineering},
            addressline={UT San Antonio},
            city={San Antonio},
            postcode={78249},
            state={TX},
            country={USA}}
\cortext[cor1]{Corresponding author}

%% Abstract
\begin{abstract}
Space habitats will face a range of disruptive events that can compromise nominal operations and pose serious risks to crew safety. For example, impact events such as micro-meteorite impacts can induce structural damage and cause thermal anomalies in the interior environment. Consequently, the Environmental Control and Life Support System (ECLSS) of the habitat must be resilient and robust against such disruptive events. To support resilience and enable onboard decision making, digital twins have emerged as a promising paradigm. Several studies have focused separately on physics-based modeling and fault detection for ECLSS. This includes data-driven, machine learning-based onboard thermal anomaly detection approaches, which offer strong performance but struggle to generalize beyond the training regime, thereby emphasizing the need for physics-based models. Recent studies have highlighted the need for modeling the thermal dynamics of habitats subjected to micro-meteorite impacts. Prior digital twin research has also emphasized the need for probabilistic representations of system states and parameters to support autonomous operation and uncertainty quantification. However, limited work has attempted to construct a digital twin for the thermal aspects of ECLSS. The objective of this work is to develop a Bayesian thermal digital twin for a habitat cyber-physical testbed experiencing thermal anomalies in its structure and interior environment as secondary effects of an impact event. We construct a coupled thermal resistance capacitance network model that captures both the physical and cyber thermal aspects of the testbed. Furthermore, we embed physics-based activation functions to automate model selection, enable adaptation, and facilitate health-state estimation. Offline Bayesian calibration is performed for the physical subsystem network using experimental temperature data. For the cyber subsystem, we identify the reduction in structural protective layer thickness as the impact sensitive parameter. After fixing the physical parameters and the insensitive parameters, we perform Bayesian inference to continuously estimate the impact-relevant cyber parameters, enabling the detection of impact location, timing, and severity. Synthetic studies are used to examine hyperparameter selection, observability, and noise effects, and the final framework is validated using experimental data from the testbed. The results show that the proposed digital twin can detect impact-induced thermal anomalies, infer impact-relevant parameters with quantified uncertainty, generate informative temperature forecasts, and support time-to-critical estimation for autonomous habitat operation.
\end{abstract}

%% Keywords

\end{frontmatter}

\section{Introduction}
\label{section: introdution}
NASA and the broader space community are preparing for crewed missions to the Moon \cite{askins2026nasa} and Mars \cite{merancy2025moon}, which will require habitation on extraterrestrial surfaces. This vision requires extraterrestrial habitats that must operate under a range of disruptive events that can compromise nominal operations and pose serious risks to crew safety. A major class of hazards in these habitats arises from micro-meteorite and orbital debris impacts, which can induce structural damage and trigger cascading failures.
Thermomechanical cyber-physical testbed studies have emulated damage to the structural protective layer through prescribed changes in its numerical model and have physically imposed the resulting thermal boundary conditions on the habitat testbed. Under these scenarios, measurable time-dependent changes were observed in structural and interior temperatures and in thermal-management operation \cite{montoya2025validation, silva2025cyber, montoya2026acyberphysical}. In addition, communication with ground-based control may experience delays of about 10 minutes for lunar missions and up to 44 minutes for Mars missions \cite{love2014autonomy}. Bandwidth constraints and extended blackout periods further limit Earth-based support. Ensuring crew safety under these conditions requires autonomous health monitoring and decision support systems that can detect, diagnose, and predict failures under significant uncertainty \cite{muralikrishnan2025systems, eshima2020failure, wu2019supporting, pischulti2026twin}. 

The Environmental Control and Life Support System (ECLSS) is a critical subsystem in space habitats that can experience many faults, including unanticipated ones, and therefore must be robust, self-aware, and self-sufficient \cite{escobar2019quantifying, gratius2024twin, eshima2020failure, hwang2023subsystemlevel, zhang2026artificial}. Digital twins, which integrate physics-based and data-driven models with real-time sensor data to provide situational awareness, state estimation, and predictive analytics \cite{nationalacademiesofsciencesFoundationalResearchGaps2023a, kapteyn2021graphical, glaessgen2012twin, ye2020twin, li2017bayesian, wang2026digital, noroozinejadfarsangi2026digital}, have emerged as a promising paradigm for enabling resilient ECLSS \cite{gratius2024twin, gratius2023learned, rollock2025impact, torralba2022estimation, george2023deep, zhang2026artificial} and space habitat \cite{zhang2026artificial, he2026spacea, noroozinejadfarsangi2026digital, pischulti2026twin, schmidt2026architecture}.

Several studies have focused on physics-based modeling \cite{torralba2022estimation} and fault detection for ECLSS subsystems \cite{hwang2023subsystemlevel}, including air leakage \cite{mirfarah2025pressure, rautela2023rapid}, impact hazards \cite{fu2025structural}, power system faults \cite{chebbo2025diagnosis}, sensor faults \cite{wang2024faulta}, and \(\text{CO}_2\) removal system faults \cite{ibrahim2026machine}. Thermal comfort and regulation are integral to ECLSS robustness \cite{benaroya2018habitats}. Researchers have also studied onboard thermal anomaly detection in space systems using data-driven and machine learning approaches \cite{he2026spacea}. Examples include deep learning-based fire detection in ECLSS \cite{xu2021learning}, onboard thermal anomaly detection using machine learning \cite{thoemel2024demonstration}, and edge-computing architectures for real-time support vector machine-based anomaly detection in space systems \cite{moreira2025computing}. While these approaches perform well in pattern recognition tasks, they often lack physical interpretability and struggle to generalize beyond the training data, particularly under unseen operating conditions \cite{wang2026digital}. Consequently, hybrid modeling approaches that combine physics-based formulations with data-driven techniques are emerging as an important direction for improving interpretability and predictive performance \cite{bencze2026aia}.

A key observation from recent studies \cite{montoya2025validation, silva2025cyber, montoya2026acyberphysical} is that thermal anomalies induced by structural damage resulting from a meteorite impact event manifest through temporal evolution rather than instantaneous changes. For instance, such anomalies may appear as gradual changes in temperature trends or slopes due to altered heat transfer mechanisms caused by compromised insulation and leakage effects. This observation highlights the need for digital twins that dynamically model impact events and capture their secondary effects on thermal dynamics rather than merely detecting the impact itself \cite{noroozinejadfarsangi2026digital}. Such a digital twin should enable probabilistic estimation of anomaly presence and severity, prediction of system evolution, and time-to-critical (TTC) assessment with quantified uncertainty, thereby providing confidence levels for its estimates and supporting informed decision-making.

Digital twin frameworks have been proposed for ECLSS, emphasizing the integration of heterogeneous subsystem models and the need for probabilistic representations of system states and parameters to support autonomous operation \cite{gratius2024twin, gratius2026multimodela}. Prior work has also demonstrated the use of probabilistic graphical models for digital twins in space habitats and highlighted the importance of uncertainty quantification and data-driven model calibration \cite{gratius2024twin, gratius2023learned, gratius2024calibration}. Limited work has attempted to construct a digital twin for the thermal aspects of ECLSS \cite{he2026spacea}. One important example is the work of Gratius et al. \cite{gratius2023learned}, where the authors construct a probabilistic graphical model-based digital twin for interior environment temperature prognosis under a temperature control failure scenario. However, existing works do not demonstrate the construction of digital twins by developing physics-based dynamical models of disruption events with adaptation capabilities and enabling the estimation of health state and TTC from multiple sensor measurements with quantified uncertainty.

In this work, we present a Bayesian thermal digital twin built on a lumped parameter model. We focus on a habitat that experiences a thermal anomaly in the interior environment as a secondary cascading effect induced by an impact event damaging the structural protective layer (SPL). We consider the Human-centered Autonomous Resilient Space Habitat (HARSH) \cite{silva2025cyber}, a cyber-physical testbed representing one realization of a smart habitat, as the physical asset for the thermal digital twin. We construct resistance capacitance (RC) thermal network models for the physical and cyber subsystems separately and then combine them to form the digital twin model. By embedding physics-based activation functions into the digital twin model, we automate model selection, enable adaptation, estimate the time, location and intensity of impact, and estimate the reduction in effective insulation capability. These activation functions also serve as health-state variables and are formulated as functions of the digital twin model parameters by leveraging the physical constraints that they must satisfy. Given system response observations, the joint probability that these parameters satisfy or violate the physical constraints can be interpreted directly as the probability of different health-state conditions. This strategy avoids introducing a separate health-state random variable and avoids assuming a priori causal interactions represented through transition probabilities, as is done in probabilistic graphical model-based digital twins \cite{gratius2024twin, gratius2024calibration}. Through these activation functions, the framework combines simultaneous inference of impact-related parameters and structural thermal health-state with automatic model selection. Automatic model selection, adaptation capabilities are research gaps identified for digital twins in space missions \cite{gratius2024twin, zhang2026artificial, wang2026digital}, because scenario-specific model selection by experts with ground-control support is often infeasible.

\begin{figure}[h!]
\centering
\includegraphics[width=1\linewidth]{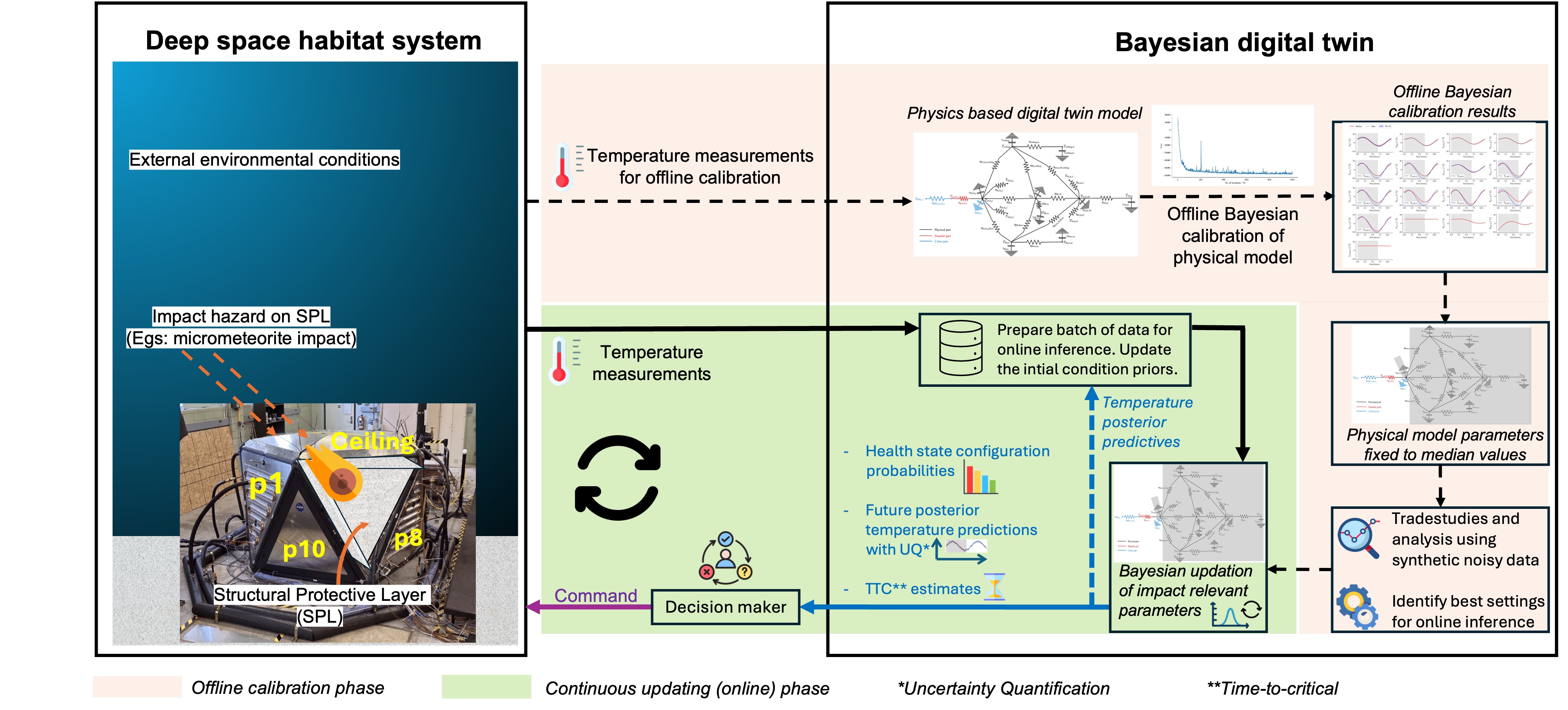}
\caption{Overview of the digital twin framework presented in this work. Here, p1, p8, and p10 denote panels 1, 8, and 10 of the habitat structure in the space habitat system \cite{silva2025cyber}.}
\label{fig: digital twin overview}
\end{figure}

An overview of the proposed digital twin framework is shown in Figure \ref{fig: digital twin overview}. The digital twin has an initial offline phase in which we calibrate the physical subsystem RC thermal network parameters using testbed temperature measurements after constructing the digital twin model. After setting these parameters to their posterior median values, we generate noisy synthetic data and perform sensitivity analysis, trade-off studies, performance analysis, and verification of the continuous updating strategy. In this phase, we also identify suitable settings for online inference on real experimental data, such as the batch size. In the online phase, we sequentially update the impact-relevant parameters of the digital twin using batches of time-series temperature data. During online inference, the digital twin yields probabilities of possible health-state configurations, future posterior temperature predictions, and TTC estimates that support informed decision-making. We validate the approach using the HARSH testbed, which enables controlled emulation of structural damage and its thermal consequences \cite{montoya2025validation, silva2025cyber}.

The main contributions of this work are as follows: (i) we develop a lumped parameter thermal network model for a cyber-physical testbed representing a smart space habitat; (ii) we construct a digital twin based on a dynamical system model for a space habitat experiencing a thermal anomaly due to an impact event; (iii) we design activation functions based on physics-based constraints on model parameters to automate model selection, enable adaptation, and estimate health state without introducing separate functionality-level state variables; and (iv) we demonstrate the proposed digital twin capabilities through controlled synthetic studies and validate them using experimental testbed data, including continuous Bayesian updating, automatic adaptation, probabilistic health-state estimation, inference of impact location, timing, and severity, temperature forecasting, and TTC estimation under varying observability and measurement-noise conditions.

The rest of the paper is organized as follows. Section \ref{section: Experiment setup} presents the experimental setup. Section \ref{section: lumped parameter model} discusses the construction of the RC thermal network model for the digital twin. Section \ref{section: probabilistic model formulation} presents the probabilistic formulation adopted for Bayesian inference. Section \ref{section: Bayesian offline calibration} discusses the offline Bayesian calibration results. Section \ref{section: synthetic example trade studies} presents the trade-off studies and analyses performed using synthetic data, and Section \ref{section: experimental validation} presents the experimental validation of the approach. Finally, Section \ref{section: concluding remarks} concludes the paper.

\section{Experimental Setup}
\label{section: Experiment setup}
HARSH was developed at the Resilient Extraterrestrial Habitats Institute at Purdue University \cite{silva2025cyber}. The testbed comprises the systems of a lunar habitat integrated with a health management system. Within the cyber-physical testing framework, the habitat systems are partitioned into cyber (numerical) and physical components. Some subsystems are represented by computational models, while others are physically realized in the laboratory. The interactions between the cyber and physical subsystems are imposed through actuation transfer systems and sensors. A real-time thermo-mechanical cyber-physical testing methodology enforces the boundary conditions \cite{montoya2023realtime,montoya2024actuator,montoya2025validation,salmeron2025temperature}. The physical subsystem of the testbed primarily consists of the structural system (ST), interior environment (IE), thermal management (TM) system, communication system, and physical sensors. The ST consists of 10 lateral panels and a ceiling resting on the floor and enclosing a bladder. The bladder contains a lightly pressurized air volume that forms the IE and keeps the bladder in contact with the inner side of the panels. All panels except the panels 5 and 10 include a thermal transfer plate (TTP) on the outer side. Each TTP contains serpentine channels through which cryogenic-chiller-conditioned heat transfer fluid circulates to impose thermal boundary conditions at the interface between the ST and the structural protective layer associated with the panel. The TTPs serve as the interface between the physical and cyber subsystems and therefore act as the transfer system. The SPL belongs to the cyber subsystem, is modeled using a finite element approach, is made of regolith material, and is subjected to external disturbances. For panels 1 to 9, a state-space thermal transfer panel model is used for control and estimation at the SPL-ST interface to enforce the interface thermal conditions \cite{montoya2024thermomechanical}. Panel 10 serves as the access door to the dome and contains one sensor. Because of the limited sensing, no thermal estimator is assigned explicitly to panel 10 \cite{montoya2024thermomechanical}. The TM system regulates the IE temperature when active. For more details, the reader is referred to \cite{silva2025cyber, montoya2024thermomechanical}.

In this work, we develop a digital twin that focuses on the thermal aspects of the IE when the testbed is subjected to an impact event causing damage to the SPL. At its core, the digital twin employs a lumped parameter thermal model to enable thermal anomaly detection, temperature prediction, and TTC estimation. We use a Bayesian approach to infer parameters and temperature predictions with quantified uncertainty. We first construct an RC thermal network model for the physical subsystem and calibrate its parameters using temperature measurement data. We then model the SPL thermal response by introducing the effective SPL thickness as an explicit parameter to represent the impact scenario. Finally, we combine the physical and cyber subsystem models and introduce the modifications required to equip the digital twin with thermal anomaly detection, temperature forecasting, and TTC prediction capabilities.

\section{Lumped Parameter Thermal Model}
\label{section: lumped parameter model}
Simulation models form a crucial component of the digital twin framework proposed by Gratius et al. for space missions \cite{gratius2024twin}. In space settings, computational resources are constrained. Reduced-order digital twins are preferred over high-fidelity physics-based models when the latter are computationally prohibitive \cite{kapteyn2021graphical, kapteyn2022physicsbased}. Lumped parameter models, as a class of reduced-order models, provide a suitable balance between physical interpretability, computational efficiency, parameter calibration, and model flexibility.
Accordingly, we adopt thermal lumped parameter modeling as the physics-based simulation component at the core of the digital twin, consistent with the framework presented in \cite{gratius2024twin}.

\begin{figure}[h!]
\centering
\includegraphics[width=1\linewidth]{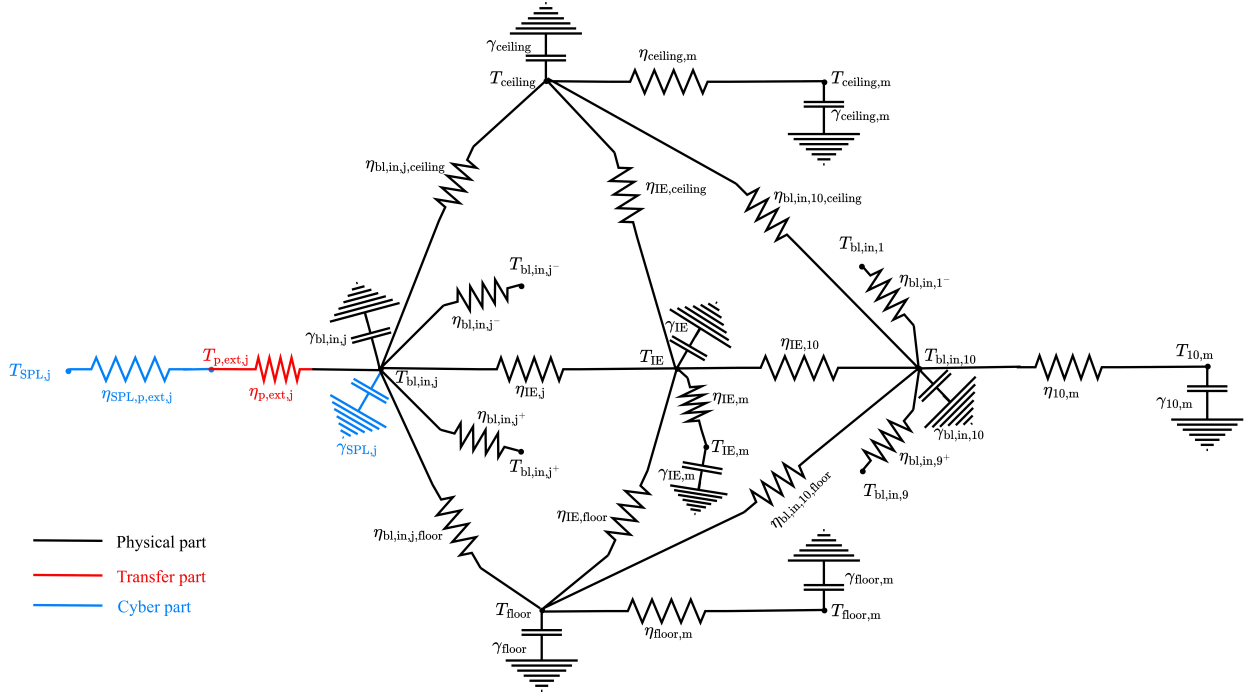}
\caption{Coupled RC thermal network used to model the thermal dynamics of the HARSH cyber-physical testbed. The network includes the physical subsystem, transfer system, and cyber subsystem components used in the digital twin formulation.} % Figure 29
\label{fig: cpt rc network}
\end{figure}

The scenario description and model construction presented in this section also align with the outline and design phases identified for constructing a digital twin for ECLSS in \cite{gratius2023learned}. Lumped parameter thermal modeling has been used for thermal control in spacecraft \cite{gaite2011analysis, garmendia2022parameters} and for thermal analysis of extraterrestrial habitats \cite{rhee2025evaluation, rhee2023development}. In this section, we describe the construction of the thermal lumped parameter model for the physical and cyber subsystems of the testbed. We then present the coupled cyber-physical thermal model used for the digital twin. To formulate the state-space equations, we adopt the general dimensionless representation of an RC thermal network proposed by Manikkan et al. in \cite{manikkan2025switch}. Figure \ref{fig: cpt rc network} shows the thermal lumped parameter network employed in this work. We assume Earth-like interior conditions and negligible radiative effects inside the habitat \cite{montoya2025validation}. Although the thermophysical properties of lunar regolith vary spatially and temporally with temperature, depth, and local solar illumination over the lunar day-night cycle \cite{hayne2017regolith, hemingway1973specific},we assume constant thermophysical properties for the SPL in the present model.

\subsection{Physical Subsystem Model}
\label{section: physical part model}
For the thermal lumped parameter model, we consider temperature nodes associated with the interior air volume and the interior surface of the bladder in contact with the panels, ceiling, and floor. We model the interior air as a single lumped mass with temperature \(T_{\text{IE}}\). The nominal pre-disruption condition considered in this work corresponds to a pressurized habitat under lunar-night conditions in which the SPL remains healthy and provides strong thermal insulation, so that \(T_{\text{IE}}\) stays approximately near steady with negligible variation before the disruption. In the testbed, the TM subsystem regulates the interior temperature when active. However, in the present lumped-parameter-model-based digital twin, we intentionally assume TM to be inactive as a study-specific modeling choice. This assumption is consistent with prior studies that start from pressurized, thermal steady-state lunar-night conditions and use TM-off configurations to isolate heat-transfer behavior and SPL disruption effects \cite{montoya2024thermomechanical}. Removing an additional heat input to the IE makes the thermal signature of impact-induced SPL damage more directly observable and thereby supports clearer evaluation of thermal anomaly detection, temperature forecasting, and prognosis. Accordingly, we do not apply a heat-source term to the IE temperature node in the physical subsystem model under the nominal condition.

We model the bladder interior surface as 12 lumped thermal masses with temperatures corresponding to the 10 side panels (\(T_{\text{bl,in,1}},...,T_{\text{bl,in,10}}\)), \(T_{\text{floor}}\), and \(T_{\text{ceiling}}\), where \(T_{\text{floor}}\) and \(T_{\text{ceiling}}\) denote the floor and ceiling temperatures, respectively. We also include four additional temperature nodes \(T_{\text{IE,m}}, T_{\text{10,m}}, T_{\text{floor,m}}, \text{ and }T_{\text{ceiling,m}}\), which interact with \(T_{\text{IE}}, T_{\text{bl,in,10}}, T_{\text{floor}}, \text{ and }T_{\text{ceiling}}\), respectively, to account for unobserved heat interactions and improve model expressivity, as recommended in \cite{manikkan2025switch}. We consider the j\(^\text{th}\) panel external surface to be at temperature \(T_{\text{p,ext,j}}\).\\
For the interior air volume, we have (Einstein summation for index j):
\begin{equation}
\label{eqn: IE temp physical}
\begin{split}
    \frac{dT_{\text{IE}}}{dt} = \gamma_{\text{IE}}\left\{
    \eta_{\text{IE,j}}(T_{\text{bl,in,j}} - T_{\text{IE}})
    +\eta_{\text{IE,ceiling}}(T_{\text{ceiling}} - T_{\text{IE}})\right.\\
    \left.
    +\eta_{\text{IE,floor}}(T_{\text{floor}} - T_{\text{IE}})
    +\eta_{\text{IE,m}}(T_{\text{IE,m}} - T_{\text{IE}})
    \right\}
    \text{ for j} = 1,\ldots,10.
\end{split}
\end{equation}
For the bladder interior surface temperature corresponding to the \(\text{j}^{\text{th}}\) panel, with \(j = 1,2,...,9\),
\begin{equation}
\label{eqn: jth bladder_in_temp physical}
\begin{split}
    \frac{dT_{\text{bl,in,j}}}{dt} =
    \gamma_{\text{bl,in,j}}\left\{
    \eta_{\text{p,j}}(T_{\text{p,ext,j}} - T_{\text{bl,in,j}})
    + \eta_{\text{IE,j}}(T_{\text{IE}} - T_{\text{bl,in,j}})\right.\\
    \left.
    +\eta_{\text{bl,in,j}^-}( T_{\text{bl,in,j}^-}-T_{\text{bl,in,j}})
    +\eta_{\text{bl,in,j}^+}( T_{\text{bl,in,j}^+}-T_{\text{bl,in,j}})\right.\\
    \left.
    + \eta_{\text{bl,in,j,ceiling}}(T_{\text{ceiling}} - T_{\text{bl,in,j}})
    +\eta_{\text{bl,in,j,floor}}(T_{\text{floor}} - T_{\text{bl,in,j}})
    \right\},
\end{split}
\end{equation}
\(\text{j}^-,\text{j}^+\) denote adjacent bladder interior surface temperature nodes. Panel 10 has fewer sensors and does not have an explicitly assigned thermal transfer control or estimator as mentioned earlier. Therefore, we do not associate an SPL with panel 10, and we treat \(T_{\text{bl,in,10}}\) differently from the other bladder interior surface temperatures. Specifically, we treat \(T_{\text{bl,in,10}}\) as an end node that interacts with another unknown temperature node \(T_{\text{10,m}}\). This choice reflects the absence of SPL at panel 10; consequently, we do not model the heat transfer mechanism beyond \(T_{\text{bl,in,10}}\) toward the exterior of the dome. The governing equation for \(T_{\text{bl,in,10}}\) is,
\begin{equation}
\label{eqn: 10th bladder_in_temp physical}
\begin{split}
    \frac{dT_{\text{bl,in,10}}}{dt} =
    \gamma_{\text{bl,in,10}}\left\{
    \eta_{\text{IE,10}}(T_{\text{IE}} - T_{\text{bl,in,10}})
    +\eta_{\text{bl,in,1}^-}( T_{\text{bl,in,1}}-T_{\text{bl,in,10}})\right.\\
    \left.
    +\eta_{\text{bl,in,9}^+}( T_{\text{bl,in,9}}-T_{\text{bl,in,10}})
    + \eta_{\text{bl,in,10,ceiling}}(T_{\text{ceiling}} - T_{\text{bl,in,10}})\right.\\
    \left.
    +\eta_{\text{bl,in,10,floor}}(T_{\text{floor}} - T_{\text{bl,in,10}})
    +\eta_{\text{10,m}}(T_{\text{10,m}}-T_{\text{bl,in,10}})
    \right\}.
\end{split}
\end{equation}
We write the state-space equations for the ceiling and floor temperature nodes as follows:
\begin{equation}
\label{eqn: floor or ceiling physical}
\begin{split}
    \frac{dT_{\text{k}}}{dt} =
    \gamma_{\text{k}}\left\{
    \eta_{\text{IE,k}}(T_{\text{IE}} - T_{\text{k}})
    +\eta_{\text{k,m}}( T_{\text{k,m}}-T_{\text{k}})\right.\\
    \left.
    + \eta_\text{bl,in,j,k}(T_\text{bl,in,j}-T_\text{k})
    +\eta_\text{bl,in,10,k}(T_\text{bl,in,10}-T_\text{k})
    \right\},
\end{split}
\end{equation}
where k \(\in\) \{floor, ceiling\}. In this model, \(T_{\text{IE}},\  T_{\text{bl,in,j}}\) for j \(\in\) \{1,3,5,7,9,10\}, \(T_{\text{ceiling}},\ T_{\text{floor}}\) are observed states. By contrast, \(T_{\text{bl,in,j}}\) for j \(\in\) \{2,4,6,8\}, \(T_{\text{IE,m}}, T_{\text{10,m}}, T_{\text{floor,m}}, T_{\text{ceiling,m}}\) are unobserved states because no sensors are assigned to them in the testbed. We treat \(T_{\text{p,ext,j}}\) for j \(\in\) \{1,2,...,9\} as input signals during Bayesian calibration of the physical subsystem model, as described in Section \ref{section: Bayesian offline calibration}. We make this choice because these measurements correspond to the farthest available sensor locations for all panels in the physical subsystem, including panels without TTP, during the heat transfer characterization test \cite{rhee2025quantitative}. When we couple the cyber and physical subsystem models to form the digital twin model, we do not include \(T_{\text{p,ext,j}}\) explicitly, as explained in Section \ref{subsection: cyber part model}.
\subsection{Cyber Subsystem Model}
\label{subsection: cyber part model}
Among the cyber subsystems, we consider the SPL because we are interested in the reduction in its effective thickness due to an impact. We represent the SPL through nine panel-wise boundary temperature nodes \(T_{\text{SPL,j}}\), for \(j=1,\ldots,9\), corresponding to the nine relevant panels. We exclude the panel 10, as explained in the previous section. In the higher-fidelity SPL finite element model, the temperature varies spatially across the depth of SPL. The SPL exterior surface temperature varies from \(100\ \text{K}\) to \(90\ \text{K}\)  \cite{montoya2024thermomechanical, montoya2026acyberphysical}. We treat \(T_{\text{SPL,j}}\) as a boundary temperature node with a prescribed constant average temperature, as shown in Figure \ref{fig: cpt rc network}, and set this temperature to \(95\ \text{K}\). Introducing dynamics for \(T_{\text{SPL,j}}\) would create an additional unobserved boundary state, thereby increasing the ill-posedness of the Bayesian inference problem. To avoid this issue while retaining the thermal inertia associated with the SPL, we assign the inverse capacitance represented by \(\gamma_{\text{SPL,j}}\) to the \(T_{\text{bl,in,j}}\) node in the coupled cyber-physical model, as illustrated in Figure \ref{fig: cpt rc network}. We do not include the temperature \(T_{\text{p,ext,j}}\) as an explicit state in the coupled cyber-physical model because the boundary condition information is available at the \(T_{\text{SPL,j}}\) node. We model the thermal conduction between \(T_{\text{SPL,j}}\) and \(T_{\text{bl,in,j}}\) using the effective parameter \(\eta_{\text{SPL,j}}\), which we obtain by combining \(\eta_{\text{SPL,p,ext,j}}\) and \(\eta_{\text{p,ext,j}}\) in series, as shown in Figure \ref{fig: cpt rc network}. From \cite{manikkan2025switch}, \(\eta_{\text{SPL,p,ext,j}}~\propto~R_{\text{SPL,p,ext,j}}^{-1}\) and \(\eta_{\text{p,ext,j}}~\propto~R_{\text{p,ext,j}}^{-1}\), where \(R_{\text{SPL,p,ext,j}}\) and \(R_{\text{p,ext,j}}\) denote thermal resistances. Therefore, for the series configuration, we have,
\begin{equation}
    \label{eqn: eta_spl_bl_in_j eqn}
    \begin{split}
    \eta_{\text{SPL,j}}^{-1} &= \eta_{\text{SPL,p,ext,j}}^{-1} + \eta_{\text{p,ext,j}}^{-1},\\
    & \implies \eta_{\text{SPL,j}} = \frac{\eta_{\text{SPL,p,ext,j}} \eta_{\text{p,ext,j}}}{\eta_{\text{SPL,p,ext,j}} +\eta_{\text{p,ext,j}}}.
    \end{split}
\end{equation}
Based on \cite{manikkan2025switch}, we have,
\begin{equation}
\label{eqn: gamma_spl_j eqn_1}
   \gamma_{\text{SPL,j}} = \frac{C}{\rho_{\text{SPL,j}} c_{\text{SPL,j}} V_{\text{SPL,j}}} ,
\end{equation}
where \(\rho_{\text{SPL,j}},~c_{\text{SPL,j}},~ V_{\text{SPL,j}} \) denote the density, specific heat capacity, and volume of the SPL corresponding to the \(\text{j}^\text{th}\) panel. \(C\) denotes the capacitance of the primary node \cite{manikkan2025switch} and is taken implicitly here. We write \(V_{\text{SPL,j}} = A_{\text{SPL,j}}l_{\text{SPL,j}}\), where \(A_{\text{SPL,j}}\) and \(l_{\text{SPL,j}}\) denote the effective cross-sectional area and thickness of the SPL, respectively. From \cite{manikkan2025switch}, 
\begin{equation}
\label{eta_spl_ttp_j eqn}
      \eta_{\text{SPL,p,ext,j}} = \frac{t_s A_{\text{SPL,j}} k_{\text{SPL,j}}}{l_{\text{SPL,j}} C},
\end{equation}
where \(t_s\) denotes the time scale and \(k_{\text{SPL,j}}\) denotes the thermal conductivity of the portion of SPL corresponding to the \(\text{j}^\text{th}\) panel. We define \(c_{\text{1,j}} = \frac{C}{\rho_{\text{SPL,j}} c_{\text{SPL,j}} A_{\text{SPL,j}}}\) and \(c_{\text{2,j}} = \frac{t_s k_{\text{SPL,j}} }{\rho_{\text{SPL,j}} c_{\text{SPL,j}} }\). Assuming constant thermophysical properties and a common effective cross-sectional area of the SPL across all nine panels, we drop the j subscript and set \(c_{\text{1,j}} = c_1\) and \(c_{\text{2,j}} = c_2\). Using the above definitions, the parameters can be expressed as \(\gamma_{\text{SPL,j}} = \frac{c_1}{l_{\text{SPL,j}}}\) and \(\eta_{\text{SPL,p,ext,j}} = \frac{c_2}{c_1l_{\text{SPL,j}}}\). Substituting these relations into Eqn. \ref{eqn: eta_spl_bl_in_j eqn} yields
\begin{equation}
\label{eqn: eta_spl_j eqn}
    \begin{split}
        \eta_{\text{SPL,j}} =\frac{c_2\eta_{\text{p,ext,j}}}{{c_2}+\eta_{\text{p,ext,j}}c_1l_{\text{SPL,j}}}.
    \end{split}
\end{equation}
We additionally incorporate the thermal-radiation interaction between \(T_{\text{SPL,j}}\) and \(T_{\text{bl,in,j}}\), modeled as \(Q_{\text{SPL,bl,in,j}} = c_3'(T_{\text{SPL,j}}^4 - T_{\text{bl,in,j}}^4)\), acting on the \(T_{\text{bl,in,j}}\) node as illustrated in Figure \ref{fig: cpt rc network}. The radiation term is evaluated using absolute temperatures in Kelvin. Accounting for thermal radiation is important because damage to the SPL can create a substantial temperature difference between \(T_{\text{SPL,j}}\) and \(T_{\text{bl,in,j}}\), thereby increasing radiative heat transfer and contributing significantly to the resulting temperature drop and thermal cascading effects.
\subsection{Coupled Cyber-Physical Thermal Model}
\label{section: cpt impact model}
We formulate the thermal model for the digital twin by combining the physical subsystem model presented in Section \ref{section: physical part model} with the cyber subsystem model presented in Section \ref{subsection: cyber part model}. We first construct separate models for the nominal and fault conditions. We then combine these nominal and fault-condition models to obtain a digital twin model with automatic model selection capabilities.

The nominal condition considered in this work corresponds to a pressurized habitat under lunar-night conditions in which the IE remains approximately in thermal equilibrium and shows negligible temperature variation because the SPL provides strong insulation. In this nominal state, we assume that no subsystem faults are present. Therefore, the SPL remains undamaged and all relevant habitat subsystems operate in their healthy baseline condition. Consistent with the assumptions adopted in the physical subsystem model, we keep the TM subsystem inactive, and thus we do not apply a direct heat-source term to the interior-environment node. The exterior environment corresponds to lunar night, and the SPL is subjected to the same exterior boundary condition described earlier in Section \ref{subsection: cyber part model}.

The fault condition corresponds to an event-driven off-nominal state initiated by a discrete impact event on the SPL. We assume that the impact causes SPL damage and reduces the effective thickness of the affected portion of the SPL and thereby reduces its effective insulation capability. In this work, we represent this reduction in effective insulation capability only through the geometric change in effective thickness, while keeping other material properties, such as thermal conductivity and specific heat capacity, unchanged. The resulting change in thermal resistance increases heat transfer across the affected region and produces thermal cascading effects in the habitat, which manifest as a drop in the IE temperature and in the bladder interior surface temperatures. We treat this temperature drop as the thermal anomaly of interest in this study. Thermal anomalies involving a sharp temperature drop or rise, or a strong change in the slope of the temperature profile, are expected in disruptive space scenarios such as micrometeoroid impacts and fire \cite{wang2024faulta}.
\subsubsection{Nominal Condition Model}
\label{section: cpt nominal condtion model}
In the nominal condition model, the governing differential equations for \(T_{\text{IE}},~ T_{\text{bl,in,10}},~T_\text{k}\) for k \(\in\) \{floor, ceiling\}, remain the same as Eqns. \ref{eqn: IE temp physical}, \ref{eqn: 10th bladder_in_temp physical}, \ref{eqn: floor or ceiling physical}, respectively. As \(\gamma_{\text{SPL,j}}\) is proportional to inverse capacitance of the SPL, the effective \(\gamma\) parameter of \(T_{\text{bl,in,j}}\) when capacitance of SPL is added to it is given by,
\begin{equation}
\label{eqn: gamma_effective T_bl_in_j eqn}
    \begin{split}
        \gamma_\text{bl,in,j,eff}& =  \frac{\gamma_\text{bl,in,j} \gamma_{\text{SPL,j}}}{\gamma_\text{bl,in,j} +\gamma_{\text{SPL,j}}}\\
        & = \frac{\gamma_\text{bl,in,j} c_1}{(\gamma_\text{bl,in,j}l_{\text{SPL,j}}+c_1)}.
    \end{split}
\end{equation}

We write the nominal-condition equation for \(T_{\text{bl,in,j}}\), with \(\text{j} = 1,2,...,9\), as
\begin{equation}
\label{eqn: jth bladder_in_temp cyber physical}
\begin{split}
    \frac{dT_{\text{bl,in,j}}}{dt}
    =
    \gamma_\text{bl,in,j,eff} \left\{
    \eta_{\text{SPL,j}}(T_{\text{SPL,j}} - T_{\text{bl,in,j}})
    + \eta_{\text{IE,j}}(T_{\text{IE}} - T_{\text{bl,in,j}})\right.\\
    \left.
    +\eta_{\text{bl,in,j}^-}( T_{\text{bl,in,j}^-}-T_{\text{bl,in,j}})
    +\eta_{\text{bl,in,j}^+}( T_{\text{bl,in,j}^+}-T_{\text{bl,in,j}})\right.\\
    \left.
    + \eta_{\text{bl,in,j,ceiling}}(T_{\text{ceiling}} - T_{\text{bl,in,j}})
    +\eta_{\text{bl,in,j,floor}}(T_{\text{floor}} - T_{\text{bl,in,j}})
    \right\}.
\end{split}
\end{equation}
We assume that thermal-radiation effects are negligible under the nominal condition and therefore omit the radiation term from the model. Retaining this term would introduce a significant temperature drop even in the absence of damage to the SPL, which is inconsistent with the near-steady nominal thermal behavior considered in this work. Substituting Eqn. \ref{eqn: eta_spl_j eqn} for \(\eta_{\text{SPL,j}}\) and Eqn. \ref{eqn: gamma_effective T_bl_in_j eqn} for \(\gamma_{\text{bl,in,j,eff}}\) into Eqn. \ref{eqn: jth bladder_in_temp cyber physical} yields
\begin{equation}
\label{eqn: jth bladder_in_temp cyber physical 2}
\begin{split}
    \frac{dT_{\text{bl,in,j}}}{dt}
    =
    \frac{\gamma_\text{bl,in,j} c_1}{(\gamma_\text{bl,in,j}l_{\text{SPL,j}}+c_1)}
    \left\{
    \frac{
    c_2\eta_{\text{p,ext,j}}
    }
    {
    c_2+\eta_{\text{p,ext,j}}c_1l_{\text{SPL,j}}
    }
    (T_{\text{SPL,j}} - T_{\text{bl,in,j}})
    + \eta_{\text{IE,j}}(T_{\text{IE}} - T_{\text{bl,in,j}})\right.\\
    \left.
    +\eta_{\text{bl,in,j}^-}( T_{\text{bl,in,j}^-}-T_{\text{bl,in,j}})
    +\eta_{\text{bl,in,j}^+}( T_{\text{bl,in,j}^+}-T_{\text{bl,in,j}})\right.\\
    \left.
    + \eta_{\text{bl,in,j,ceiling}}(T_{\text{ceiling}} - T_{\text{bl,in,j}})
    +\eta_{\text{bl,in,j,floor}}(T_{\text{floor}} - T_{\text{bl,in,j}})
    \right\}.
\end{split}
\end{equation}
Eqn. \ref{eqn: jth bladder_in_temp cyber physical 2} gives the governing equation for the bladder interior surface temperature corresponding to the \(\text{j}^{\text{th}}\) panel, with \(\text{j} = 1,2,...,9\), under the nominal condition model.

\subsubsection{Fault Condition Model}
\label{section: cpt fault condition model}
In this section, we extend the lumped parameter model presented in Section \ref{section: cpt nominal condtion model} to a fault condition in which damage to the SPL is caused by an impact event. We represent this SPL damage as a reduction in effective thickness from the nominal value \(l_{\text{SPL,j}}\) to a damaged value \(l_{\text{SPL,j,d}}\) at the impact time \(t_{\text{impact,j}}\). Consistent with the assumptions adopted by Montoya et al. in \cite{montoya2025validation}, we keep all thermophysical properties unchanged and represent the damage only through the geometric reduction in effective SPL thickness.

The functional forms of the equations for \(T_{\text{IE}},~ T_{\text{bl,in,10}},~T_\text{k}\), with k \(\in\) \{floor, ceiling\}, remain the same as in the nominal condition model because the dynamics of these nodes do not explicitly depend on the SPL thickness parameter. We write the fault-condition equation for \(T_{\text{bl,in,j}}\), with \(\text{j} = 1,2,...,9\), as
\begin{equation}
\label{eqn: jth bladder_in_temp cyber physical 2 fault condition}
\begin{split}
    \frac{dT_{\text{bl,in,j}}}{dt}
    =
    \frac{\gamma_\text{bl,in,j} c_1}{(\gamma_\text{bl,in,j}l_{\text{SPL,j,d}}+c_1)}
    \left\{
    \frac{
    c_2\eta_{\text{p,ext,j}}
    }
    {
    c_2+\eta_{\text{p,ext,j}}c_1l_{\text{SPL,j,d}}
    }
    (T_{\text{SPL,j}} - T_{\text{bl,in,j}})
    + \eta_{\text{IE,j}}(T_{\text{IE}} - T_{\text{bl,in,j}})\right.\\
    \left.
    +c_3'(T_{\text{SPL,j}}^4 - T_{\text{bl,in,j}}^4)+ \eta_{\text{bl,in,j}^-}( T_{\text{bl,in,j}^-}-T_{\text{bl,in,j}})
    +\eta_{\text{bl,in,j}^+}( T_{\text{bl,in,j}^+}-T_{\text{bl,in,j}})\right.\\
    \left.
    + \eta_{\text{bl,in,j,ceiling}}(T_{\text{ceiling}} - T_{\text{bl,in,j}})
    +\eta_{\text{bl,in,j,floor}}(T_{\text{floor}} - T_{\text{bl,in,j}})
    \right\}.
\end{split}
\end{equation}
Eqn. \ref{eqn: jth bladder_in_temp cyber physical 2 fault condition} is obtained by replacing \(l_{\text{SPL,j}}\) with \(l_{\text{SPL,j,d}}\) in Eqn. \ref{eqn: jth bladder_in_temp cyber physical 2} and incorporating the thermal-radiation term. Inclusion of radiation term allows the fault-condition model to capture the pronounced temperature drop and associated thermal cascading effects observed following SPL damage.

\subsubsection{Digital Twin Model}
\label{section: digital twin model}
We want the digital twin to switch from the nominal condition model to the fault condition model once an impact affects the SPL and induces a thermal anomaly. To achieve this, we combine the nominal and fault-condition models by introducing a time-of-impact parameter \(t_{\text{impact,j}}\).
Let \(f_1\) and \(f_{1,d}\) denote the right-hand sides of Eqns. \ref{eqn: jth bladder_in_temp cyber physical 2} and \ref{eqn: jth bladder_in_temp cyber physical 2 fault condition}, respectively. We then express \(T_{\text{bl,in,j}}\), for \(\text{j} = 1,2,...,9\), in the digital twin as
\begin{equation}
    \label{eqn: jth bladder_in_temp cyber physical 2 combined}
    \frac{dT_{\text{bl,in,j}}}{dt} = f_1 + H(t-t_{\text{impact,j}})\left \{f_{1,d}-f_1\right \},
\end{equation}
where \(H(\cdot)\) is the Heaviside step function. This function ensures that the nominal condition model applies before impact and the fault condition model applies after impact. Accordingly, the second term in Eqn. \ref{eqn: jth bladder_in_temp cyber physical 2 combined} represents the deviation from the nominal condition model toward the fault condition model. Automatic model selection is one of the research gaps identified for digital twins in space missions \cite{gratius2024twin}, because scenario-specific model selection by experts with ground-control support is often infeasible. Recent studies have explored model selection and discovery by leveraging domain knowledge and physics-based constraints. For example, Manikkan et al. enable Bayesian discovery of lumped parameter RC thermal networks by exploiting the non-negativity constraints of model parameters \cite{manikkan2025switch}. Gratius et al. use domain knowledge to prune nodes in a Bayesian network more effectively than purely statistical heuristics \cite{gratius2025bayesian}. More broadly, the digital twin literature identifies adaptability to an evolving physical twin as a key requirement \cite{kapteyn2022digital, nationalacademiesofsciencesFoundationalResearchGaps2023a, wang2026digital}. To enable automatic model selection between the nominal and fault condition models, and thereby provide adaptation capability for the digital twin, we introduce an indicator function \(I_j = I(l_{\text{SPL,j}},l_{\text{SPL,j,d}},t_{\text{impact,j}})\) such that,
\begin{equation}
    \label{eqn: jth spl node temperature equation c1 c2 combined with indicator fn}
    \frac{dT_{\text{bl,in,j}}}{dt} = f_1 + I_j \left \{f_{1,d}-f_1\right \}.
\end{equation}
\(I_j\) takes the value 1 when damage to the portion of SPL corresponding to the \(\text{j}^{\text{th}}\) panel occurs due to an impact, and 0 otherwise. Thus, \(I_j\) represents the functionality-level health-state of the portion of the SPL associated with the \(\text{j}^{\text{th}}\) panel.

This indicator function enables automatic selection between the nominal and fault condition models through Bayesian inference of \(l_{\text{SPL,j}},\ l_{\text{SPL,j,d}}\), and \(t_{\text{impact,j}}\) using temperature measurements from the testbed. To formulate \(I_j\), we identify the relevant constraints associated with these parameters. We begin by expanding and simplifying \(f_{1,d}-f_1\). This yields the following simplified form:
\begin{equation}
\label{eqn: df1 simplified form}
\begin{split}
f_{1,d}-f_1
=
\Delta l_{\text{SPL,j}} f_2,
\end{split}
\end{equation}
where,
\begin{equation}
\label{eqn: df1 expanded form part}
\begin{split}
f_2 = \Bigg\{
&
\frac{
\gamma_{\text{bl,in,j}}
c_1c_2
\eta_{\text{p,ext,j}}
\left(
T_{\text{SPL,j}}-T_{\text{bl,in,j}}
\right)f_3
}
{
f_4(l_{\text{SPL,j}})f_4(l_{\text{SPL,j,d}})
}
+
\frac{\gamma_{\text{bl,in,j}}^2c_1}{f_5(l_{\text{SPL,j}})f_5(l_{\text{SPL,j,d}})}
\\
&
\Big[
\eta_{\text{bl,in,j}^-}
( T_{\text{bl,in,j}^-}-T_{\text{bl,in,j}})
+\eta_{\text{bl,in,j}^+}
( T_{\text{bl,in,j}^+}-T_{\text{bl,in,j}})
\\
&
+
\eta_{\text{IE,j}}(T_{\text{IE}} - T_{\text{bl,in,j}})
+ \eta_{\text{bl,in,j,ceiling}}
(T_{\text{ceiling}} - T_{\text{bl,in,j}})
+\eta_{\text{bl,in,j,floor}}
(T_{\text{floor}} - T_{\text{bl,in,j}})
\Big]
\\
&
+
\frac{\gamma_{\text{bl,in,j}}c_1c_3
\left(
T_{\text{SPL,j}}^4-T_{\text{bl,in,j}}^4
\right)}{f_5(l_{\text{SPL,j,d}})}
\Bigg\},
\end{split}
\end{equation}
\(f_3 = c_2\gamma_{\text{bl,in,j}}+\eta_{\text{p,ext,j}}c_1\gamma_{\text{bl,in,j}} \left(l_{\text{SPL,j}}+ l_{\text{SPL,j,d}}\right)+c_1^2\eta_{\text{p,ext,j}}\), 
\(f_4 (l_{\text{SPL,j}})= (\gamma_{\text{bl,in,j}}l_{\text{SPL,j}}+c_1)(c_2 + \eta_{\text{p,ext,j}}c_1l_{\text{SPL,j}})
\),
\(f_5 (l_{\text{SPL,j}}) =  \gamma_{\text{bl,in,j}} l_{\text{SPL,j}} + c_1\), \(c_3 = c_3'/ \Delta l_{\text{SPL,j}}\) and, \(\Delta l_{\text{SPL,j}} = l_{\text{SPL,j}} - l_{\text{SPL,j,d}} \geq 0\). In the numerical implementation, we treat \(c_3 \geq 0\) as the explicit radiation-related parameter instead of \(c_3'\). This avoids computing \(c_3'/\Delta l_{\text{SPL,j}}\), which can become numerically problematic when \(\Delta l_{\text{SPL,j}}\) is close to zero. Equivalently, the effective radiation coefficient in the fault-condition model can be interpreted as \(c_3'=\Delta l_{\text{SPL,j}}c_3\). Inspired by the entropy-switch concept in \cite{manikkan2025switch}, we use the non-negativity constraint on \(\Delta l_{\text{SPL,j}}\) as one of the constraints that define the indicator function. We also impose the constraint \( t_l \leq t_{\text{impact,j}} \leq t_h\), where \(t_l\) and \(t_h\) denote the lower and upper bounds of the time interval covered by the testbed measurements used for Bayesian inference. This constraint enforces that, if the impact occurs within the measurement window, then the inferred time of impact must also lie within that window. By incorporating these physics-based constraints, we encode the model with scenario-specific domain knowledge, which is an important requirement for digital twins in space missions \cite{gratius2024twin}. We express the indicator function as
\begin{equation}
    \label{eqn: impact indicator function}
    I_j =H(t-t_{\text{impact,j}})\left \{H(t_{\text{impact,j}}-t_l) - H(t_{\text{impact,j}}-t_h) \right \}H(\Delta l_{\text{SPL,j}}).
\end{equation}
In this way, we express the health-state variable as a function of the digital twin model parameters by leveraging the physics-based constraints that these parameters must satisfy. Given system response observations, the joint probability that these parameters satisfy or violate the physical constraints can be interpreted directly as the probability of different health-state configurations. This strategy avoids introducing a separate health-state random variable and avoids assuming a priori causal interactions represented through transition probabilities, as is done in probabilistic graphical model-based digital twins \cite{gratius2024twin, gratius2024calibration}.

Inspired by \cite{manikkan2025switch}, in the numerical implementation we use \(S(\Delta l_{\text{SPL,j}})\) instead of \(H(\Delta l_{\text{SPL,j}})\Delta l_{\text{SPL,j}}\) in the product \( I_j \left\{f_{1,d}-f_1\right\}\) when substituting Eqns. \ref{eqn: df1 simplified form}--\ref{eqn: impact indicator function} into Eqn. \ref{eqn: jth spl node temperature equation c1 c2 combined with indicator fn}. Here, \(S(x)=\max(0,x)\) denotes the standard ReLU function, which enforces the non-negativity constraint on \(\Delta l_{\text{SPL,j}}\) in a numerically convenient manner. Consequently, Eqn. \ref{eqn: jth spl node temperature equation c1 c2 combined with indicator fn} can be expressed as
\begin{equation}
    \label{eqn: jth spl node temperature equation c1 c2 combined with indicator fn and relu fun}
    \frac{dT_{\text{bl,in,j}}}{dt} = f_1 + H(t-t_{\text{impact,j}})\left \{H(t_{\text{impact,j}}-t_l) - H(t_{\text{impact,j}}-t_h) \right \}S(\Delta l_{\text{SPL,j}})f_2.
\end{equation}
With this parametrization, the radiation contribution is also activated through \(S(\Delta l_{\text{SPL,j}})\), so that the effective radiation coefficient is zero in the absence of inferred SPL thickness reduction and increases with the inferred damage magnitude.
Having established the lumped parameter thermal model of the digital twin, we now formulate the associated Bayesian inverse problem. This probabilistic formulation enables both offline calibration and online inference with quantified uncertainty.

\section{Probabilistic Model Formulation}
\label{section: probabilistic model formulation}
%%% Paraphrase this section a bit.
Calibration of simulation models, together with quantified uncertainty, is one of the key research directions identified for enabling digital twins in space missions \cite{gratius2024twin}. It is also one of the four main phases outlined for constructing a digital twin for ECLSS \cite{gratius2023learned}. In this work, we adopt a Bayesian approach because it naturally supports uncertainty quantification during model calibration.

We first perform Bayesian offline calibration of the physical subsystem model parameters. After this offline calibration, we fix the calibrated parameters at their posterior median values and emulate the continuous Bayesian updating of the digital twin model parameters for online thermal anomaly detection, temperature forecasting, and TTC estimation for the IE temperature, which we refer to as online inference. This section presents the general inverse problem formulation used for both offline calibration and online inference in the present work.\\
Eqns. \ref{eqn: IE temp physical} - \ref{eqn: floor or ceiling physical} for offline Bayesian calibration, and Eqns. \ref{eqn: IE temp physical}, \ref{eqn: 10th bladder_in_temp physical}, \ref{eqn: floor or ceiling physical}, and \ref{eqn: jth spl node temperature equation c1 c2 combined with indicator fn and relu fun} for online inference can be written in the following general form as a continuous-time deterministic dynamical system:  
\begin{equation}
    \label{eqn: general conti det dynamic system}
    \Dot{x}(t) = f(x(t),t;\theta,u(t)),
\end{equation} 
where $t$ denotes time, $x(t) \in \mathbb{R}^{n_x}$ is the temperature state vector, and $f$ is the vector field defined by the right-hand sides of the above set of equations. The parameter vector $\theta \in \mathbb{R}^{n_\theta}$ contains all inferred parameters. For offline Bayesian calibration, \(\theta\) consists of \(\gamma_{\text{IE}},~\gamma_{\text{floor}},~\gamma_{\text{ceiling}},~\gamma_{\text{bl,in,10}}, ~\gamma_{\text{bl,in,j}},~\gamma_{\text{k,m}},~\eta_{\text{IE,j}},~\eta_{\text{IE,q}}, ~\eta_{\text{IE,10}}
,~\eta_{\text{p,ext,j}},~\eta_{\text{bl,in,j}^-},\) \(~\eta_{\text{bl,in,j}^+},~\eta_{\text{k,m}},~\eta_\text{bl,in,j,q},~\eta_\text{bl,in,10,q}\) for j \(\in\) \{1, ... , 9\}, k \(\in\) \{IE,10, floor, ceiling\}, q \(\in\) \{floor, ceiling\}. For online inference, we treat \(c_1\), \(c_2\), and \(c_3\) as fixed parameters and infer only the impact-relevant parameters \(l_{\text{SPL,j}}\), \(\Delta l_{\text{SPL,j}}\), and \(t_{\text{impact,j}}\) for \(j \in \{1,2,\ldots,9\}\). This choice is motivated by the sensitivity analysis presented in Appendix \ref{appdx: sensitivity analysis}, which demonstrates that fixing \(c_1\), \(c_2\), and \(c_3\) yields a more physically interpretable sensitivity structure and avoids confounding with the impact-relevant parameters. We denote the control or input signals by $u(t)$. We assume that $u(t)$ is known exactly and that the system has no latent process noise. Accounting for uncertainty in \(u(t)\) or for latent process noise would require a stochastic system model, which lies beyond the scope of the present work. For offline Bayesian calibration, we take \(T_{\text{p,ext,j}}\) as \(u(t)\). For online inference, we treat the isothermal boundary node temperature \(T_{\text{SPL,j}}\) as \(u(t)\).

We combine the latent physical process model with an emission model that generates discrete-time noisy temperature observations, 
\begin{equation}
    \label{eqn: emission model equation}
    y(t_p) = Hx(t_p) + \epsilon(t_p),
\end{equation}  
where \(H_{n_y \times n_x}\) is a measurement mask that selects the observed states at each observation time $t_p$, and $\epsilon(t_p)$ denotes independent, identically distributed Gaussian noise. The observed data form the time series $y = (y(t_1), y(t_2), \ldots, y(t_N))$. We infer the posterior over the unknown quantities, namely the parameters $\theta$, initial conditions $x_0$, and observation-noise variances $\sigma^2 = \text{diag}(\sigma_1^2, \ldots, \sigma_{n_y}^2)$, using Bayes' rule,
\begin{equation}
\label{eqn: Bayes' rule}
     p(\theta,x_0,\sigma \mid y) = \frac{p(y \mid \theta,x_0,\sigma)\, p(\theta,x_0,\sigma)}{\int p(y \mid \tilde{\theta},\tilde{x}_0,\tilde{\sigma})\,p(\tilde{\theta},\tilde{x}_0,\tilde{\sigma})d\tilde{\theta} d\tilde{x}_0 d\tilde{\sigma}}.
\end{equation}  
The likelihood is defined as  
\begin{equation}
     \label{eqn: likelihood function}
     p(y \mid \theta,x_0,\sigma) = \prod_{p=1}^N p(y(t_p) \mid \theta,x_0, \sigma),
\end{equation}  
where \(y(t_p) \mid \theta,x_0, \sigma \sim \mathcal{N}(H\hat{x}(t_p;\theta,x_0),\sigma^2)\). Here, \(\mathcal{N}()\) denotes multivariate normal distribution, and \( \hat{x}(t_p;\theta,x_0) \) denotes the solution of the dynamical system in Eqn \ref{eqn: general conti det dynamic system}. We compute \( \hat{x}(t_p;\theta,x_0) \) using a differentiable ODE solver. Specifically, we use Dormand-Prince's 5/4 method \cite{dormand1980family}, implemented in the Python package diffrax \cite{kidgerNeuralDifferentialEquations2021a}.

With the general probabilistic framework in place, we first apply it to the offline calibration of the physical subsystem model parameters. The resulting calibrated parameter values then provide the basis for the synthetic studies and subsequent online inference.
\section{Offline Bayesian Calibration of the Physical Subsystem Model}
\label{section: Bayesian offline calibration}
In this section, we discuss the offline calibration of the physical subsystem model parameters presented in Section \ref{section: physical part model}. This model has \(n_x = 17\) states, of which \(n_y = 9\) are observed, as described in Section \ref{section: physical part model}. In total, we estimate 98 parameters, comprising 72 dimensionless model parameters, 17 initial-condition parameters, and 9 observation-noise parameters.

For reliable calibration of a state-space model, the data should span the relevant operating range. The input signal must excite different system modes so that the data remain informative for parameter identification. In addition, the experimental data should contain minimal measurement noise. We obtain the temperature dataset for offline calibration by circulating cryogenic-chiller-conditioned heat transfer fluid through the TTP. The data-generation process follows the first thermal characterization test in \cite{rhee2024development}. We impose a sinusoidal profile on the chiller temperature setpoint, which causes the TTP exterior temperature to vary from \(-55\celsius\) to \(40 \celsius\). We use the average temperature measurements on the exterior surfaces of panels 1 to 9 as the input signals \(T_{\text{p,ext,j}}\) for calibration of the physical subsystem model. The panels equipped with TTP primarily drive the IE temperature variation. This variation then drives the temperatures of the bladder interior surfaces and of the internal and external surfaces associated with panels without TTP. We keep the TM inactive so that the temperatures can evolve freely under the imposed boundary excitation.

We recorded a total of 42101 measurements at 1 s intervals. For inference, we subsample the dataset by retaining every 15th point. From this subsampled dataset, we discard the first 100 data points to avoid initial abrupt fluctuations and to obtain uniform initial conditions. We normalize each variable to the range [-1,1] using the global maximum absolute value of the observations as the scaling factor. We use the first 60\% of the data for calibration and the remaining 40\% for posterior predictive evaluation. First, this split allows the calibration regime to contain major variations, such as the crest of the sine wave and the transition toward the trough, which helps the model learn the system dynamics. Second, it places the upward trend from the trough to the crest, which is not fully present in the calibration regime, within the posterior predictive evaluation regime. This distinct trend in the evaluation data helps assess whether the calibrated model can effectively predict the system dynamics beyond those used during calibration.

We specify the priors as follows. We set \( p(x_0) = \mathcal{N}(0.5\textbf{1}_{n_x},0.04\textbf{I}_{n_x})\), where \( \textbf{1}_{n_x} \) denotes the unit vector of size \( n_x \) and \( \textbf{I}_{n_x} \) denotes the identity matrix of size \( n_x \). For each observation noise parameter, we use an independent \(\text{Exp}(2) \) prior where Exp(r) denotes the exponential distribution with rate r. For \(\theta_i \in \theta\), we adopt a truncated normal prior \(\text{TN}(\mu_{\theta_i}=5,\sigma_{\theta_i}=30,l_{\theta_i}=0, h_{\theta_i})\) as \( p(\theta_i) \). For \(\theta_i \in \{ \gamma_{\text{IE}},~\gamma_{\text{floor}},~\gamma_{\text{ceiling}},~\gamma_{\text{bl,in,j}},~\gamma_{\text{k,m}}\}\), with j \(\in\) \{1, ... , 10\} and k \(\in\) \{IE,10, floor, ceiling\}, we set \(h_{\theta_i} = 15\). For all other \(\theta_i\), we set \(h_{\theta_i} = 25\).

\begin{figure}[h!]
\centering
\includegraphics[width=1\linewidth]{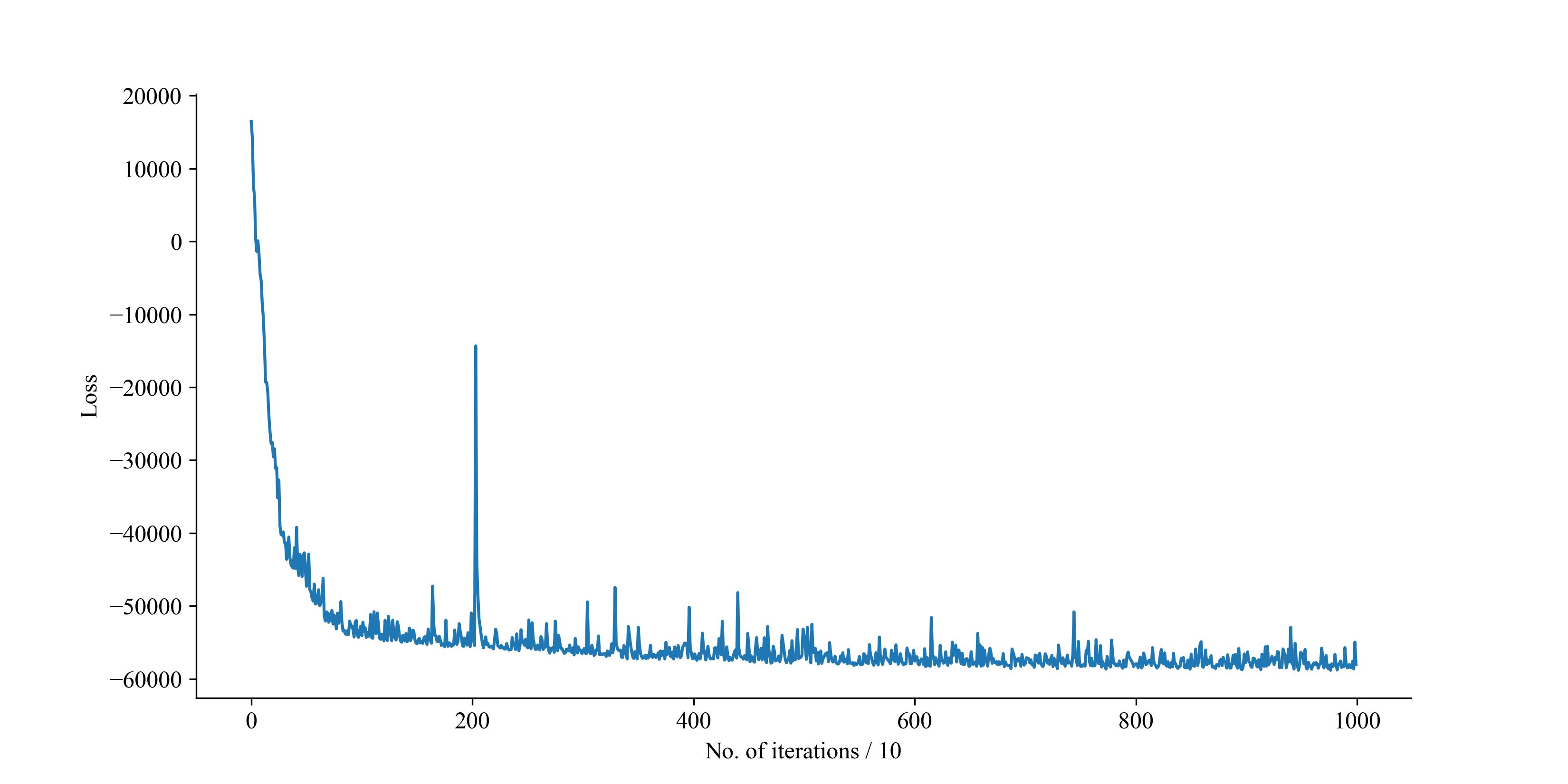}
\caption{Variational inference loss during offline Bayesian calibration of the physical subsystem model parameters.}
\label{fig: offline calibration svi losses}
\end{figure}

\begin{figure}[h!]
\centering
\includegraphics[width=1\linewidth]{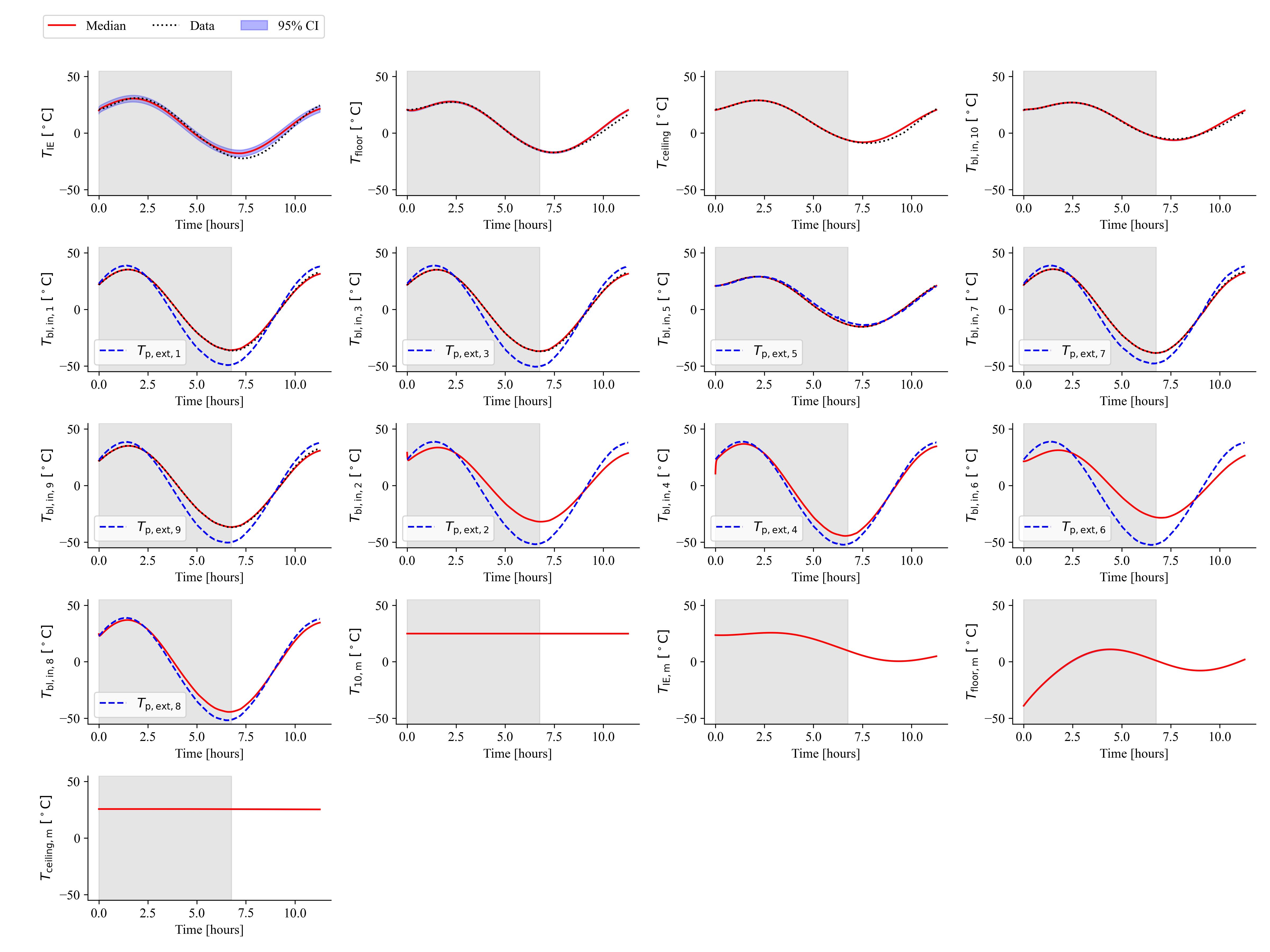}
\caption{Temperature posterior predictives obtained after Bayesian offline calibration of the physical subsystem model parameters. The grey-shaded region denotes the portion of data used for Bayesian inference.}
\label{fig: offline calibration results}
\end{figure}

We use variational inference to approximate the posterior distributions. Specifically, we use the block neural autoregressive flow \cite{de2020block} guide available in the NumPyro \cite{phanComposableEffectsFlexible2019} Python package. We run the inference algorithm for 10000 iterations as this resulted to be sufficient to verify the loss convergence. Figure \ref{fig: offline calibration svi losses} shows the loss evolution. By the end of the iterations, the loss exhibits minimal variation, which indicates that the method has reached a good posterior approximation.
Figure \ref{fig: offline calibration results} shows the temperature posterior predictives. For the observed states, the tight 95\% credible intervals contain the measurements in the training region, and the predictions at unseen time instants beyond the training regime align very well with the measurements. These results indicate successful calibration of the model parameters and show that the model captures the system dynamics well. For the unobserved states, although we cannot assess predictive accuracy directly, the predictions remain qualitatively consistent with the expected behavior. For example, \({T_\text{bl,in,2}}\) shows variation similar to that of \({T_\text{bl,in,1}}\), while \({T_\text{bl,in,3}}\) follows the trend of the corresponding input signal \({T_\text{p,ext,3}}\). We observe similar behavior for the other unobserved states \({T_\text{bl,in,4}},{T_\text{bl,in,6}}\), and \({T_\text{bl,in,8}}\). Therefore, the inferred trajectories of the unobserved states are physically plausible and consistent with the learned model dynamics. However, because no corresponding measurements are available for these states, their accuracy cannot be independently verified using the current dataset. We do not interpret the dynamics of the additional states \(T_{\text{10,m}},~T_{\text{IE,m}},~T_{\text{floor,m}},~T_{\text{ceiling,m}}\) because they need not have direct physical interpretability. 

The offline calibration results confirm that the physical subsystem model captures the measured thermal dynamics with sufficient accuracy. We therefore proceed to synthetic trade-off studies to analyze the behavior of the digital twin under controlled impact scenarios and to identify suitable online inference settings.

\section{Synthetic Examples with Trade-off Studies}
\label{section: synthetic example trade studies}
In this section, we emulate online inference with the digital twin to identify suitable hyperparameters using synthetic temperature data. We also study the effects of varying system observability and observation noise.

\begin{figure}[h!]
\centering
\includegraphics[width=1\linewidth]{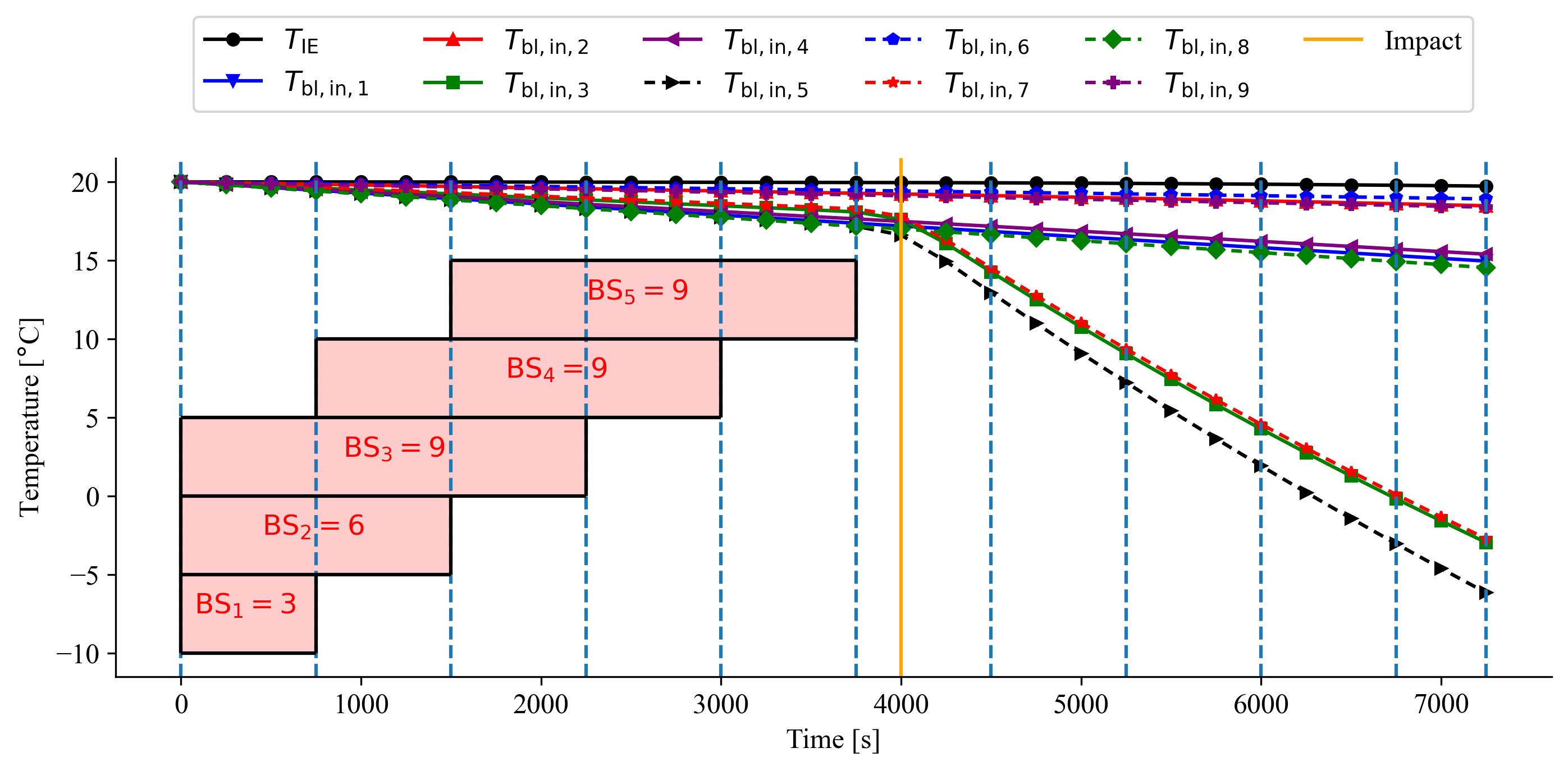}
\caption{Synthetic temperature trajectories generated without noise using the digital twin model for the impact scenario considered in the synthetic studies. The impact damages the SPL portions associated with panels 3, 5, and 7.}
\label{fig: synthetic data plot}
\end{figure}

We use the model presented in Section \ref{section: digital twin model} to infer the health-state of the SPL online and to perform temperature prediction and TTC estimation with quantified uncertainty. The digital twin model parameters include the calibrated physical subsystem model parameters together with \(c_1,~c_2,~c_3,~l_{\text{SPL,j}},t_{\text{impact,j}},\Delta l_{\text{SPL,j}}\) for j = 1,2,...,9. We fix the physical subsystem model parameters at their posterior median values obtained from the offline calibration. We set \(l_{\text{SPL,j}} = l_{\text{SPL,j,true}}\), consistent with the finite element model value of 0.2 used in \cite{montoya2025validation}. In the synthetic examples, we take \(\Delta l_{\text{SPL,j}} = 0.15\) and a common \(t_\text{impact} = 4000\) s for all panels associated with damaged portions of the SPL. We set the time scale to \(t_s = 7500\) s. To ensure numerical stability and avoid divergence of the ODE solver, we scale the \(\eta\) parameters of the physical subsystem model by 0.01 and the \(\gamma\) parameters by 0.1. We then perform a Sobol sensitivity analysis to examine the influence of \(c_1\), \(c_2\), and \(c_3\) relative to the impact-relevant parameters \(l_{\text{SPL,j}}\) and \(\Delta l_{\text{SPL,j}}\). The results, presented in Appendix \ref{appdx: sensitivity analysis}, indicate that when all five parameters are varied simultaneously, the system response is strongly dominated by \(c_1\), which introduces confounding with \(\Delta l_{\text{SPL,j}}\) in the post-impact regime. In contrast, fixing \(c_1\), \(c_2\), and \(c_3\) yields a physically meaningful sensitivity structure in which \(l_{\text{SPL,j}}\) predominantly controls the pre-impact response and \(\Delta l_{\text{SPL,j}}\) predominantly controls the post-impact response. Based on these findings, we treat \(c_1\), \(c_2\), and \(c_3\) as fixed constants during online inference and infer only the impact-relevant parameters \(l_{\text{SPL,j}}\) and \(\Delta l_{\text{SPL,j}}\). We set \(c_1 = 0.025,~c_2=1.\) and \(c_3 = 0.1\). For the selected values of \(l_{\text{SPL,j}}\), \(\Delta l_{\text{SPL,j}}\), and \(t_s\), the chosen parameter values and scaling factors produce temperature-drop magnitudes that are representative of those observed in the real testbed experiments. We selected the values of \(c_1\), \(c_2\), \(c_3\) and scaling factors based on trial-and-error studies. Consequently, the results presented in this work should be interpreted as conditional on these chosen parameter values. Estimating \(c_1\), \(c_2\), and \(c_3\) either through a dedicated calibration procedure prior to online inference or jointly with the impact-relevant parameters during online inference is beyond the scope of the present study and represents a direction for future work. We generate synthetic data using the previously described ODE solver with a time step of 250 s. To preserve differentiability, we approximate the Heaviside function in Equation \ref{eqn: impact indicator function} using the following parameterized sigmoid function \(\hat{H}(x) = 1/(1+e^{-ax})\) where we set \(a=100\) in this work. Figure \ref{fig: synthetic data plot} shows the synthetic temperature data generated by the digital twin model. For the synthetic examples, we initialize all 17 temperature states of the digital twin model at \(20^\circ\text{C}\) representing the nominal condition.

In the synthetic examples, the impact damages portions of the SPL associated with panels 3, 5, and 7. This damage reduces the effective thickness of the affected SPL regions and thereby reduces their effective insulation capability. The resulting cold flux then causes the temperatures of the corresponding bladder interior surfaces to drop sharply. The steep decrease in \(T_{\text{bl,in,3}},~T_{\text{bl,in,5}},~T_{\text{bl,in,7}}\) after \(t_{\text{impact,j}}\) in Figure \ref{fig: synthetic data plot} confirms this expected behavior.

\subsection{Time-to-Critical Estimation}
\label{section:ttc estimation}
To achieve the vision of autonomous space habitats, what-if analysis and decision-making under uncertainty are essential capabilities. As shown in Figure \ref{fig: digital twin overview}, the digital twin can support the decision-making by serving as a source of structured and up-to-date knowledge about the system state \cite{gratius2024twin}. TTC estimation based on future state predictions with quantified uncertainty can further support more informed decision-making. For example, \cite{mirfarah2025pressure} estimates pressure trajectories and uses a Kalman filter to compute TTC during an impact scenario, while \cite{gratius2023learned} estimates temperature time-to-failure using a probabilistic graphical model digital twin under a temperature-controller-failure scenario.

In deep space settings, the habitat must maintain a healthy and comfortable IE to support crew safety and mission success. Several organizations have proposed acceptable ranges for interior-environment variables such as temperature, pressure, humidity, oxygen, and \(\text{CO}_2\) levels based on crew comfort, productivity, and safety \cite{rhee2025quantitative}. These standards include the Deep Space Habitability Design Guidelines developed from the NASA NextSTEP Phase 2 ground test program \cite{gernhardt2019deep} and the Baseline Values and Assumptions Document \cite{ewert2022life}. A comprehensive summary of these ranges is provided in \cite{rhee2025quantitative}.

In this work, we use the TTC estimation to demonstrate the capability of the digital twin to support what-if analysis. Specifically, we ask: what is the remaining time for \(T_{\text{bl,in,j}}\) to reach the critical value \(T_{\text{bl,in,j,crit}}\) if a damage is reported on the SPL associated with the \(\text{j}^\text{th}\) panel? We obtain posterior temperature predictions with quantified uncertainty through online inference and use them to estimate TTC for \(T_{\text{bl,in,j}}\). TTC is defined as the difference between the time at which \(T_{\text{bl,in,j}}\) reaches \(T_{\text{bl,in,j,crit}}\) and the current time step. Mathematically, TTC at the current time step \(t_k\) can be expressed as \cite{lei2018healtha},
\begin{equation}
    \label{eqn: ttc expression}
    \text{TTC}(t_k) = \text{inf}(t:T_{\text{bl,in,j},t+t_k} \leq T_{\text{bl,in,j,crit}})
\end{equation}
where inf(\(\cdot\)) denotes the inferior limit, and \(T_{\text{bl,in,j},t+t_k}\) is the predicted \(T_{\text{bl,in,j}}\) at time step \(t+t_k\). In this work, we take \(T_{\text{bl,in,j,crit}} = -1\celsius\). We choose this value as a representative threshold to demonstrate the TTC estimation capability of the digital twin. In the synthetic examples and in the validation using experimental data, we report TTC estimates starting from the first successful detection. We make this choice because, under the nominal condition, TTC estimates are generally unavailable or uninformative, since the temperature does not decrease sufficiently to reach \(T_{\text{bl,in,crit}}\) within the prediction horizon considered.

Having defined the TTC metric used in this work, we next describe how the digital twin parameters are updated continuously during online inference. This sequential updating framework is central to anomaly detection, posterior prediction, and TTC estimation.

\subsection{Continuous Bayesian Updating}
\label{section: online inference setup}
Here, we describe the approach used for the continuous Bayesian updating of the digital twin. We update the digital twin parameters sequentially using batches of data. We define the batch size used for the j\(^{\text{th}}\) inference, \(\text{BS}_{\text{j}}\), as the number of observations per state, which is the same for all observed states. We denote the start and end times of the data used for inference by \(t_l\) and \(t_h\), respectively. The batch size \(\text{BS}_{\text{j}}\) varies from \(\text{BS}_{\text{min}}\) to \(N_{\text{BS}} \text{BS}_{\text{min}}\), where \(\text{BS}_{\text{min}}\) and \(N_{\text{BS}}\) are prescribed parameters. The digital twin starts the updating process with \(\text{BS}_{\text{1}} = \text{BS}_{\text{min}}\). For the j\(^{\text{th}}\) inference with \(1< \text{j} \leq N_{\text{BS}}\), we set \(\text{BS}_{\text{j}} = \text{j} \cdot \text{BS}_{\text{min}}\). For all \(\text{j} > N_{\text{BS}}\), we fix \(\text{BS}_{\text{j}} = N_{\text{BS}}\text{BS}_{\text{min}}\). Figure \ref{fig: synthetic data plot} illustrates the batch sizes for the first five inferences for the case \(\text{BS}_{\text{min}} = 3\) and \(N_{\text{BS}} = 3\).

We specify the priors as follows. For \(\text{j} \in \{1,...,9\}\), we set \(p(t_{\text{impact,j}}) = U(0,1)\), where \(t=0\) and \(t=1\) correspond to the dimensionless start and end times of the entire synthetic data. We adopt this prior because the constraint \(t_l \leq t_{\text{impact,j}} \leq t_h\) is already encoded in the digital twin through the indicator function in Eqn. \ref{eqn: impact indicator function}. If an impact occurs between \(t_l\) and \(t_h\), the posterior of \(t_{\text{impact,j}}\) narrows and concentrates within \([t_l,t_h]\); otherwise, it remains broadly distributed over \([0,1]\). For \(l_{\text{SPL,j}}\), we use a truncated normal prior concentrated at the true value:
\(p(l_{\text{SPL,j}}) = \text{TN}(\mu_{l_{\text{SPL,j}}}=l_{\text{SPL,j,true}},\sigma_{l_{\text{SPL,j}}}=10^{-3},l_{l_{\text{SPL,j}}}=l_{\text{SPL,j,true}}-2 \times 10^{-4} , h_{l_{\text{SPL,j}}}=l_{\text{SPL,j,true}}+2 \times 10^{-4})\).
This prior is effectively equivalent to fixing \(l_{\text{SPL,j}}\) at its true value \(l_{\text{SPL,j,true}}\). However, we retain it as a prior because explicitly fixing \(l_{\text{SPL,j}}\) to its true value reduces inference accuracy. For \(\Delta l_{\text{SPL,j}}\), we use the truncated normal prior
\(\text{TN}(\mu_{\Delta l_{\text{SPL,j}}}=0,\sigma_{\Delta l_{\text{SPL,j}}}=5l_{\text{SPL,j,true}},l_{\Delta l_{\text{SPL,j}}}=- \infty , h_{\Delta l_{\text{SPL,j}}}=l_{\text{SPL,j,true}})\).
The choice \(\mu_{\Delta l_{\text{SPL,j}}}=0\) reflects the prior expectation that impact on the SPL is a rare event. We set the standard deviation to \(5l_{\text{SPL,j,true}}\) to obtain a broad prior. Because \(\Delta l_{\text{SPL,j}} \leq l_{\text{SPL,j}}\), we set the upper limit to \(l_{\text{SPL,j}}\). We set the lower limit to \(-\infty\) because negative values are permitted and are mapped to zero through the ReLU function as mentioned in Section \ref{section: digital twin model}. This prior on \(\Delta l_{\text{SPL,j}}\) is similar in spirit to the mixture prior used in \cite{manikkan2025switch} for a non-negative parameter. For \(x_0\), we use \( p(x_0 \mid \text{inference 1}) = \mathcal{N}(0.9\textbf{1}_{n_x},0.16\textbf{I}_{n_x})\) in the first inference. For subsequent inferences, we use the posterior predictives from the past inferences to construct the initial-condition prior for the present inference, as indicated by the vertical blue dashed line in Figure \ref{fig: digital twin overview}. Specifically, we use \( p(x_0 \mid \text{j}^{\text{th}}\text{ inference},\text{j} > 1) = \mathcal{N}(x_{0,\text{median,j}^-},0.04\textbf{I}_{n_x})\), where \(x_{0,\text{median,j}^-}\) denotes the posterior predictive median state vector obtained from the \(\text{k}^{\text{th}}\) inference, with \(k = \max(0,\text{j}-N_{\text{BS}})\), and evaluated at the start time \(t_l\) of the \(\text{j}^{\text{th}}\) inference.

We perform all computations on a 14-inch Apple MacBook Pro with 32 GB RAM and an Apple M1 Pro chip. For all online inferences, we use a Markov Chain Monte Carlo (MCMC)-based approach. Specifically, we use the No-U-Turn Sampler (NUTS) \cite{hoffmanNoUTurnSamplerAdaptively2014} implemented in NumPyro \cite{phanComposableEffectsFlexible2019}. Each chain uses 1500 samples, including 500 warmup samples and 1000 retained samples. We run three chains in total. From the 3000 retained samples across all chains, we discard the first 500 samples and then thin every other sample, which leaves 750 posterior samples for analysis.

With the continuous updating strategy specified, we now introduce the metrics used to evaluate digital twin performance. These metrics quantify anomaly detection accuracy, computational efficiency, and the practical usefulness of the resulting predictions.

\subsection{Performance Metrics}
\label{subsection: metrics section}
We define the following metrics to evaluate the performance of the digital twin in inferring the thermal health-state of the SPL, predicting temperatures, and estimating TTC. Let \(\bS = \{I_j\}\) for j \(\in \{1,...,9\}\) denote the health-state configuration of the SPL portions associated with the nine habitat panels. Under the nominal condition, all elements of \(\bS\) are zero. Let \(\bS^{\text{hp,j}}\) denote the \(\text{j}^\text{th}\) highest-posterior-probability configuration among those obtained from the posteriors of \(\theta = \{ l_{\text{SPL,j}}, \Delta l_{\text{SPL,j}}, t_{\text{impact,j}}\}\) for j \(\in \{1,...,9\}\). A posterior sample of \(\bS\) is generated by evaluating the corresponding \(I_j\) values for each posterior sample of \(\theta\), thereby yielding a posterior distribution over possible health-state configurations.
\begin{itemize}
    \item[1)] \textit{Execution time} (\(t_\text{exec}\)): This metric is the cumulative runtime required to complete the Bayesian inference and therefore reflects the computational cost.
    \item[2)] \textit{Results Available Time} (\(t_\text{res}\)): This metric is the time at which the Bayesian inference is completed. We estimate \(t_\text{res}\) as \(t_\text{start} + t_{\text{exec}}\), where \(t_\text{start}\) is the time at which the Bayesian inference starts.
    \item[3)] \textit{Inference start time} (\(t_\text{start}\)): This metric is the time at which the Bayesian inference starts. We take \(t_\text{start} = t_h\) for the first inference and \(t_\text{start} = \text{max}(t_h, t_\text{res,j-})\) for the \(\text{j}^\text{th}\) inference. Here, \(t_\text{res,j-}\) is the \(t_\text{res}\) corresponding to the \((\text{j}-1)^\text{th}\) inference for \(\text{j}>0\). We neglect latencies associated with processes other than inference, such as posterior predictive computation, and TTC estimation.
    \item[4)] \textit{First Detection Time} (FDT): We define FDT as the \(t_\text{res}\) corresponding to the inference at which the first successful detection occurs. In the fully observed scenario, this time corresponds to the first post-impact detection with no false positives and no false negatives. In the partially observed scenario, false negatives are expected when \(T_{\text{bl,in,j}}\) corresponding to a panel with damaged SPL is not observed. Therefore, we consider a detection successful if it has no false positives and at least one true positive, even when false negatives are present. Because the success criterion is relaxed under partial observability, FDT should be interpreted as first reliable anomaly indication rather than full damage-configuration recovery.
    \item[5)] \textit{Configuration Accuracy} (CA): This metric measures the accuracy of \(\bS^{\text{hp,1}}\) relative to the true configuration \(\bS^{\text{true}}\), given by
    \begin{equation}
        \label{eqn: config accuracy}
         \text{CA} = \frac{TP + TN}{TP+TN+FP+FN} \times 100\%       
    \end{equation}
    where \(TP,TN,FP,\text{ and }FN\) denote the numbers of true positives, true negatives, false positives, and false negatives computed for \(\bS^{\text{hp,1}}\) with respect to \(\bS^{\text{true}}\).
\end{itemize} 
Using these metrics, we first determine suitable values of the online inference hyperparameters. We therefore begin with a study on the choice of batch-size-related parameters for continuous updating.

\subsection{Batch-Size Hyperparameter Study}
\label{section: BSmin, Nmin estimation}
\begin{figure*}[h!]
\captionsetup[subfigure]{justification=centering}
\centering
\begin{subfigure}{\textwidth}
  \centering
  \includegraphics[width=1\linewidth]{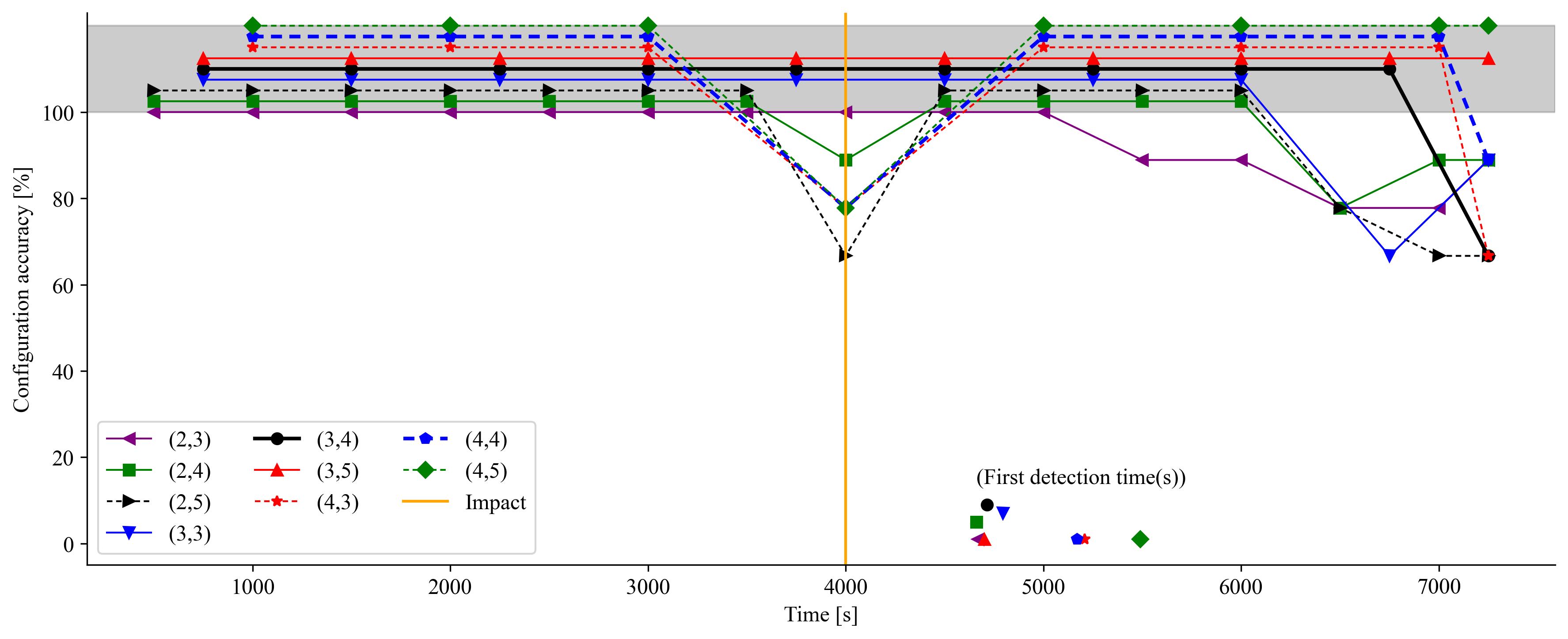}
  \caption{}
  \label{fig: synthetic data ca variation}
\end{subfigure}
\begin{subfigure}{\textwidth}
  \centering
  \includegraphics[width=1\linewidth]{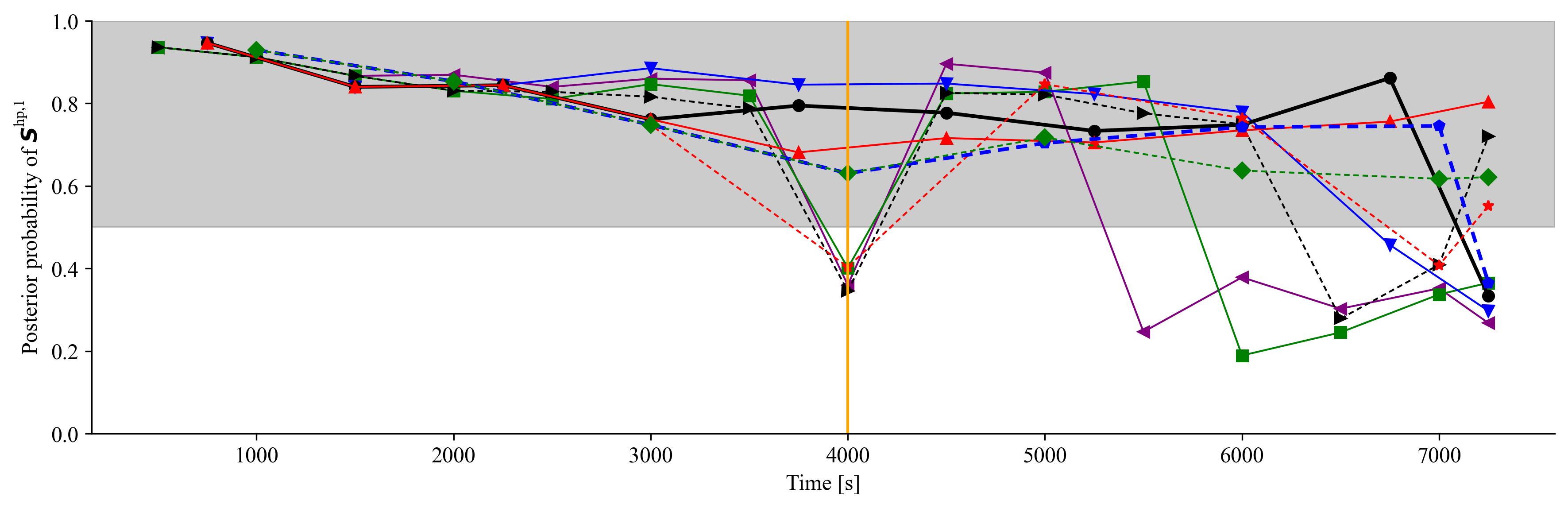}
  \caption{}
  \label{fig: synthetic data 1st config variation}
\end{subfigure}
\begin{subfigure}{\textwidth}
  \centering
  \includegraphics[width=1\linewidth]{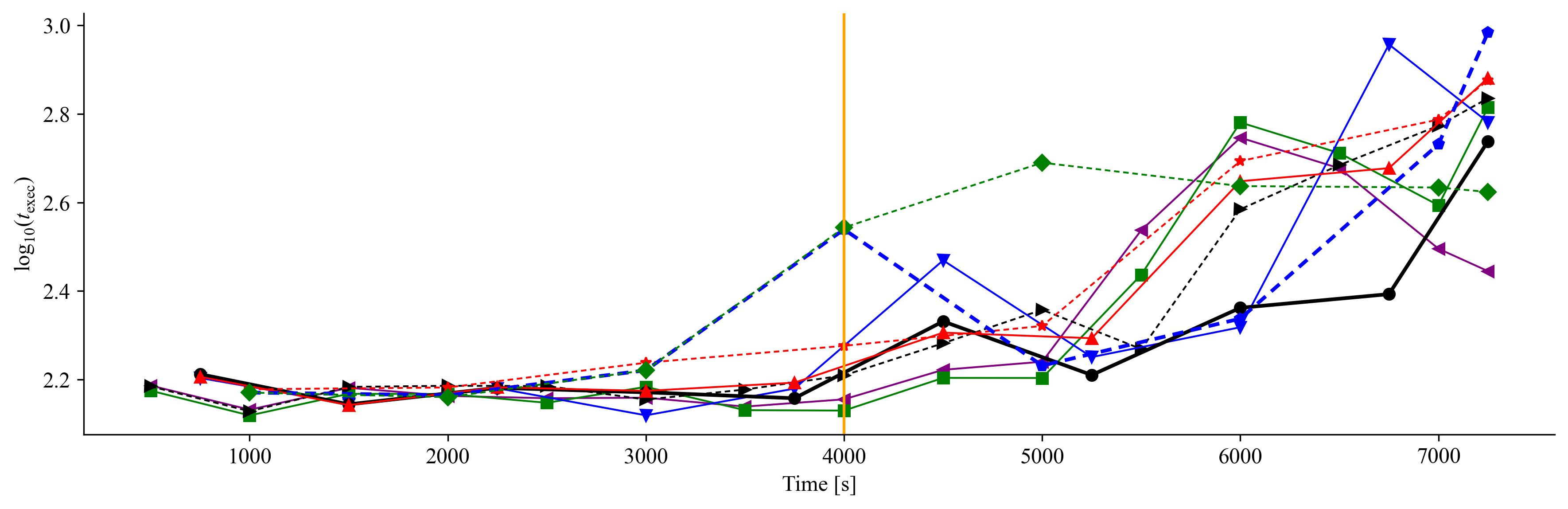}
  \caption{}
  \label{fig: synthetic data execution time variation}
\end{subfigure}
%\captionsetup{justification=centering}
\caption{Performance of the digital twin for different \((\text{BS}_{\text{min}}, N_{\text{BS}})\) combinations in the synthetic study: (a) configuration accuracy of the highest-posterior-probability health-state configuration \(S^{\text{hp,1}}\), (b) posterior probability \(P(S^{\text{hp,1}} \mid y)\), and (c) execution time \(t_{\text{exec}}\). The grey-shaded region in (a) denotes 100\% accuracy, and that in (b) denotes the interval [0.5,1].}
\label{fig: synthetic data ca, 1st config prob, execution time variations}
\end{figure*}

To identify suitable values of \(\text{BS}_{\text{min}}\) and \(N_{\text{BS}}\), we performed continuous updating of the digital twin parameters for all \((\text{BS}_{\text{min}},~N_{\text{BS}})\) combinations in \{2,3,4\} \(\times\) \{3,4,5\}. We treat \(T_{\text{IE}},~ T_{\text{bl,in,1}},~T_{\text{bl,in,3}},~T_{\text{bl,in,5}},~T_{\text{bl,in,7}},\) and \(T_{\text{bl,in,9}}\) as the observed temperatures. We add Gaussian white noise to all observations with standard deviation equal to 0.5\% of \(\bar{T}\), where \( \bar{T} = 20 \celsius\) is the scaling factor applied to the dataset. In the synthetic example, the impact damages portions of the SPL associated with panels 3, 5, and 7 at \(t_\text{impact}\). Therefore, for \(t \leq t_\text{impact}\), \(\bS^\text{true} = \{0,0,0,0,0,0,0,0,0\}\), whereas for \(t>t_\text{impact}\), \(\bS^\text{true} = \{0,0,1,0,1,0,1,0,0\}\). Figure \ref{fig: synthetic data ca variation} shows how the CA values vary with time for the different parameter combinations. We plot the CA values at the corresponding \(t_h\) of the data batch used for each inference. We plot the FDT values at the bottom of the same figure.

\begin{figure}[h!]
\centering
\includegraphics[width=1\linewidth]{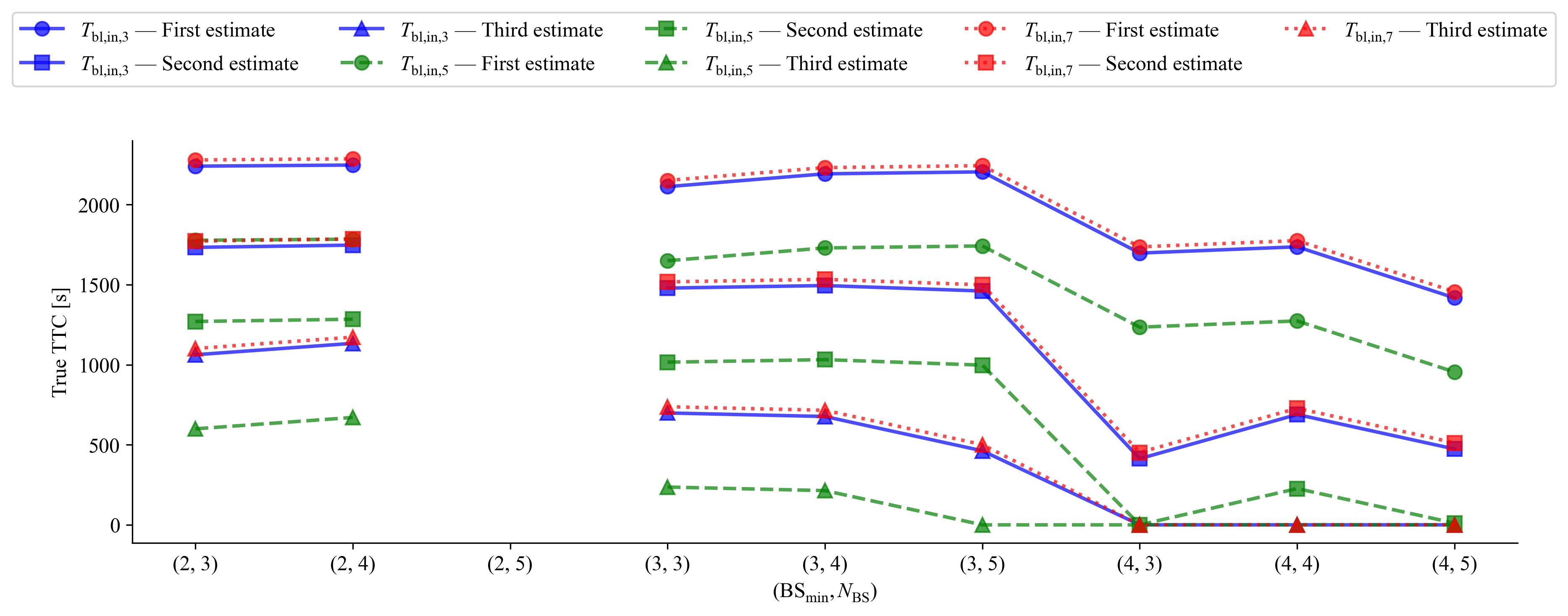}
\caption{True TTC estimates obtained from the synthetic \(T_{\text{bl,in,j}}\) observations for panels with SPL damage in the batch-size hyperparameter study.}
\label{fig: true ttc estimates bs_min nbs study}
\end{figure}

For \(t_h \leq t_{\text{impact}}\), the \((2,3)\) combination and all combinations with \(\text{BS}_{\text{min}} = 3\) achieve a 100\% CA value throughout the inference sequence, indicating perfect agreement between \(\bS^{\text{hp,1}}\) and \(\bS^{\text{true}}\). Among the remaining combinations, all except \((2,5)\) successfully identify the onset of the anomaly without generating false positives for the batch with \(t_h = t_{\text{impact}}\), despite the batch containing only partial information about the impact. In contrast, the \((2,5)\) combination produces a false-positive detection at \(t_h = t_{\text{impact}}\). Because this combination does not satisfy the anomaly-detection success criterion, we do not report
corresponding FDT and TTC estimates. Once the impact occurs, all combinations yield 100\% CA for the immediate inference with \(t_h > t_{\text{impact}}\), which indicates successful detection of the thermal anomaly. This successful detection persists for one to four subsequent inferences until the data batch no longer contains sufficient pre- and post-impact temperature variation for reliable detection. Once the data contain only post-impact information, the digital twin updating becomes unreliable. By that stage, recovery or repair actions should ideally have been taken, and the digital twin should be reset after the system recovers. Figure \ref{fig: synthetic data ca variation} shows that the CA values begin to drop near the end of the time window, where the data contain only post-impact information. The earliest FDT is obtained by (2,4) at 4660 s, whereas the latest is obtained by (4,5) at 5489s.

Figure \ref{fig: synthetic data 1st config variation} shows the variation of \(P(\bS^\text{hp,1} \mid y)\), the posterior probability of \(\bS^\text{hp,1}\). During the nominal condition, all combinations produce values above 0.5 and close to 1. Near \(t_\text{impact}\), the probability begins to decrease. For the combinations (2,3), (2,4), (2,5), and (4,3), \(P(\bS^\text{hp,1} \mid y)\) falls below 0.5, which indicates higher uncertainty and lower confidence in \(\bS^\text{hp,1}\). After \(t_\text{impact}\), the probabilities increase and remain above 0.5 when accurate detection occurs. As with CA, the probability values decrease again toward the end. This posterior probability provides a quantitative measure of confidence in the inferred health-state configuration. Such confidence measures are important for building trust in digital twin outputs before triggering inspection, repair, or model-reset actions. In autonomous habitat operation, these probabilities support risk-informed decision-making by distinguishing high-confidence detections from uncertain cases in which additional data or delayed intervention may be preferable.

Figure \ref{fig: synthetic data execution time variation} shows the variation of \(\log t_{\text{exec}}\) for all combinations. The \(t_{\text{exec}}\) values remain of the same order across all combinations during the nominal condition, indicating comparable computational requirements prior to the impact event. Following the impact, the execution times increase and exhibit oscillatory behavior. Consistent with this trend, the variability of \(t_{\text{exec}}\) across combinations is minimal during the nominal condition but becomes more pronounced after the impact. Initially, the digital twin selects the nominal condition model, in which the parameters \(\Delta l_{\text{SPL,j}}\) and \(t_{\text{impact,j}}\) are inactive. As a result, their posteriors remain close to the priors because the data contain no information to update them. The remaining parameter \(l_{\text{SPL,j}}\) has a very narrow prior, so its posterior also changes little. Consequently, posterior exploration is relatively fast during the nominal condition, which yields lower computational cost. Once the impact occurs, the data contain information about the anomaly, and the NUTS algorithm must explore a richer parameter space. In the first three inferences, even though the batch size increases, \(t_{\text{exec}}\) does not show a clear increase with batch size. Thus, the switching mechanism embedded in the digital twin not only selects the model automatically, but also makes the computational time scale with the complexity of the selected model.

Figure \ref{fig: true ttc estimates bs_min nbs study} shows the TTC estimates obtained using the synthetic \(T_{\text{bl,in,j}}\) observations for \(\text{j} = 3,5,7\), corresponding to the panels whose associated SPL regions are damaged by the impact event. We report the first estimate at FDT and the second and third estimates at \(t_\text{res}\) of the next two inferences. The \((2,5)\) combination is excluded from this analysis because it produces a false-positive detection, as discussed previously. For a given \(T_\text{bl,in,j}\) and a combination, the TTC estimates decrease from the first to the third estimate, as expected. For a given estimate, the values corresponding to \(T_{\text{bl,in,3}}\) and \(T_{\text{bl,in,7}}\) are similar, whereas those corresponding to \(T_{\text{bl,in,5}}\) are consistently smaller. Across all \(T_{\text{bl,in,j}}\), the combinations \((2,4)\) yields the largest true TTC values for the first, second, and third estimates. The first estimates for \((2,3)\), \((3,3)\), \((3,4)\), and \((3,5)\) are close to that of \((2,4)\), which is consistent with their relatively early FDT values. In contrast, \((4,5)\) yields the smallest TTC values for the first estimate, while \((4,3)\) yields the smallest TTC values for the second estimate. For the third estimate, the combinations \((4,3),~(4,4)\), and \((4,5)\) all result in zero TTC values. Larger \(t_{\text{exec}}\) values increase \(t_{\text{res}}\) and can therefore reduce the available TTC when the result becomes available. For a given estimate, the variation across combinations follows a broadly similar pattern for all \(T_{\text{bl,in,j}}\). However, the ranking of combinations changes between successive estimates because both the FDT and the computational requirements evolve differently across combinations.

Overall, the \((3,4)\) combination exhibits the best performance. It maintains a CA value of 100\% and \(P(\bS^\text{hp,1} \mid y) > 0.5\) across all inferences except the last. The computational time remains relatively low after the impact. The first TTC estimate is comparable to the highest value among all combinations, while the subsequent estimates remain reasonably large and non-zero. Therefore, we select \(\text{BS}_{\text{min}}=3,~N_{\text{BS}}=4\) as the optimal values for further analysis and evaluation on experimental data.

Figure \ref{fig: (3,3) posterior predictives} in Appendix \ref{appendix: synthetic example plots} shows the posterior predictives across all observed states and inferences for the combination (3,4). Figures \ref{fig: (3,3) t_impact_j posterior distributions} and \ref{fig: (3,3) dl_reg_j posterior distributions} in the same appendix show the corresponding posterior distributions of \(t_\text{impact,j}\) and \(\Delta l_{\text{SPL,j}}\) across the inferences. Up to inference 9, the training predictives of all states remain very close to the observed measurements, and the uncertainty band remains very tight. Up to inference 5, the training window does not contain the impact, so the digital twin selects the nominal condition model. This behavior is evident from the posterior distributions of \(t_\text{impact,j}\) and \(\Delta l_{\text{SPL,j}}\), which remain close to their priors. The digital twin therefore generates future predictions using the nominal condition model. For \(T_\text{IE}\), \(T_{\text{bl,in,1}}\), and \(T_{\text{bl,in,9}}\), the future predictions agree completely with the observed data. For \(T_{\text{bl,in,3}}\), \(T_{\text{bl,in,5}}\), and \(T_{\text{bl,in,7}}\), the future predictions agree only for \(t < t_\text{impact}\). For \(t > t_\text{impact}\), the predictions deviate because the training window does not yet contain information about the impact and the model remains nominal. In inferences 6 to 9, the training window contains the impact, and the digital twin accurately infers the impact parameters associated with the fault condition model. In Figure \ref{fig: (3,3) t_impact_j posterior distributions}, the posterior distributions of \(t_\text{impact,3},~t_\text{impact,5},~t_\text{impact,7}\) are highly concentrated around the true value. Likewise, the posterior distributions of \(\Delta l_{\text{SPL,3}},\Delta l_{\text{SPL,5}},\Delta l_{\text{SPL,7}}\) are concentrated around the true values without extending into the negative region. The future predictions then agree very well with the observations, as shown in Figure \ref{fig: (3,3) posterior predictives}. Inferences 9 and 10 fall in the regime where digital twin updating is no longer reliable, as discussed earlier. These results verify that the digital twin framework has been implemented correctly, which \cite{gratius2023learned} identifies as an important requirement. The results also confirm that the digital twin can support diagnosis and prognosis through Bayesian inference, which is a key characteristic identified for space-mission digital twins in \cite{gratius2024twin}.

After identifying suitable hyperparameter values, we next investigate how digital twin performance changes as the available measurements are reduced. This observability study helps assess robustness under practically relevant sensing constraints.
\begin{figure*}[h!]
\captionsetup[subfigure]{justification=centering}
\centering
\begin{subfigure}{\textwidth}
  \centering
  \includegraphics[width=1\linewidth]{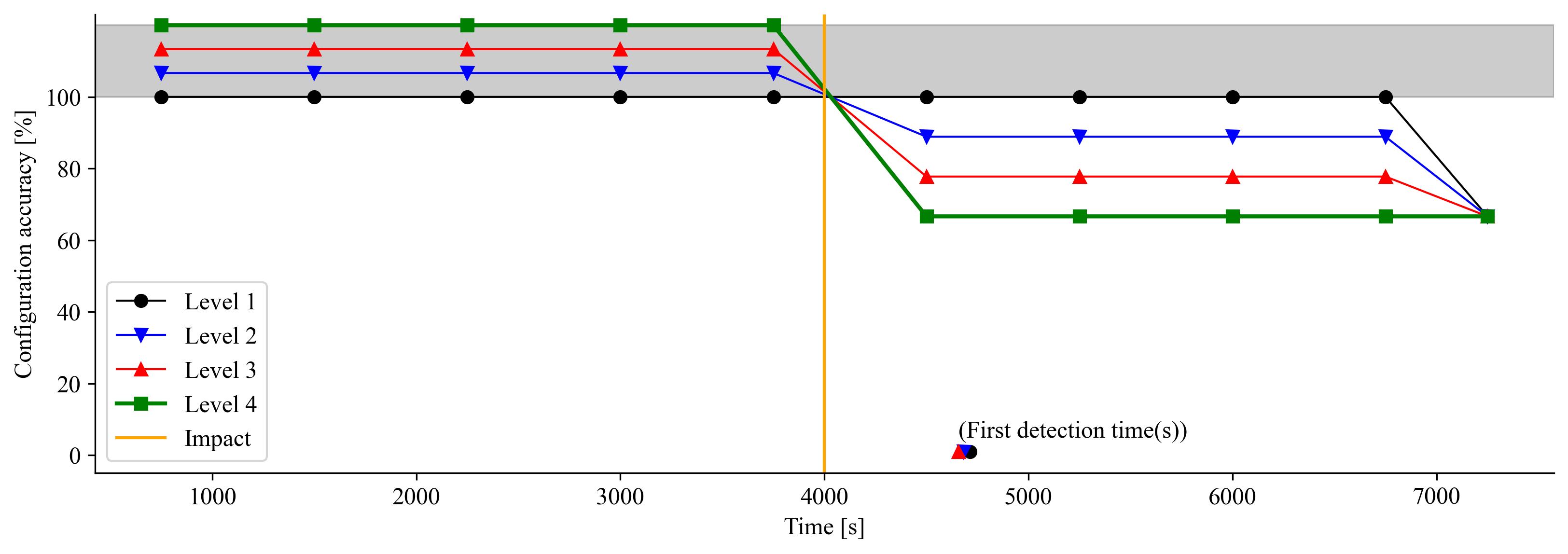}
  \caption{}
  \label{fig: synthetic data ca variation part obs}
\end{subfigure}
\begin{subfigure}{\textwidth}
  \centering
  \includegraphics[width=1\linewidth]{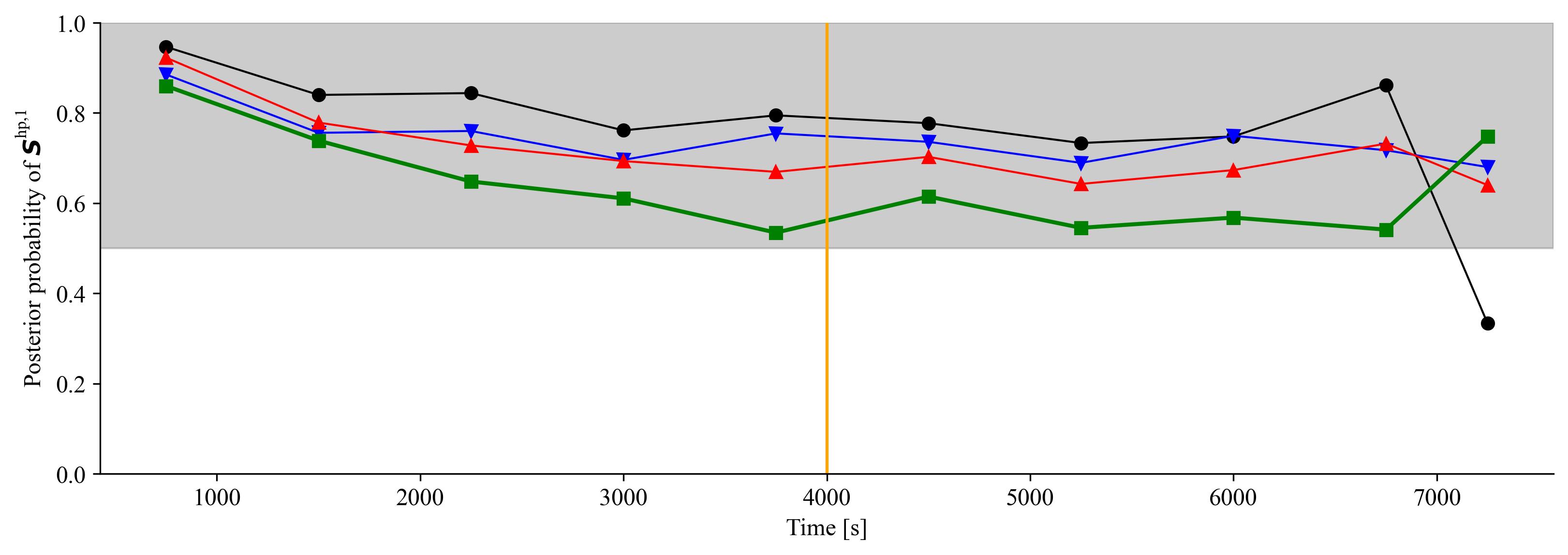}
  \caption{}
  \label{fig: synthetic data part obs 1st config prob variation}
\end{subfigure}
\begin{subfigure}{\textwidth}
  \centering
  \includegraphics[width=1\linewidth]{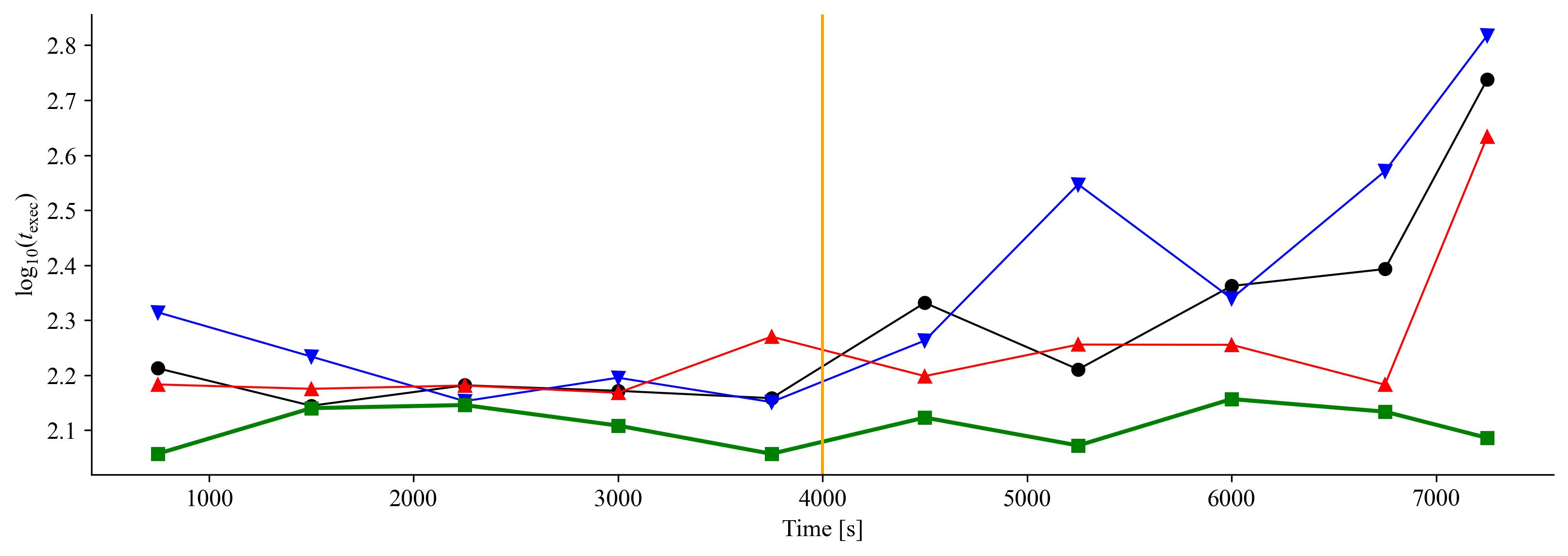}
  \caption{}
  \label{fig: synthetic data part obs execution time variation}
\end{subfigure}
%\captionsetup{justification=centering}
\caption{Performance of the digital twin under different observability levels in the synthetic study: (a) configuration accuracy of the highest-posterior-probability health-state configuration \(S^{\text{hp,1}}\), (b) posterior probability \(P(S^{\text{hp,1}} \mid y)\), and (c) execution time \(t_{\text{exec}}\). The grey-shaded region in (a) denotes 100\% accuracy, and that in (b) denotes the interval [0.5,1].}
\label{fig: synthetic data part obs CA, 1st config prob, execution time variations}
\end{figure*} 
\subsection{Observability Analysis}
\label{section: varying observability study}
Here, we present the performance of the digital twin under varying levels of system observability. Studying different observability scenarios is important because, in practical habitat operation, only a limited subset of temperatures may be available due to sensor placement, missing measurements, or subsystem instrumentation constraints. Sensor failures can further reduce observability \cite{wang2024faulta, zhang2026artificial}. Evaluating the digital twin under different observability levels therefore helps us assess how robust the Bayesian inference remains for anomaly detection and prognosis when the available system information is reduced.

We consider four observability levels. The observed-state sets for levels 1 to 4 are \(\{T_{\text{IE}},~T_{\text{bl,in,1}},~T_{\text{bl,in,3}},~T_{\text{bl,in,5}}\), \(T_{\text{bl,in,7}},~T_{\text{bl,in,9}}\}\), \(\{T_{\text{IE}},~T_{\text{bl,in,1}},~T_{\text{bl,in,3}},~T_{\text{bl,in,7}},~T_{\text{bl,in,9}}\}\), \(\{T_{\text{IE}},~T_{\text{bl,in,1}},~T_{\text{bl,in,7}},~T_{\text{bl,in,9}}\}\), and \(\{T_{\text{IE}}\}\), respectively. Level 1 is the same setting used in the previous section. Relative to level 1, level 2 excludes \(T_{\text{bl,in,5}}\), and level 3 excludes \(T_{\text{bl,in,3}}\) and \(T_{\text{bl,in,5}}\). As in the previous section, we add Gaussian white noise with standard deviation equal to 0.5\% of \(\bar{T}\) to all observations. The impact damages the portions of the SPL associated with panels 3, 5, and 7 at \(t_\text{impact}\), and therefore \(\bS^\text{true} = \{0,0,1,0,1,0,1,0,0\}\) after impact.

Figure \ref{fig: synthetic data ca variation part obs} shows the variation of CA across the different observability levels. Before the impact event, the CA values remain at 100\% for all observability settings. Following the impact event, and excluding the final inference, the CA value remains at 100\% for level 1 because \(T_{\text{bl,in,j}}\) corresponding to all panels with damaged SPL are observed. Consequently, the digital twin is able to identify the anomaly accurately, consistent with the results presented in the previous section. For the remaining observability levels, the CA values decrease after the impact and then remain approximately constant until inference 9. The magnitude of this reduction increases progressively from level 2 to level 4, indicating a degradation in damage-localization capability as the amount of available temperature information decreases. As described in Section \ref{subsection: metrics section}, for levels 2 and 3, a successful detection is the first post-impact detection with no false positives. For both levels 2 and 3, the first successful detection occurs in inference 6, with \(\bS^\text{hp,1}, P(\bS^\text{hp,1} \mid y)\) equal to \(\{0,0,1,0,0,0,1,0,0\}, 0.74\) and \(\{0,0,0,0,0,0,1,0,0\}, 0.70\), respectively. 

The FDT values corresponding to levels 1 through 3 are relatively similar, indicating comparable anomaly-detection responsiveness despite the reduction in observability. In contrast, no successful detection is achieved for level 4, where only \(T_{\text{IE}}\) is observed. Beginning with inference 6, the digital twin consistently predicts a nominal health-state configuration and therefore fails to identify any true positive damage location. As a result, no FDT value is reported for level 4. This finding highlights the importance of distributed temperature measurements for reliable thermal-anomaly detection and damage localization.

Figure \ref{fig: synthetic data part obs 1st config prob variation} shows the variation of \(P(\bS^\text{hp,1} \mid y)\). Across all inferences, \(P(\bS^\text{hp,1} \mid y)\) is highest on average for level 1 and decreases progressively as the number of observed states is reduced. This behavior reflects the increased uncertainty in health-state estimation under reduced observability. Figure \ref{fig: synthetic data part obs execution time variation} shows the execution-time results. The \(t_{\text{exec}}\) values remain small and exhibit little variability during the nominal condition, mirroring the behavior observed in the previous study. Because level 4 utilizes only a single observed state, it consistently produces the lowest execution times. Following the impact event, the execution times for levels 1 to 3 increase as the inference problem becomes more challenging. In contrast, the execution times for level 4 remain nearly constant because the digital twin continues to use the nominal-condition model and does not transition to the fault-condition model.
\begin{figure}[h!]
\centering
\includegraphics[width=1\linewidth]{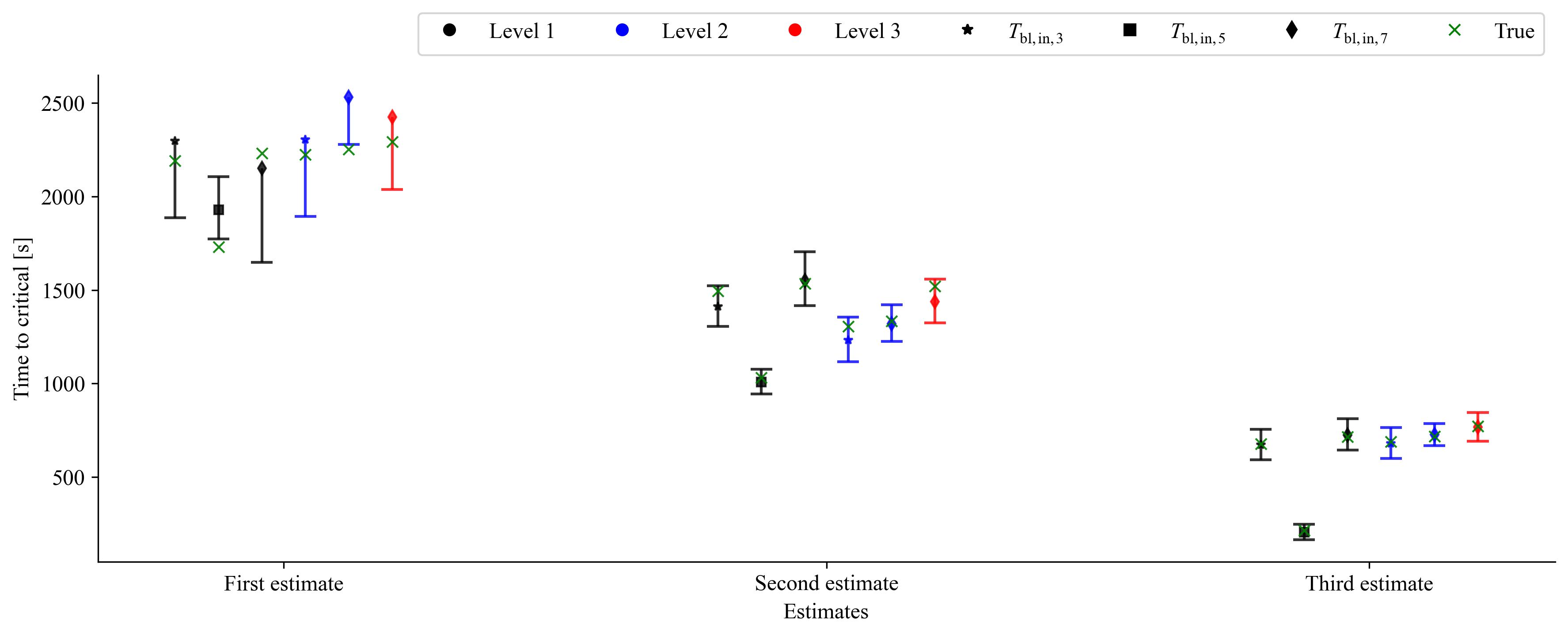}
\caption{Posterior TTC estimates for different observability levels, shown using 95\% credible intervals. The true TTC is computed from the noisy synthetic \(T_{\text{bl,in,j}}\) measurements.}
\label{fig: synthetic egs part obs ttc decomposition}
\end{figure}

Figure \ref{fig: synthetic egs part obs ttc decomposition} shows the TTC estimates derived from posterior future predictions of \(T_{\text{bl,in,j}}\) for different observability levels. The figure reports the 2.5, 50, and 97.5 percentile values of the posterior TTC distribution together with the corresponding true TTC estimates computed from the noisy synthetic observations. We report these TTC estimates for the inference at which successful detection first occurs and for the next two inferences. For each observability level, TTC estimates corresponding to the observed \(T_{\text{bl,in,j}}\) states are reported. We do not report TTC estimates for level 4 because no successful detection occurs. Most of the first TTC estimates do not have a 97.5 percentile value because the corresponding posterior predictive trajectories do not cross the \(T_{\text{bl,in,j,crit}}\) within the prediction horizon considered. Nevertheless, the true TTC values remain close to the resulting partial or complete credible intervals. For the second and third TTC estimates, complete 95\% credible intervals are obtained, and the posterior median estimates are very close to the corresponding true values. As expected, the TTC estimates progressively decrease from the first to the third estimate as the elapsed time since the impact event increases.

The observability study highlights the role of available measurements in anomaly detection and prognosis. We now complement this analysis by examining the effect of different observation-noise levels on digital twin performance.
\subsection{Observation Noise Analysis}
\label{section: varying noise study}
As a final synthetic study, we evaluate the continuous updating of the digital twin under varying levels of observation noise for all observability levels considered earlier. This noise analysis is important because temperature measurements in habitat testbeds are inevitably affected by sensor noise, estimation error, and other experimental uncertainties. By varying the noise level, we assess the robustness of parameter inference, anomaly detection, and temperature prediction, as well as the associated computational requirements, when the measured thermal response is imperfect.

We add Gaussian white noise with standard deviations of 0.05\%, 0.5\%, 5\%, and 50\% of \(\bar{T}\) to all observed temperatures.
\clearpage
\begin{figure*}[h!]
\captionsetup[subfigure]{justification=centering}
\centering
\begin{subfigure}{\textwidth}
  \centering
  \includegraphics[width=0.9\linewidth]{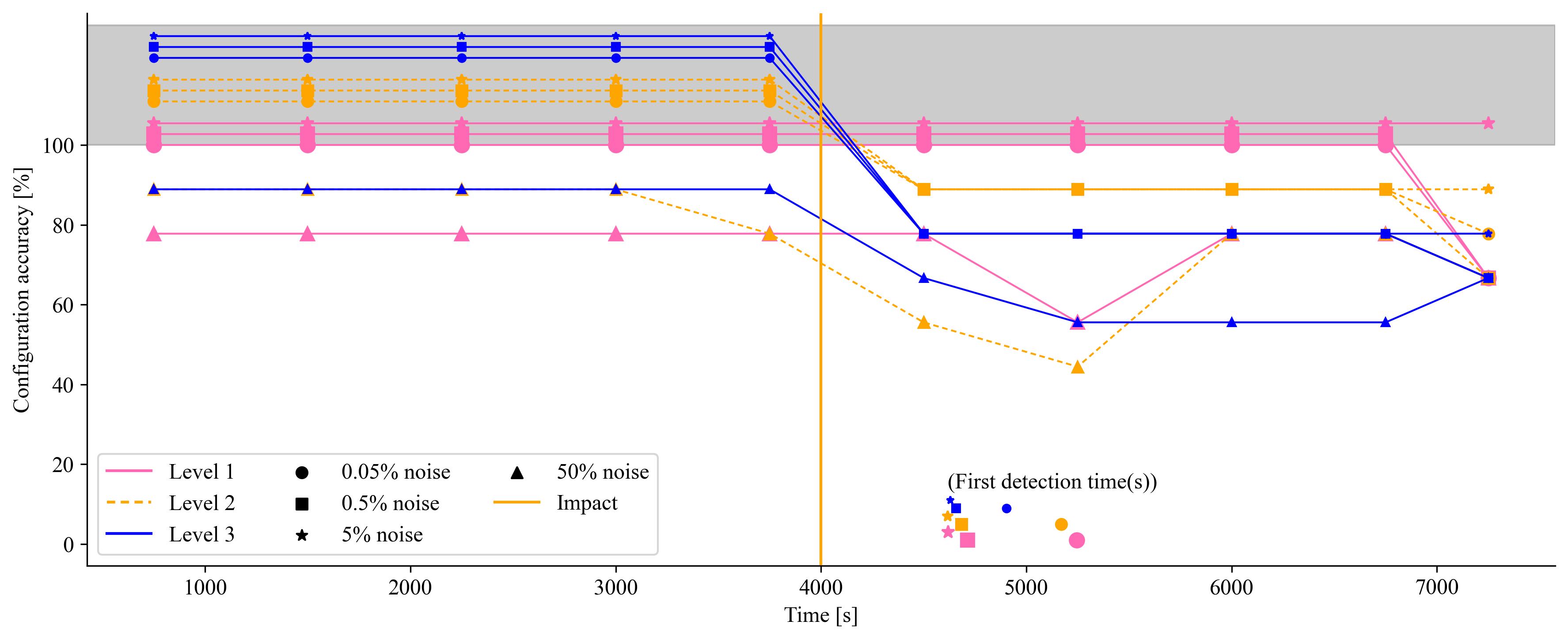}
  \caption{}
  \label{fig: synthetic data ca variation noise analysis}
\end{subfigure}
\begin{subfigure}{0.9\textwidth}
  \centering
  \includegraphics[width=1\linewidth]{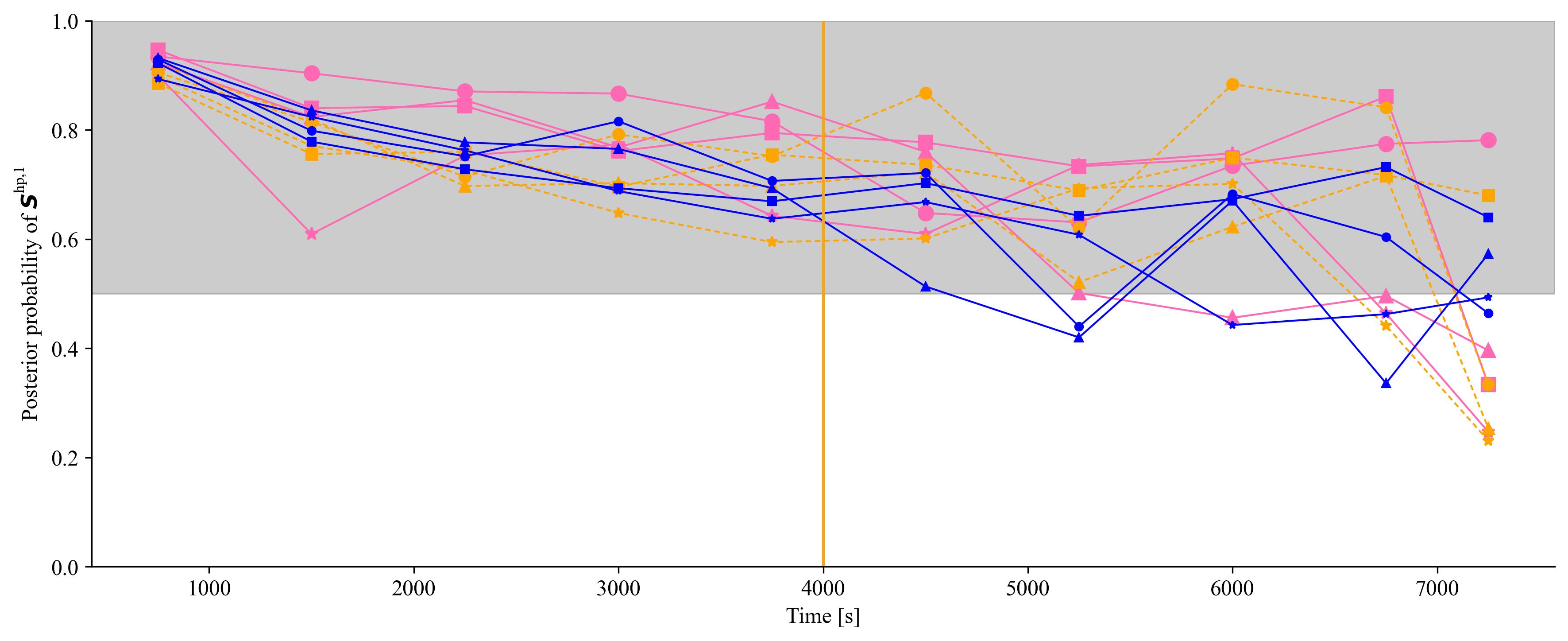}
  \caption{}
  \label{fig: synthetic data noise analy 1st config prob variation}
\end{subfigure}
\begin{subfigure}{0.9\textwidth}
  \centering
  \includegraphics[width=1\linewidth]{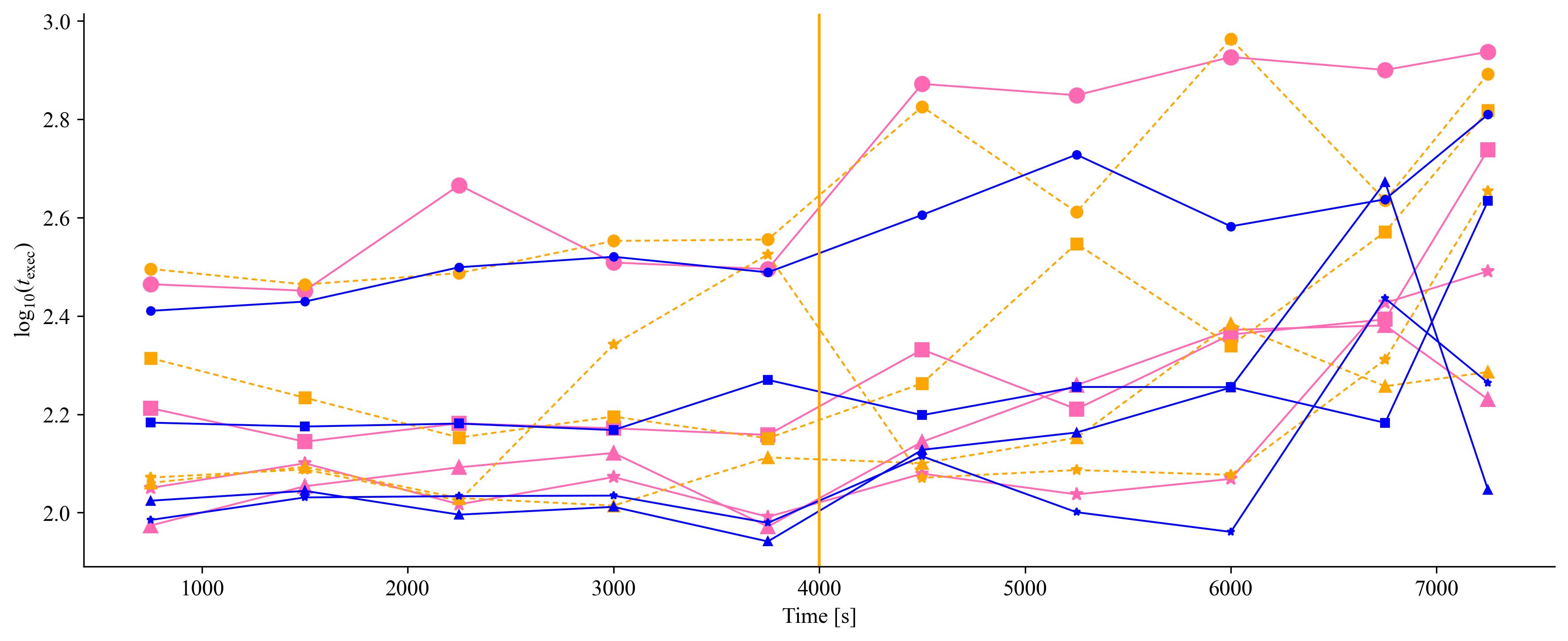}
  \caption{}
  \label{fig: synthetic data noise analy execution time variation}
\end{subfigure}
%\captionsetup{justification=centering}
\caption{Performance of the digital twin under different observation-noise levels in the synthetic study: (a) configuration accuracy of the highest-posterior-probability health-state configuration \(S^{\text{hp,1}}\), (b) posterior probability \(P(S^{\text{hp,1}} \mid y)\), and (c) execution time \(t_{\text{exec}}\). The grey-shaded region in (a) denotes 100\% accuracy, and that in (b) denotes the interval [0.5,1].}
\label{fig: synthetic data noise analysis CA, 1st config prob, execution time variations}
\end{figure*}
\clearpage
Figure \ref{fig: synthetic data ca variation noise analysis} shows the variation of CA. For all observability levels, the CA remains at 100\% during the nominal condition for noise levels up to 5\%. Following the impact event, level 1 maintains a CA value of 100\% for most inferences across noise levels between 0.05\% and 5\%, demonstrating robustness to measurement noise under full observability. In contrast, levels 2 and 3 experience reductions in CA over the same noise range, consistent with the observability-dependent degradation observed in the previous section. By contrast, the 50\% noise level yields low CA values during both nominal and fault conditions for all observability levels. Figure \ref{fig: synthetic data noise analy 1st config prob variation} shows the variation of \(P(\bS^\text{hp,1} \mid y)\). During the nominal condition, \(P(\bS^\text{hp,1}\mid y)\) remains confined to the interval \([0.5,1]\) for all observability and noise levels, indicating relatively low uncertainty in the inferred health-state configuration. Following the impact event, the probability values corresponding to the 5\% and 50\% noise levels frequently fall outside this interval, particularly during the later inferences. Across all observability levels, a gradual decrease in \(P(\bS^\text{hp,1}\mid y)\) is observed during the transition from the nominal condition to the fault condition, reflecting the increased uncertainty associated with damage identification after the impact. The effect of observation noise is also evident in the execution time. For a given observability level, \(t_{\text{exec}}\) generally decreases as the noise level increases from 0.05\% to 50\%, as shown in Figure \ref{fig: synthetic data noise analy execution time variation}. At the 50\% noise level, the execution times exhibit a mildly oscillatory behavior. When the noise level is very low, such as 0.05\%, the likelihood becomes sharply concentrated, which can make the posterior geometry harder for NUTS to explore and can increase computational time. When the noise level is very high, such as 50\%, the data contain less useful information, which leads to ill-posedness and degrades the estimation accuracy.

Figure \ref{fig: synthetic egs noise estimations level 1} in Appendix \ref{appendix: synthetic example plots} shows the variation of the 95\% posterior credible intervals of the observation noise for different noise levels under level 1 observability. For the first three noise levels, the credible intervals remain concentrated near the true value for most states and inferences. By contrast, for the 50\% noise level, the credible intervals lie far from the true value. Because the data points exhibit very large variance in this case, only a few data points are used to infer the model parameters using a reasonably low value of observation noise. This situation produces a large discrepancy between the estimated and actual observation noise.

The synthetic studies verify the implementation of the digital twin framework and clarify its behavior under controlled conditions. We now evaluate the same framework on experimental data from the HARSH testbed to assess its performance in a realistic setting.
\section{Experimental Validation}
\label{section: experimental validation}
In this section, we test the digital twin on experimental data obtained from the HARSH testbed described in Section \ref{section: Experiment setup} to evaluate its performance under realistic conditions for thermal anomaly detection, temperature prediction, and TTC estimation. For the experimental validation, we retain the same initial conditions adopted in the synthetic examples.
\begin{figure*}[h!]
\captionsetup[subfigure]{justification=centering}
\centering
\begin{subfigure}{\textwidth}
  \centering
  \includegraphics[width=0.9\linewidth]{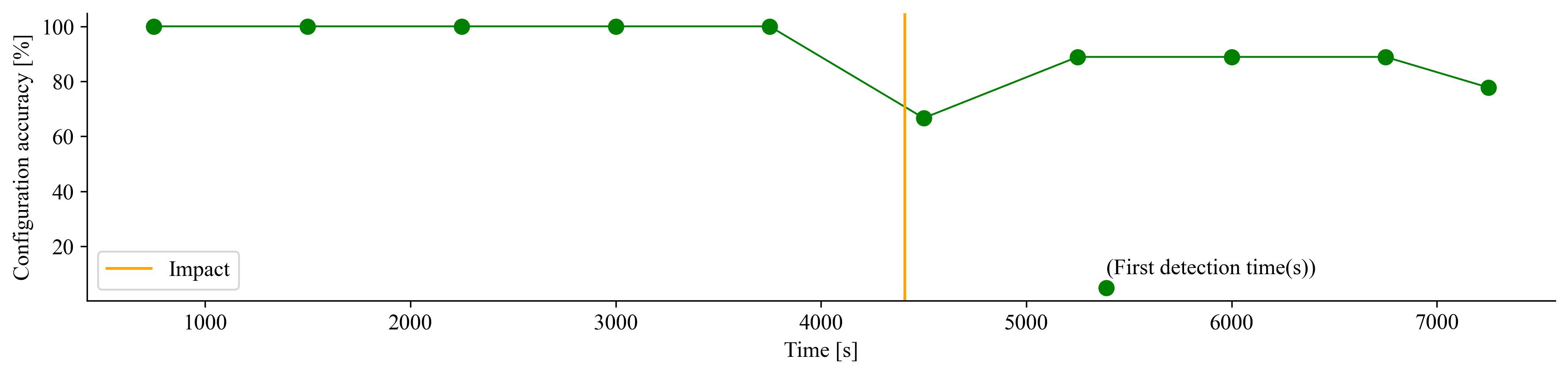}
  \caption{}
  \label{fig: expt data ca variation}
\end{subfigure}
\begin{subfigure}{0.9\textwidth}
  \centering
  \includegraphics[width=1\linewidth]{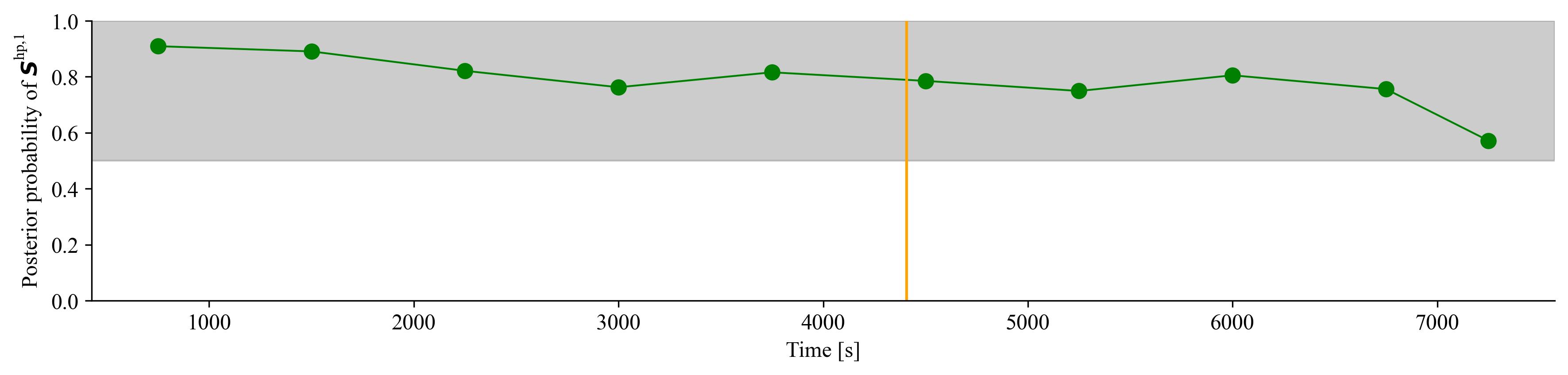}
  \caption{}
  \label{fig: expt data 1st config prob variation}
\end{subfigure}
\begin{subfigure}{0.9\textwidth}
  \centering
  \includegraphics[width=1\linewidth]{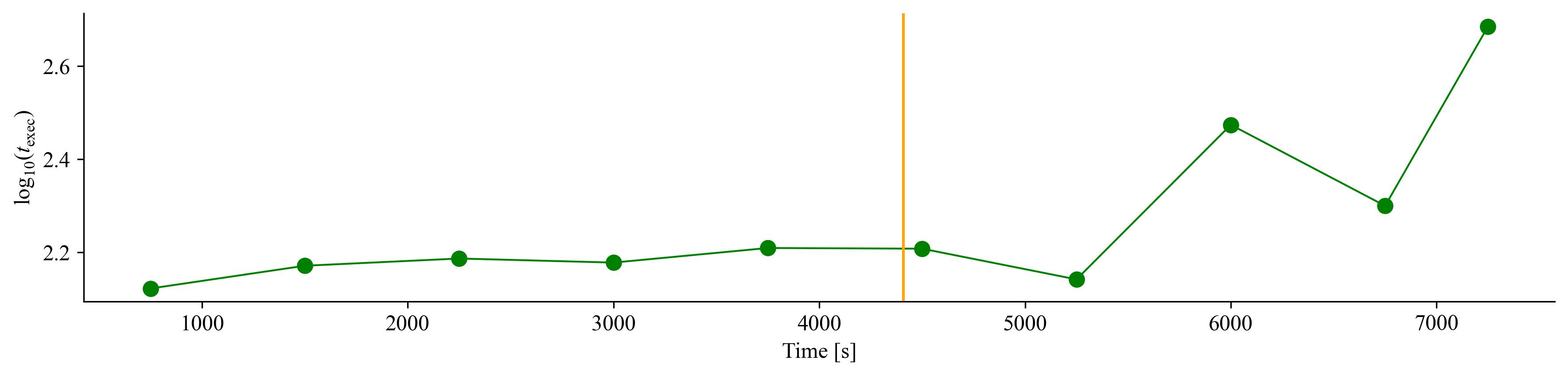}
  \caption{}
  \label{fig: expt data execution time variation}
\end{subfigure}
%\captionsetup{justification=centering}
\caption{Performance of the digital twin across inferences for the experimental validation case: (a) configuration accuracy of the highest-posterior-probability health-state configuration \(S^{\text{hp,1}}\), (b) posterior probability \(P(S^{\text{hp,1}} \mid y)\), and (c) execution time \(t_{\text{exec}}\). The grey-shaded region in (b) denotes the interval [0.5,1].}
\label{fig: expt data CA, 1st config prob, execution time variations}
\end{figure*}
We consider an impact scenario in which portions of the SPL corresponding to panels 7, 8, and 9 are damaged at 4408 s, resulting in an 87.5\% reduction in effective thickness, which corresponds to \(\Delta l_\text{SPL,j} = 0.175\) for \(j = 7,8,9\). These values correspond to known experimental scenarios reported in \cite{montoya2024thermomechanical}. We use the priors described in Section \ref{section: online inference setup} together with \(\text{BS}_{\text{min}}=3,~N_{\text{BS}}=4\). As noted in Section \ref{section: physical part model}, \(T_\text{bl,in,8}\) is not observed, which makes this case comparable to a partially observed scenario in the synthetic examples.

Figure \ref{fig: expt data ca variation} shows the CA values across the inferences. The CA remains at 100\% during the nominal condition and drops near the impact. It increases again once successful detection occurs in inference 7, where \(\bS^\text{hp,1} = \{0,0,0,0,0,0,1,0,1\}\), and it remains the same for inferences 8 and 9. Because \(T_\text{bl,in,8}\) is not observed, the digital twin does not report damage on the SPL corresponding to panel 8. Figure \ref{fig: expt data 1st config prob variation} shows the variation of \(P(\bS^\text{hp,1} \mid y)\). This probability remains within the [0.5,1] interval, which indicates good confidence. Gratius et al. propose that onboard computation should be completed faster than communication with ground control to provide an operational time advantage \cite{gratius2024calibration, gratius2026multimodela}. Figure \ref{fig: expt data execution time variation} shows the corresponding execution times. For the nominal-condition and anomaly-detection inferences, the execution time ranges from 2 to 7 minutes. These times are shorter than representative Mars-scale round-trip communication delays, which can reach approximately 44 minutes, but exceed the few-second delays associated with lunar missions \cite{mirfarah2025study}. Moreover, computational time alone does not determine whether onboard or ground-based execution is preferable, because the total response latency also includes data collection, data transmission, inference-result transmission, and subsequent decision and recovery actions which are not considered in this work. Ground-based computation is advantageous only when improvements in inference quality or information content outweigh the added communication delay \cite{mirfarah2025study}.

Figure \ref{fig: expt data posterior predictives} in the Appendix \ref{appendix: expt data plots} presents the posterior predictives, while Figures \ref{fig: expt data t_impact_j posterior distributions} and \ref{fig: expt data dl_reg_j posterior distributions} in the same appendix show the corresponding posterior distributions of \(t_\text{impact,j}\) and \(\Delta l_{\text{SPL,j}}\) across the inferences. Across all inferences, the training predictives include the experimental data. Up to inference 5, the training window does not contain the impact, and the digital twin therefore selects the nominal condition model. Consistently, the posterior distributions of \(t_\text{impact,j}\) and \(\Delta l_{\text{SPL,j}}\) remain close to their priors. For \(T_\text{IE}\), \(T_{\text{bl,in,1}}\), \(T_{\text{bl,in,3}}\), and \(T_{\text{bl,in,5}}\), the future predictions do not fully agree with the observed data at the beginning, but they improve as the batch size increases across the inferences. A similar pattern is observed for \(T_{\text{bl,in,7}}\) and \(T_{\text{bl,in,9}}\) for \(t < t_\text{impact}\). For \(t > t_\text{impact}\), the predictions do not match because the training window does not yet contain impact information and the digital twin continues to use the nominal condition model up to inference 6. From inference 7, where the training window contains impact information, the posterior samples of the inferred parameters are concentrated near the true value. In Figure \ref{fig: expt data t_impact_j posterior distributions}, the posterior distributions of \(t_\text{impact,7}\) and \(t_\text{impact,9}\) are concentrated near the true value. The posteriors of \(\Delta l_\text{SPL,7}\) and \(\Delta l_\text{SPL,9}\) show similar behavior. Figure \ref{fig: expt data noise estimation levels} in Appendix \ref{appendix: expt data plots} shows the variation of the 95\% posterior credible intervals for the estimated observation noise. The inferred noise levels are comparable to the 0.5\% to 5.0\% range observed in the synthetic examples.

\begin{figure}[h!]
\centering
\includegraphics[width=1\linewidth]{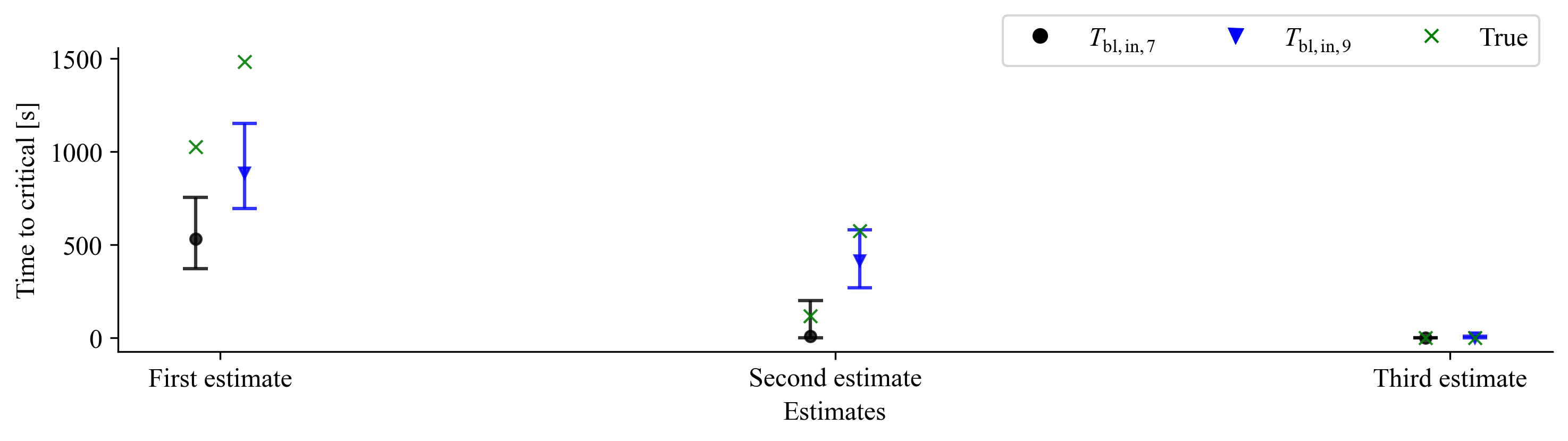}
\caption{Posterior TTC estimates for the experimental validation case, shown using 95\% credible intervals. The true TTC is computed from the measured \(T_{\text{bl,in,j}}\) values.}
\label{fig: expt data ttc decomposition}
\end{figure}

Figure \ref{fig: expt data ttc decomposition} demonstrates the credible intervals of the posterior TTC estimates together with the corresponding true values. We report these TTC estimates for the inference at which successful detection first occurs and for the subsequent two inferences. For the first TTC estimate, the inferred credible interval lies below the corresponding true TTC value, resulting in an underestimation. This behavior can be traced to the deviation between the posterior future predictions and the observed temperature trajectory during inference 7, as illustrated in Figure \ref{fig: expt data posterior predictives} in Appendix \ref{appendix: expt data plots}. As additional observations become available, the predictive accuracy improves, and the credible intervals corresponding to the subsequent TTC estimates successfully contain the associated true TTC values. In addition, the size of the credible intervals decreases for the second and third estimates relative to the first estimate. These results demonstrate that the digital twin can provide meaningful TTC estimates for this highly coupled system.

In the present digital twin, the offline-calibrated quantities and the model parameters \(c_1\), \(c_2\), and \(c_3\) are represented using single point estimates. The fact that the posterior samples of the impact-relevant parameters remain concentrated near their true values during anomaly detection suggests that the offline calibration procedure and the use of point estimates are adequate for the scenarios considered in this work. Nevertheless, the selected values of \(c_1\), \(c_2\), and \(c_3\) should not be interpreted as optimal values that maximize anomaly-detection or prognostic performance. In a fully autonomous space habitat setting, these parameters would ideally be inferred jointly with the impact-relevant parameters during online operation, thereby allowing the digital twin to continuously adapt to changing system conditions and modeling uncertainties \cite{wang2026digital}. Developing computationally efficient strategies to enable such joint online inference constitutes an important avenue for future work.

\section{Conclusions}
\label{section: concluding remarks}
We developed a Bayesian thermal digital twin for an extraterrestrial habitat cyber-physical testbed subjected to an impact event on the SPL and experiencing a thermal anomaly in the IE. The digital twin enables thermal anomaly detection, temperature prediction, and TTC estimation under an impact-driven disruption scenario. We constructed lumped parameter thermal network models for the physical and cyber subsystems and coupled them to form the digital twin. In the digital twin formulation, we combined the nominal and fault-condition models through physics-based activation functions to automate model selection and enable adaptation. We designed these activation functions from the physical constraints that the digital twin parameters must satisfy, and they also serve as health-state variables. Through them, the framework enables posterior sampling of the different health-state configurations that the system may experience. We formulated continuous updating of the digital twin as continuous Bayesian updating of the impact-relevant parameters in the cyber subsystem after offline Bayesian calibration of the physical subsystem model. We then carried out synthetic trade-off studies to examine the effects of batch size, system observability, and observation noise. The results showed that the proposed digital twin can identify the onset of thermal cascading effects induced by damage to the SPL, recover impact-related parameters with quantified uncertainty, provide informative future temperature predictions, and support TTC estimation in both synthetic and experimental settings. The framework also maintained useful performance under partial observability and realistic levels of measurement noise. At the same time, several limitations remain. Constructing health-state variables by leveraging physics-based constraints on model parameters may not be straightforward for all problems. The NUTS based inference required to obtain accurate posterior health-state configurations can be computationally expensive, which in turn limits the frequency of useful TTC estimates and can render some estimates unavailable. In addition, we did not consider uncertainty in the input signals and latent process noise. This work does not address digital-twin life-cycle management aspects, such as maintaining reliable long-term operation, updating the twin following major system changes such as degradation due to aging, or resetting and reinitializing the twin after anomaly detection. These aspects are important for sustained autonomous operation. Future work will focus on developing computationally more efficient Bayesian inference strategies that can provide more frequent and operationally useful estimates. Expanding the digital twin to a coupled temperature, pressure, and air concentration model to provide a more informative representation of the health state will also be considered.

\section*{CRediT authorship contribution statement}
\textbf{Sreehari Manikkan:} Conceptualization, Methodology, Investigation, Software, Validation, Visualization, Writing - original draft.
\textbf{Seungho Rhee:} Data curation, Investigation, Writing - review and editing.
\textbf{Herta Montoya:} Data curation, Investigation, Writing - review and editing.
\textbf{Davide Ziviani:} Supervision, Investigation, Funding acquisition, Writing - review and editing.
\textbf{Shirley J. Dyke:} Supervision, Investigation, Funding acquisition, Writing - review and editing.
\textbf{Ilias Bilionis:} Supervision, Conceptualization, Methodology, Investigation, Funding acquisition, Writing - review and editing. 

\section*{Declaration of competing interest}
The authors declare no competing interest.

\section*{Data availability}
Datasets will be made publicly available once the paper has been accepted for publication.

\section*{Declaration of generative AI and AI-assisted technologies in the writing process}
During the preparation of this work the author(s) used Chat GPT in order to correct spelling, grammatical, and syntactical errors. After using this tool/service, the author(s) reviewed and edited the content as needed and take(s) full responsibility for the content of the published article.

\section*{Acknowledgements}
\label{section: Acknowledgement}
This work was supported by a Space Technology Research Institutes Grant from National Aeronautics and Space Administration’s Space Technology Research Grants Program [grant number 80NSSC19K1076].

%\printbibliography
\bibliographystyle{elsarticle-num}
\bibliography{references}

\begin{appendices}
\section{Sensitivity Analysis}
\label{appdx: sensitivity analysis}
\begin{figure*}[h!]
\captionsetup[subfigure]{justification=centering}
\centering
\begin{subfigure}{\textwidth}
  \centering
  \includegraphics[width=1\linewidth]{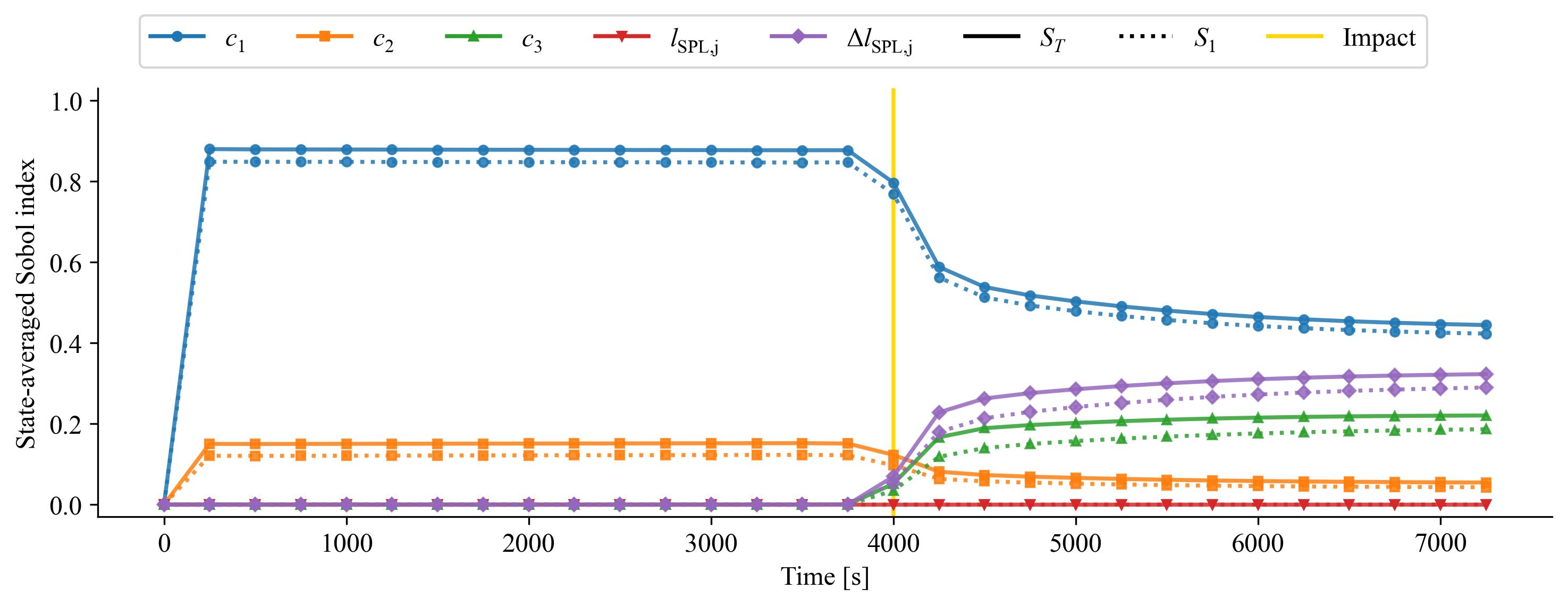}
  \caption{}
  \label{fig: synthetic data gsa all}
\end{subfigure}
\begin{subfigure}{\textwidth}
  \centering
  \includegraphics[width=1\linewidth]{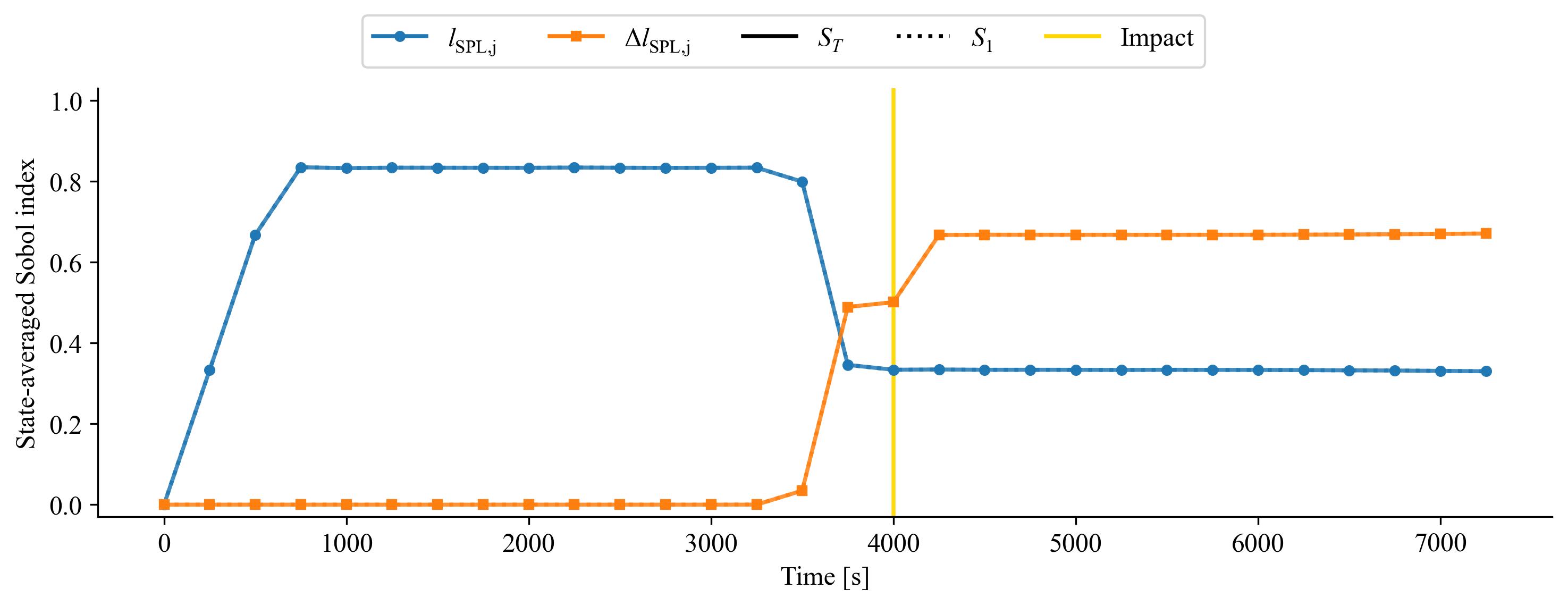}
  \caption{}
  \label{fig: synthetic data gsa fdd only}
\end{subfigure}
%\captionsetup{justification=centering}
\caption{State-averaged first-order \(S_1\) and total-order \(S_T\) Sobol indices over time using synthetic impact scenario data. (a) When the parameters \(c_1,~c_2\) and \(c_3\) are varied jointly with the parameters \(l_{\text{SPL,j}}\) and \(\Delta l_{\text{SPL,j}}\). (b) When only \(l_{\text{SPL,j}}\) and \(\Delta l_{\text{SPL,j}}\) parameters are varied with \(c_1 = 0.025,~c_2=1\) and \(c_3=0.1\). \(t_\text{impact,j}\) is fixed to 4000s.}
\label{fig: synthetic data gsa all fdd_only}
\end{figure*}

We perform a Sobol sensitivity analysis to investigate whether the parameters \(c_1\), \(c_2\), and \(c_3\) can be inferred jointly with the impact-relevant parameters \(l_{\text{SPL,j}}\) and \(\Delta l_{\text{SPL,j}}\) for \(\text{j}=1,\ldots,9\). Figure \ref{fig: synthetic data gsa all fdd_only} summarizes the results. The analysis uses the parameter ranges \(c_1,~c_2,~c_3 \in [0,0.1]\), \(l_{\text{SPL,j}} \in [0.199,0.2]\), and \(\Delta l_{\text{SPL,j}} \in [0,0.199]\). To isolate the effects of these parameters, \(t_{\text{impact,j}}\) is fixed at 4000 s, allowing the sensitivity structure to be examined separately in the pre-impact and post-impact regimes.

When all parameters are varied simultaneously, as shown in Figure \ref{fig: synthetic data gsa all}, the state-averaged Sobol indices indicate that \(c_1\) dominates the model response throughout most of the pre-impact interval and remains a major contributor after the impact. In the pre-impact regime, the total-order and first-order Sobol indices of \(c_1\) are nearly identical and close to unity, indicating that uncertainty in \(c_1\) directly explains most of the output variance. After the impact, the influence of \(\Delta l_{\text{SPL,j}}\) increases, but \(c_1\) remains higher and comparable in importance. Consequently, the observed temperature response cannot be uniquely attributed to the \(\Delta l_{\text{SPL,j}}\) parameter when \(c_1\) is also uncertain. The small Sobol indices associated with \(l_{\text{SPL,j}}\), \(c_2\), and \(c_3\) indicate that the temperature response is relatively insensitive to these parameters.

When \(c_1\), \(c_2\), and \(c_3\) are fixed to 0.025, 1, and 0.1, respectively, the resulting sensitivity structure becomes physically interpretable, as shown in Figure \ref{fig: synthetic data gsa fdd only}. Before the impact, the response is dominated by \(l_{\text{SPL,j}}\), which governs the baseline thermal dynamics. After the impact, the dominant sensitivity shifts from \(l_{\text{SPL,j}}\) to \(\Delta l_{\text{SPL,j}}\). Furthermore, the close agreement between the first-order and total-order Sobol indices indicates that the effects of \(l_{\text{SPL,j}}\) and \(\Delta l_{\text{SPL,j}}\) are largely direct rather than interaction-driven. These results motivate fixing \(c_1\), \(c_2\), and \(c_3\) during online inference before estimating \(l_{\text{SPL,j}}\) and \(\Delta l_{\text{SPL,j}}\).

\section{Synthetic Example Plots}
\label{appendix: synthetic example plots}
\begin{figure}[ht!]
\centering
\includegraphics[width=0.88\linewidth]{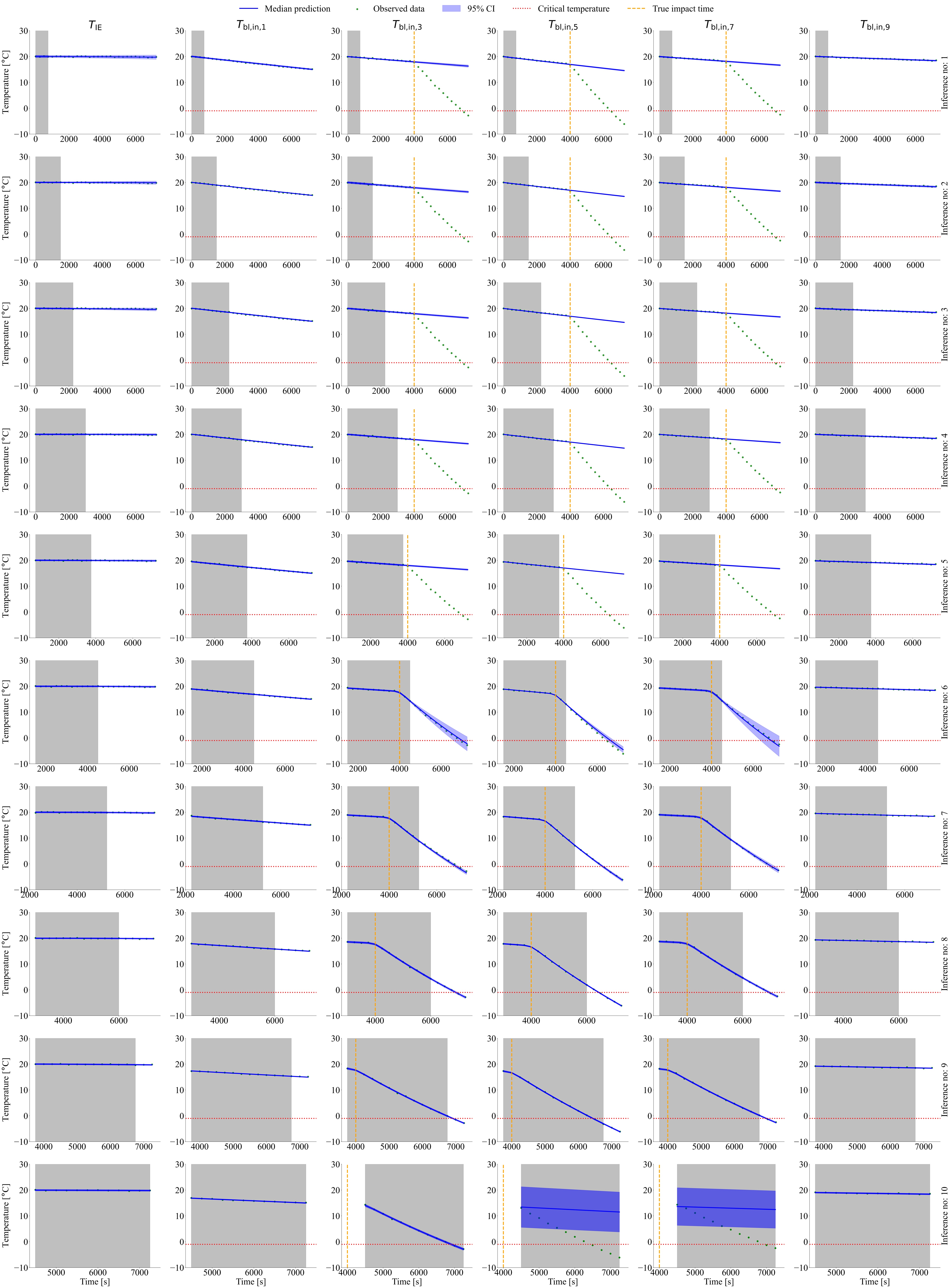}
\caption{Posterior predictive distributions of the observed temperatures across inferences for the \((\text{BS}_{\text{min}}, N_{\text{BS}}) = (3,4)\) combination in the synthetic study. The grey-shaded region represents the portion of the data used for Bayesian inference. }
\label{fig: (3,3) posterior predictives}
\end{figure}
\clearpage
\begin{figure}[ht!]
\centering
\includegraphics[angle=270,width=0.95\linewidth]{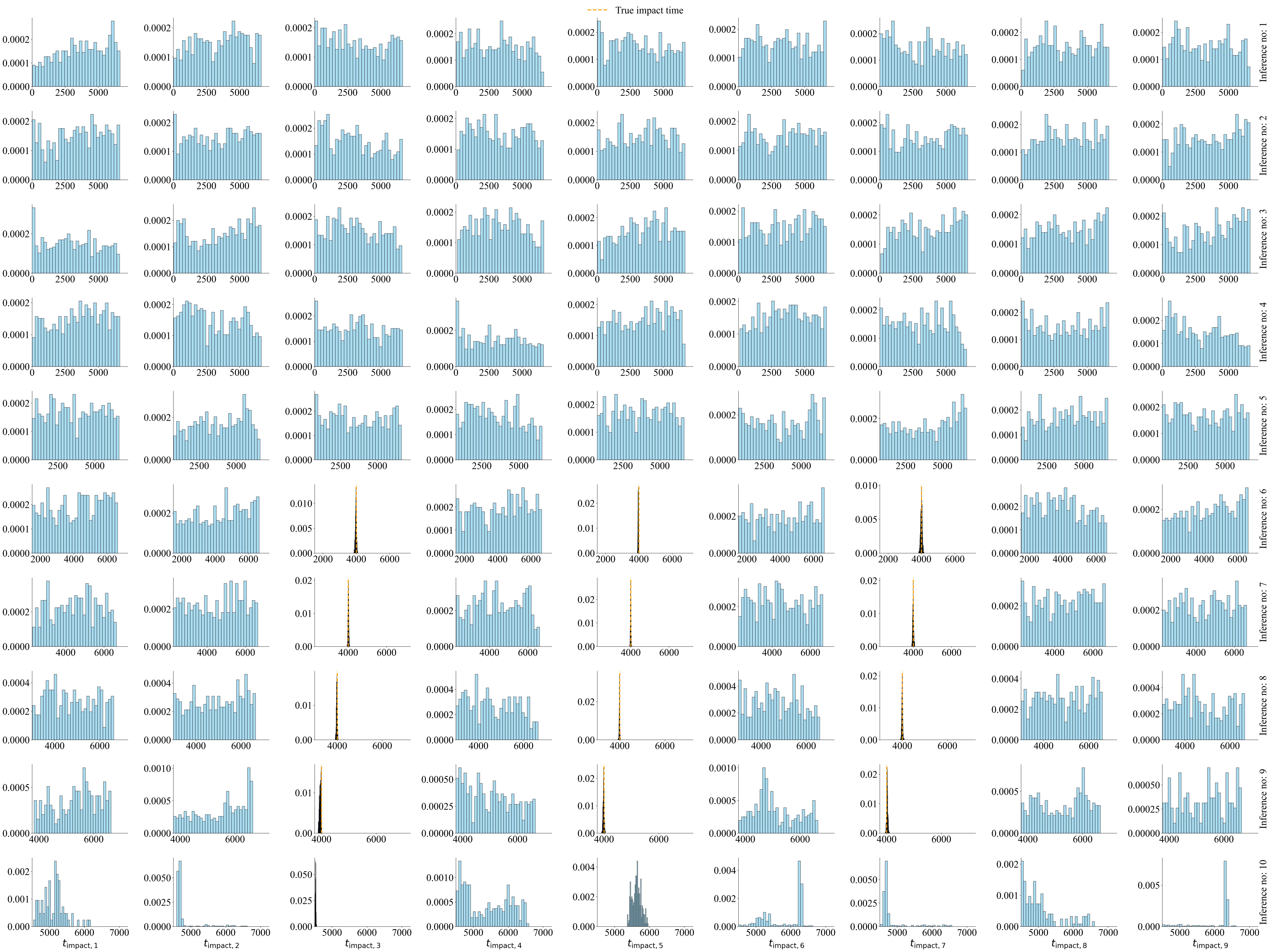}
\caption{Posterior distributions of the impact-time parameters \(t_{\text{impact,j}}\) across inferences for the \((\text{BS}_{\text{min}}, N_{\text{BS}}) = (3,4)\) combination in the synthetic study.}
\label{fig: (3,3) t_impact_j posterior distributions}
\end{figure}
\clearpage
\begin{figure}[ht!]
\centering
\includegraphics[angle=270,width=0.95\linewidth]{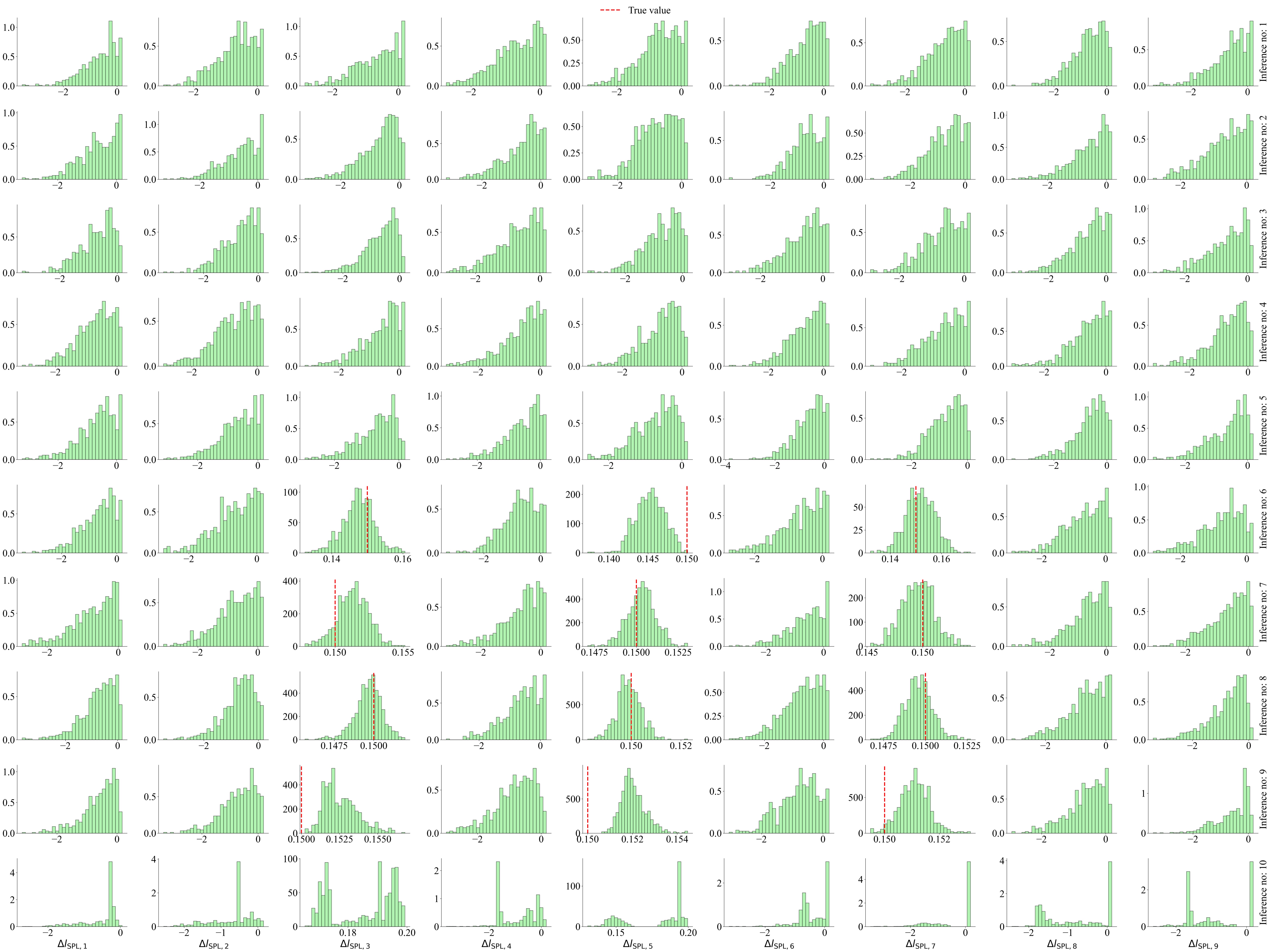}
\caption{Posterior distributions of the effective-thickness-reduction parameters \(\Delta l_{\text{SPL,j}}\) across inferences for the \((\text{BS}_{\text{min}}, N_{\text{BS}}) = (3,4)\) combination in the synthetic study.}
\label{fig: (3,3) dl_reg_j posterior distributions}
\end{figure}
%\clearpage
\begin{figure*}[ht!]
\captionsetup[subfigure]{justification=centering}
\centering
\begin{subfigure}{\textwidth}
  \centering
  \includegraphics[width=0.9\linewidth]{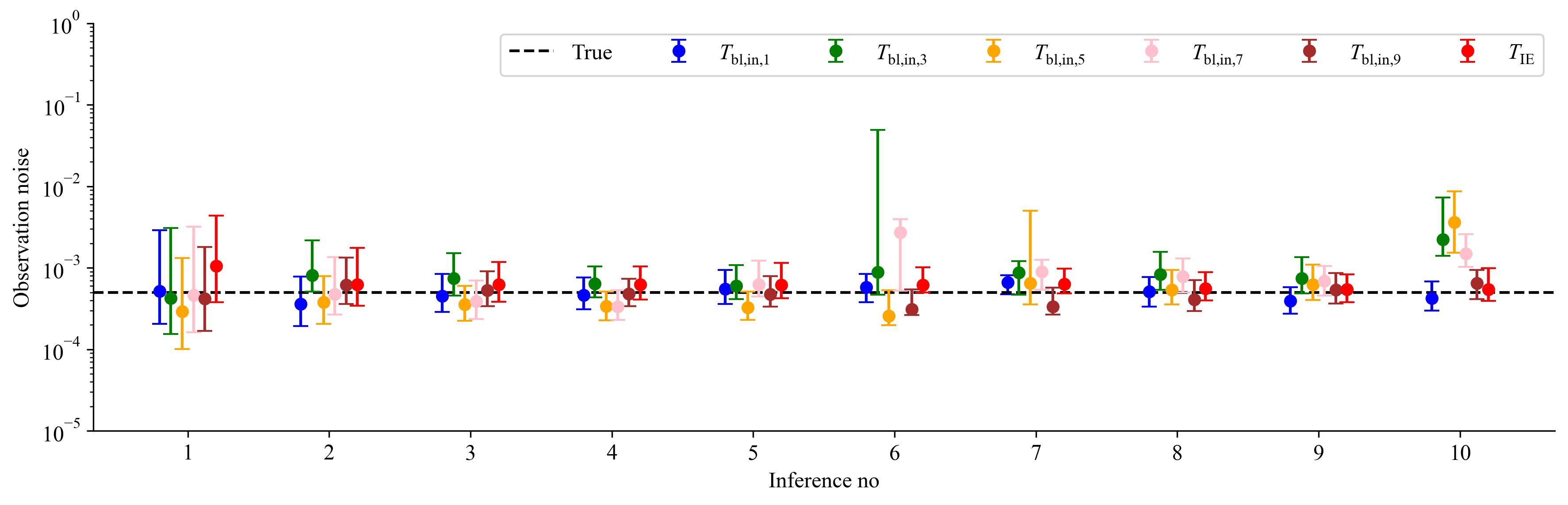}
  \caption{}
  \label{fig: synthetic data noise analy noise est 0.01 level 1}
\end{subfigure}
\begin{subfigure}{0.9\textwidth}
  \centering
  \includegraphics[width=1\linewidth]{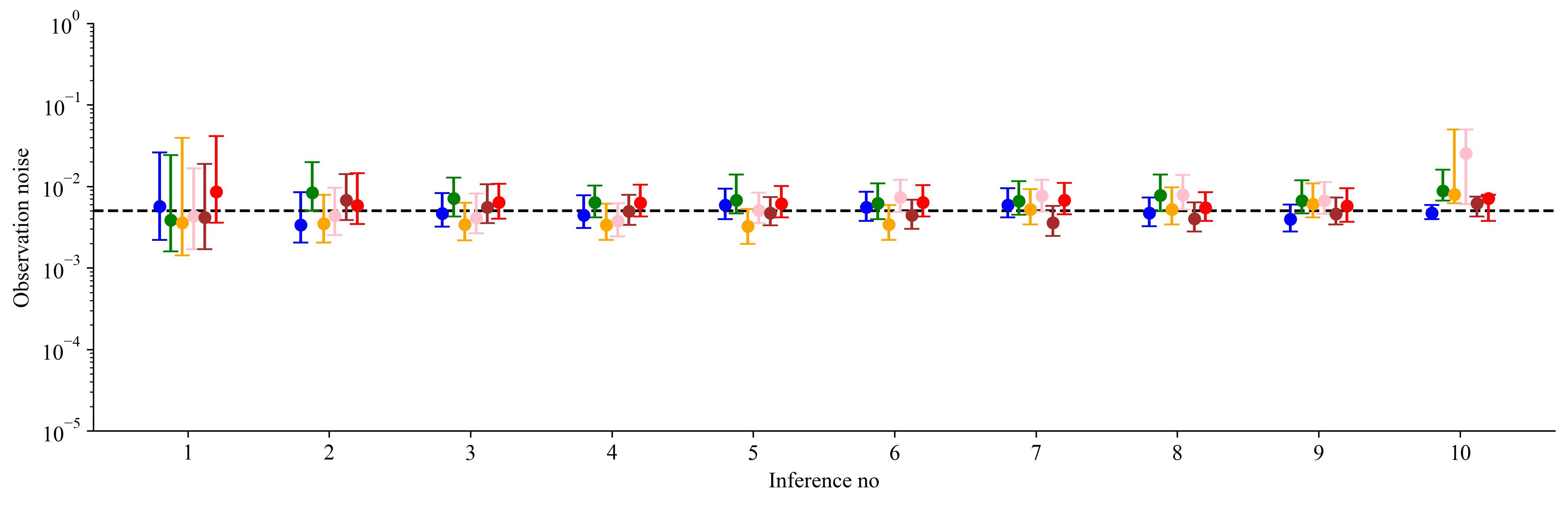}
  \caption{}
  \label{fig: synthetic data noise analy noise est 0.1 level 1}
\end{subfigure}
\begin{subfigure}{0.9\textwidth}
  \centering
  \includegraphics[width=1\linewidth]{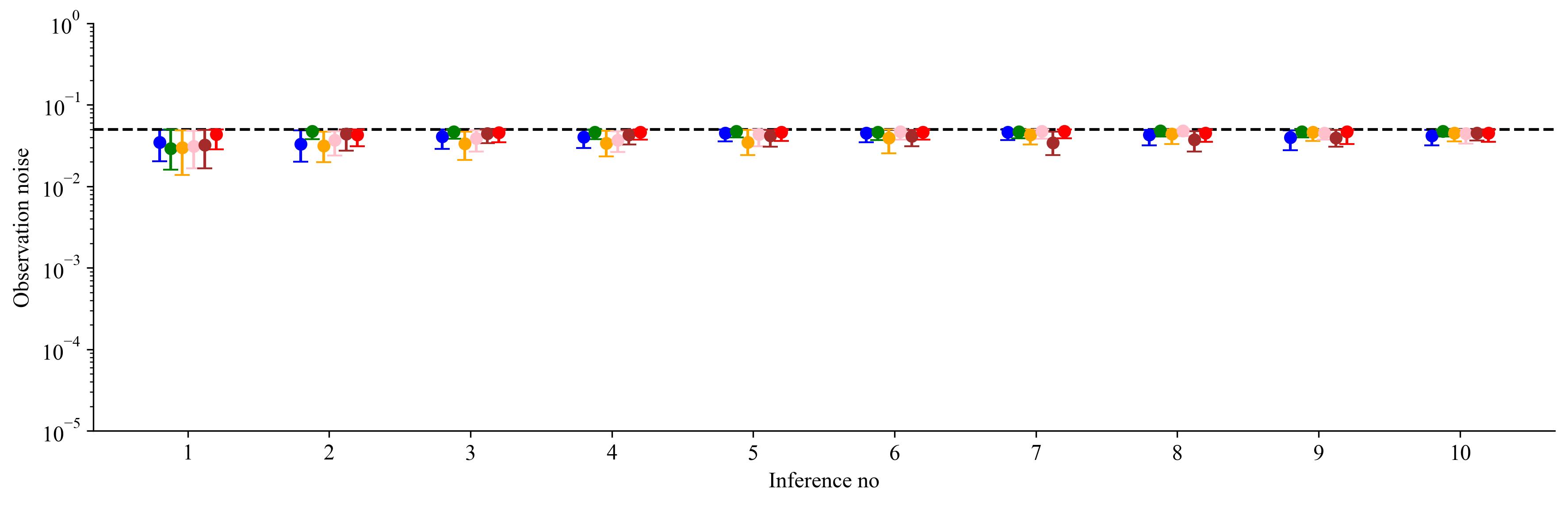}
  \caption{}
  \label{fig: synthetic data noise analy noise est 1 level 1}
\end{subfigure}
\begin{subfigure}{0.9\textwidth}
  \centering
  \includegraphics[width=1\linewidth]{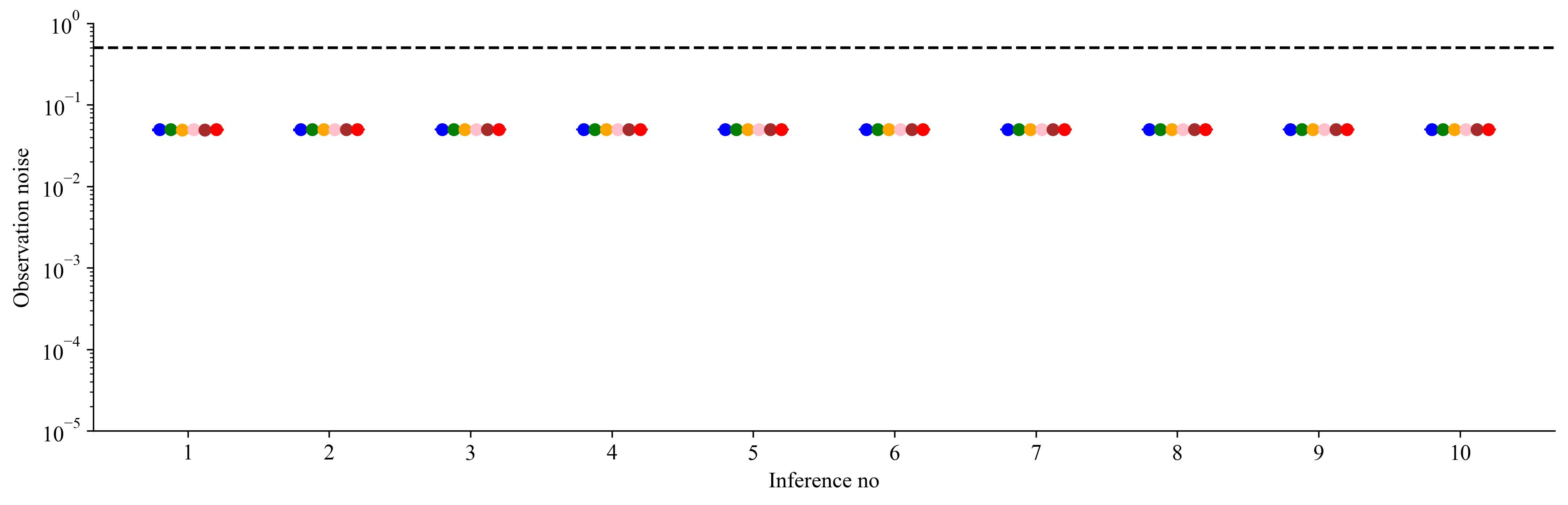}
  \caption{}
  \label{fig: synthetic data noise analy noise est 10 level 1}
\end{subfigure}
%\captionsetup{justification=centering}
\caption{Estimated observation-noise levels across inferences for level 1 observability in the synthetic noise study, shown as 95\% posterior credible intervals for different imposed observation-noise levels.}
\label{fig: synthetic egs noise estimations level 1}
\end{figure*} 
\clearpage
\section{Experimental Validation Plots}
\label{appendix: expt data plots}

\begin{figure}[ht!]
\centering
\includegraphics[width=0.88\linewidth]{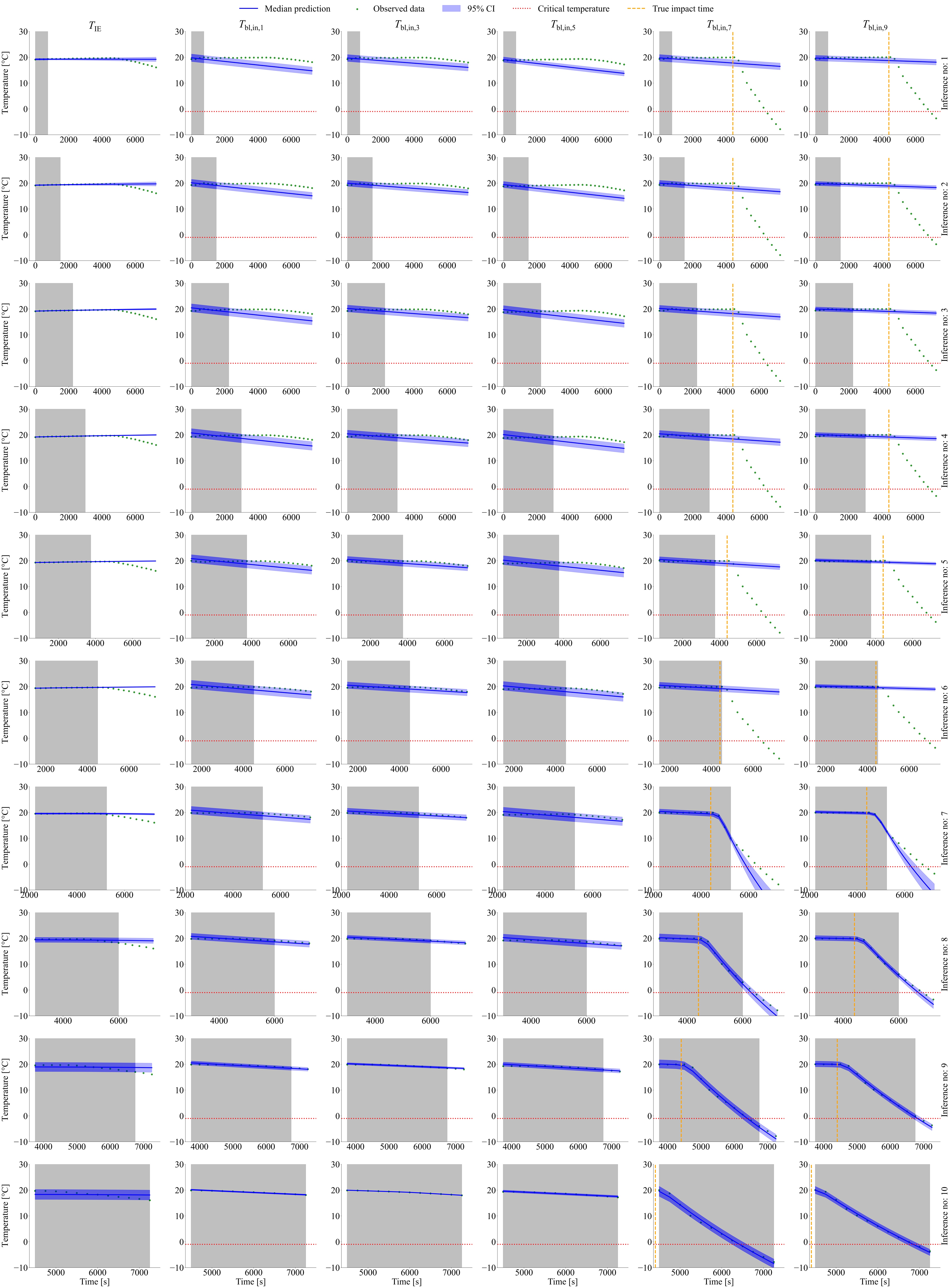}
\caption{Posterior predictive distributions of the observed temperatures across inferences for the experimental validation case. The grey-shaded region represents the portion of the data used for Bayesian inference. }
\label{fig: expt data posterior predictives}
\end{figure}
\clearpage
\begin{figure}[ht!]
\centering
\includegraphics[angle=270,width=0.95\linewidth]{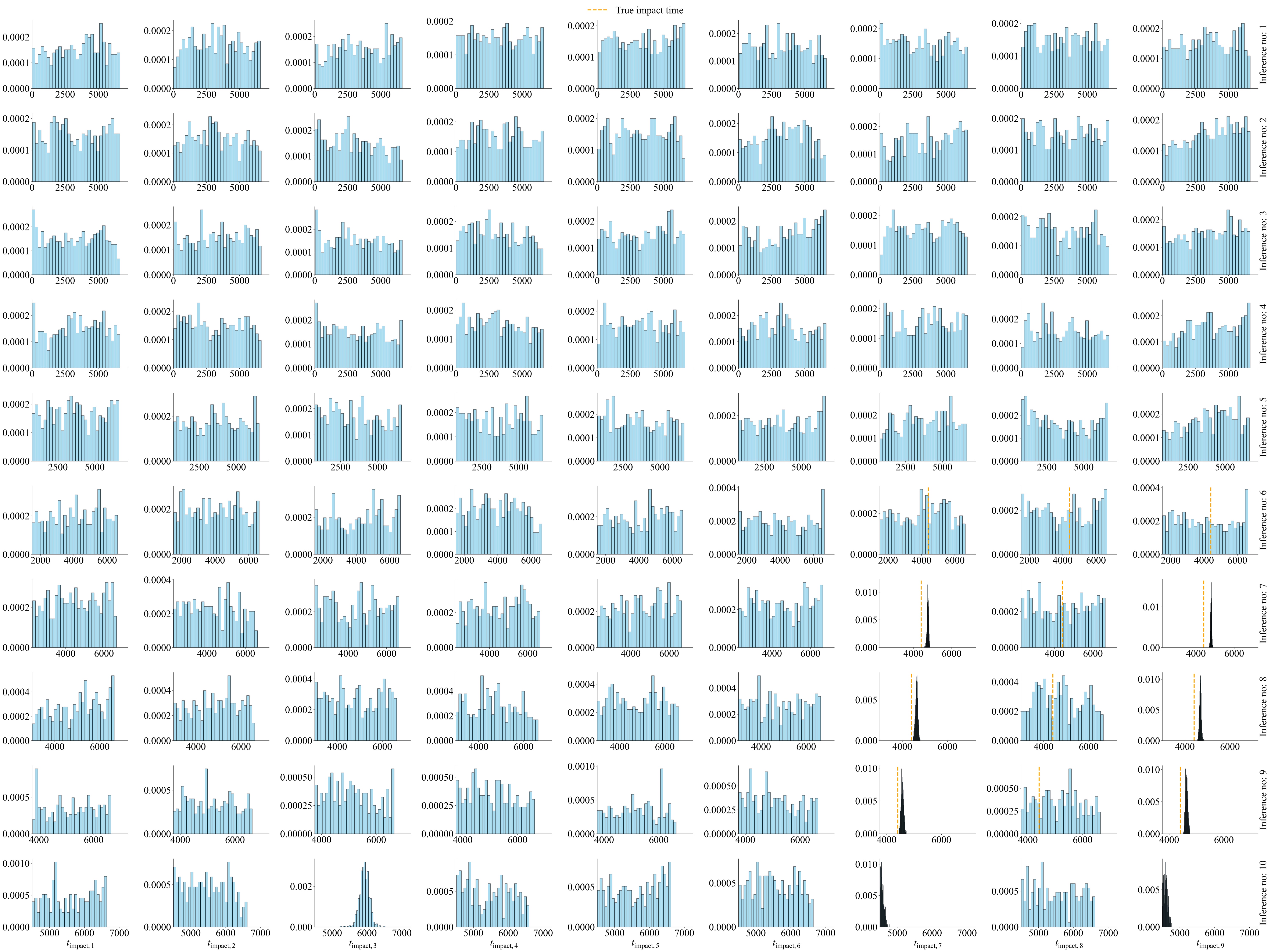}
\caption{Posterior distributions of the impact-time parameters \(t_{\text{impact,j}}\) across inferences for the experimental validation case.}
\label{fig: expt data t_impact_j posterior distributions}
\end{figure}
\clearpage
\begin{figure}[ht!]
\centering
\includegraphics[angle=270,width=0.95\linewidth]{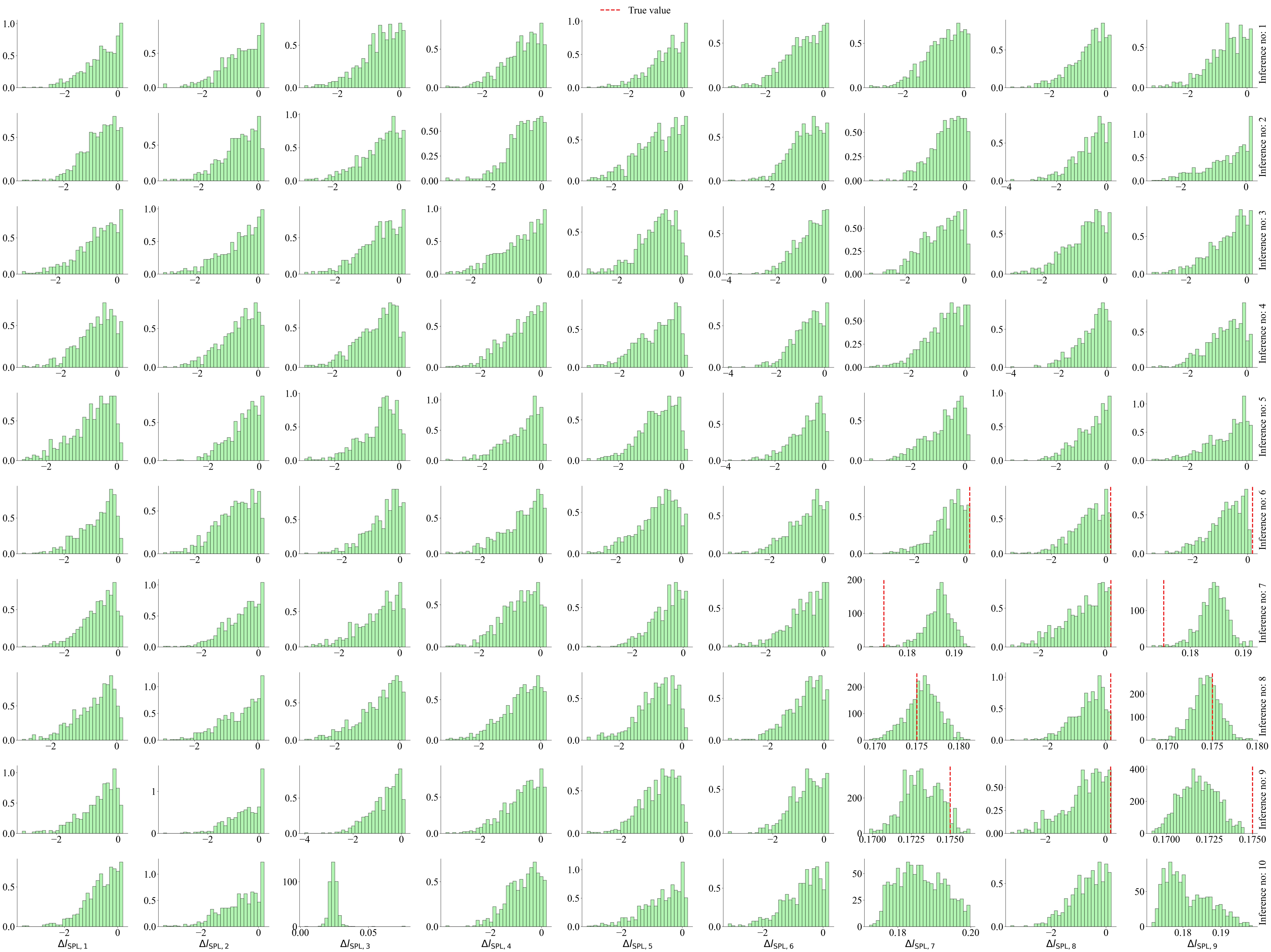}
\caption{Posterior distributions of the effective-thickness-reduction parameters \(\Delta l_{\text{SPL,j}}\) across inferences for the experimental validation case.}
\label{fig: expt data dl_reg_j posterior distributions}
\end{figure}

\begin{figure}[ht!]
\centering
\includegraphics[width=0.88\linewidth]{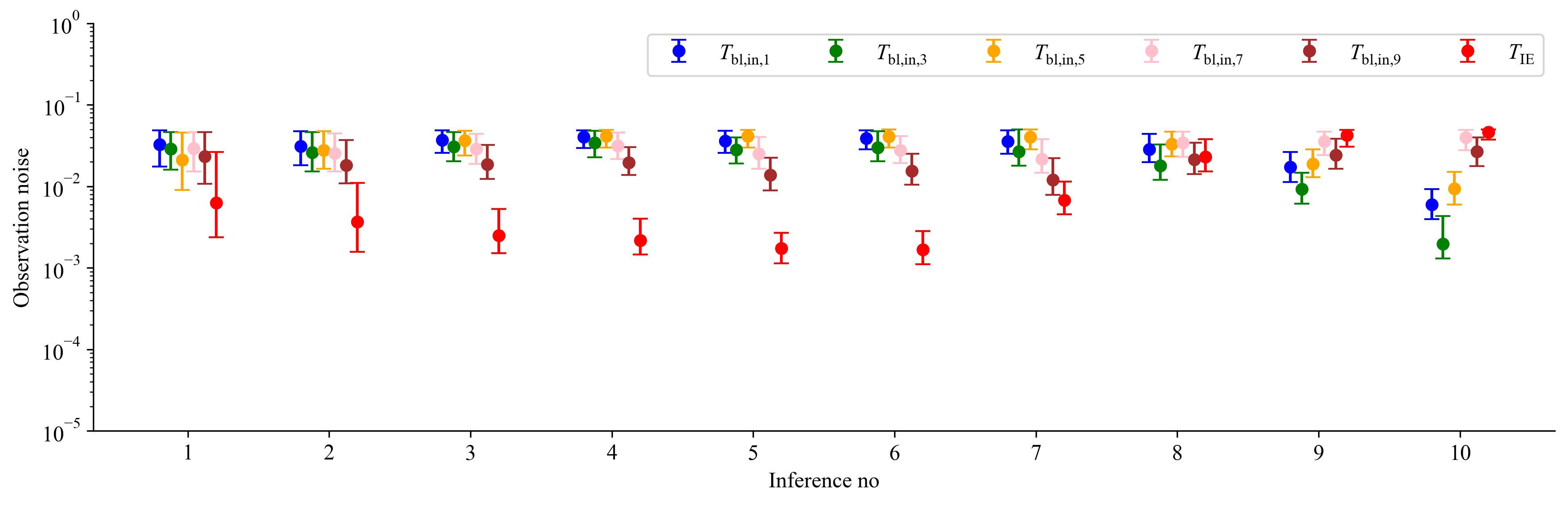}
\caption{Estimated observation-noise levels across inferences for the experimental validation case, shown as 95\% posterior credible intervals.}
\label{fig: expt data noise estimation levels}
\end{figure}

\end{appendices}

\end{document}